\documentclass[english,prb,preprint]{revtex4-1}
\pdfoutput=1
\usepackage[T1]{fontenc}
\usepackage[latin9]{inputenc}
\usepackage{color}
\usepackage{array}
\usepackage{float}
\usepackage{units}
\usepackage{multirow}
\usepackage{amsmath}
\usepackage{graphicx}

\makeatletter

\providecommand{\tabularnewline}{\\}

\setcitestyle{super}
\usepackage{babel}

\@ifundefined{showcaptionsetup}{}{%
 \PassOptionsToPackage{caption=false}{subfig}}
\usepackage{subfig}
\makeatother

\usepackage{babel}
\begin{document}
\title{A Pariser-Parr-Pople Model Based Study of Optoelectronic Properties
of Phenacenes}
\author{{\normalsize{}Deepak Kumar Rai}}
\email{dkriitb@gmail.com}

\author{Alok Shukla}
\email{shukla@phy.iitb.ac.in}

\affiliation{Department of Physics, Indian Institute of Technology Bombay, Powai,
Mumbai 400076, India}
\begin{abstract}
{\normalsize{}In this paper we present a computational study of linear
optical absorption in phenacene class of polyaromatic hydrocarbons.
For the purpose, we have employed a correlated-electron methodology
based upon configuration-interaction (CI) approach, and the Pariser-Parr-Pople
(PPP) $\pi$-electron model Hamiltonian. The molecules studied range
from the smallest one with three phenyl rings (phenanthrene), to the
largest one with nine phenyl rings. These structures can also be seen
as finite-sized hydrogen-passivated armchair graphene nanoribbons
of increasing lengths. Our CI calculations reveal that the electron-correlation
effects have a profound influence not just on the peak locations,
but also on the relative intensity profile of the computed spectra.
We also compare our phenacene results with isomeric oligo-acenes,
and find that in all the cases former have a wider optical gap than
the latter. Available experiments based upon optical absorption and
electron-energy-loss-spectroscopy (EELS) are in very good agreement
with our results. }{\normalsize\par}
\end{abstract}
\maketitle

\section{Introduction}

Over last several decades, polycyclic aromatic hydrocarbons (PAHs)
have been studied extensively using theoretical and experimental methods,
because of their importance in several fields such as physics, chemistry,
environmental science, astrophysics and biology.\textcolor{black}{\citep{clar1964polycyclic,dwek1997detection,bostrom2002cancer,Abdulazeez2017}
} PAHs, with a planar structure, are $\pi-$conjugated systems, known
for strong response to external fields, thus with potentially numerous
device applications.\textcolor{black}{\citep{okamoto2014extended,witte2004growth}
} Oligoacenes, particularly tetracene and pentacene are among the
most studied PAHs because of their possible applications in the field
of optoelectronic devices, particularly light-emitting diodes, and
photovoltaic cells.\textcolor{black}{\citep{cicoira2007organic,raghunath2006electronic}
} Phenacene oligomers, which are nothing but the isomers of oligo-acenes
of the same length, have also been found to be very useful in the
field of the device application, particularly in fabrication of organic
field-effect transistors (FETs).\citep{phenacene-review-yamashita,kubozono2014transistor,shimo2016synthesis,okamoto2013photochemical,okamoto2014extended,komura2012characteristics,9phenacene-fet}
Furthermore, it was reported that one of the phenacene oligomers,
namely picene, exhibits high-temperature superconducting behavior
when doped by alkali metals.\citep{picene-superconductivity} Phenacenes
are similar to oligoacenes in that both are composed of fused benzene
rings, while differing from each other in the way the fused rings
are arranged. In oligoacenes, the fused rings are arranged in a straight
manner leading to $D_{2h}$ symmetry, while in phenacenes, they are
arranged in a zigzag manner resulting in the point group is $C_{2h}$
if the number of rings is even, and $C_{2v}$ for odd number of rings.
A phenacene oligomer with $n$ fused rings is called {[}$n${]}phenacene,
and it is obvious that the minimum possible value for $n=3$. For
$3\leq n\leq6$, phenacene oligomers are named phenanthrene ($n=3$),
chrysene ($n=4$), picene ($n=5$), and fulminene ($n=6$), while
for $n>6$, they are referred as {[}$n${]}phenacene. 

Motivated by potential device applications of phenacenes, in this
work we undertake a systematic computational study of their electronic
structure, low-lying excited states, and linear optical response.
Such a study is necessary not just for understanding the optical properties
of individual oligomers, but also for obtaining insights into the
transport properties of these materials in the crystalline phase,
which consists of nothing but individual molecules held together by
van-der-Waals binding. Given the fact that the phenancenes are $\pi$-electron
systems, we have employed our Pariser-Parr-Pople model based electron-correlated
methodology for this study\citep{springer-chapter}. The oligomers
studied in this work range from phenanthrene to {[}9{]}phenacene,
and our results are found to be in excellent agreement with the experimental
measurements, wherever available. Because {[}$n${]}phenacene is isomeric
with acene-$n$, we also compare present results with the ones obtained
for polyacenes in an earlier work from our group,\citep{sony-acene-lo}
with the aim of understanding the role of geometry on the optical
properties of these materials. We find that for each value of $n$
considered in this work, optical absorption of spectra of two classes
of materials are qualitatively different, and that the optical gap
of {[}$n${]}phenacene is significantly larger than that of acene-$n$.
This suggests that the optical absorption spectroscopy can be used
to distinguish between isomeric phenancenes and acenes.

The remainder of the paper is structured as follows. In Section \ref{sec:Schematic_Diagram_PQDs}
we present schematic diagrams of phenacenes, and discuss their point
group symmetries, and related consequences. This is followed in Section
\ref{sec:theory} by a brief discussion of the theoretical methodology
adopted in this work. Next, in Section \ref{sec:results_PQDs} we
present and discuss the results of our calculations, followed by conclusions
and outlook in Section \ref{sec:conclusions_PQDS}.

\section{Molecular structure and point group symmetry }

\label{sec:Schematic_Diagram_PQDs}

In Fig. \ref{fig:Schematic-diagram-of PQDs}, we present the schematic
diagrams of {[}$n${]}phenacenes considered in this work, along with
their point group symmetries. We take the conjugation direction (long
axis) to be $x$ axis, and the perpendicular direction (short axis)
to be $y$ axis, so that all oligomers lie in the $x-y$ plane, with
a uniform C-C bond length of 1.4 \AA, and all the edge carbon atoms
are assumed to be passivated by hydrogen atoms. {[}$n${]}phenacene,
just like acene-$n$, has $4n+2$ carbon atoms, as also the same number
of $\pi-$electrons. The point group symmetry of phenanthrene ($14$
carbon atoms), picene ($22$ carbon atoms), {[}7{]}phenacene ($30$
carbon atoms), and {[}9{]}phenacene ($38$ carbon atoms) is $C_{2v}$,
with $1^{1}A_{1}$ being the ground state. On the other hand, the
point group symmetry of chrysene ($18$ carbon atoms) fulminene ($26$
carbon atoms) and {[}8{]}phenacene ($34$ carbon atoms) is $C_{2h}$,
with $1^{1}A_{g}$ being the ground state. As per electric-dipole
selection rules, the symmetries of the one-photon excited states are
$^{1}A_{1}$ ($y$ polarized) and $^{1}B_{1}$ ($x$ polarized) for
$C_{2v}$ molecules, and $^{1}B_{u}$ ($xy-$polarized) for $C_{2h}$
molecules. 
\begin{flushleft}
\begin{figure}[H]
~~~~~~~~~~~~~~~~~~~~~~\subfloat[Phenanthrene-($C_{2v}$) ]{\begin{raggedright}
\includegraphics[scale=0.105]{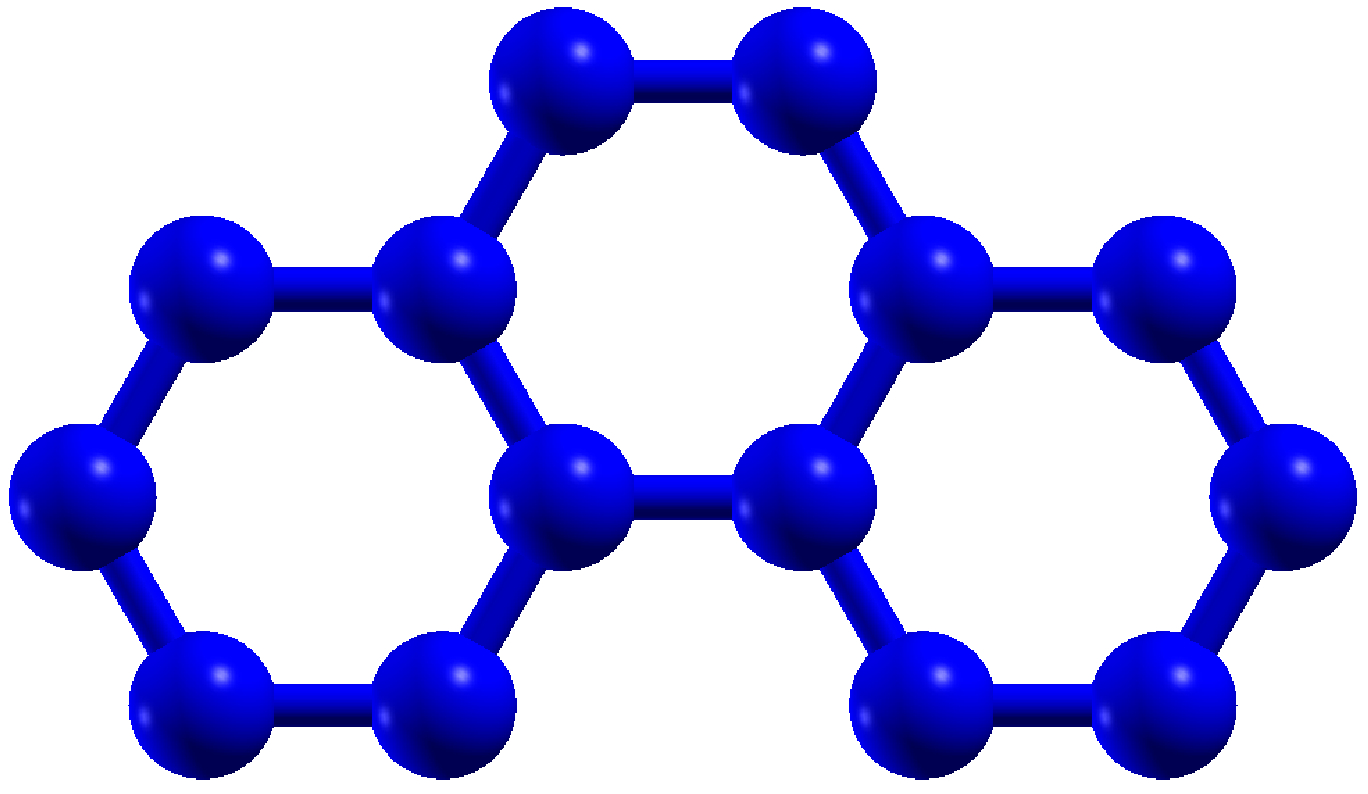}
\par\end{raggedright}
}~~~~~~~\subfloat[Chrysene-($C_{2h}$) ]{\includegraphics[scale=0.1]{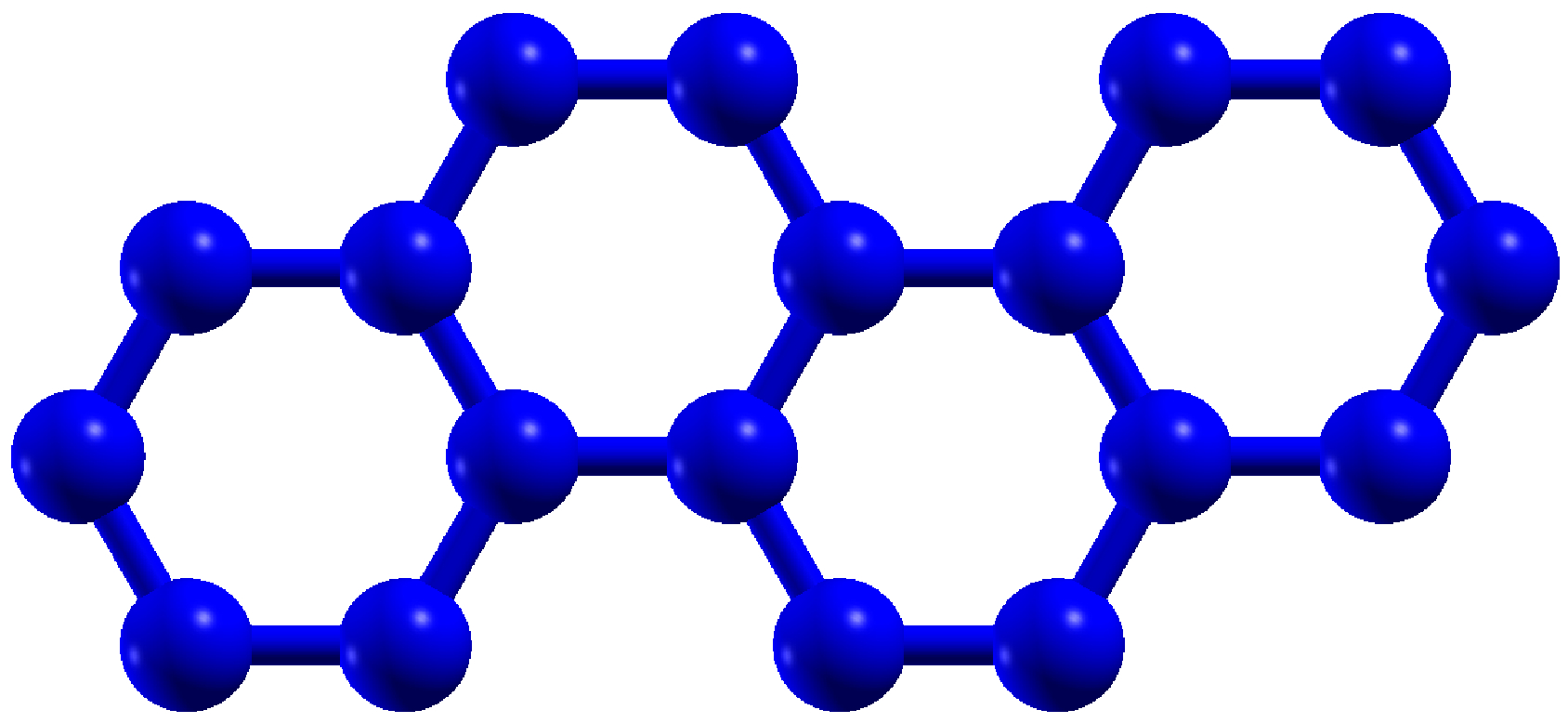}}~~~~~~~\subfloat[Picene-($C_{2v}$) ]{\includegraphics[scale=0.12]{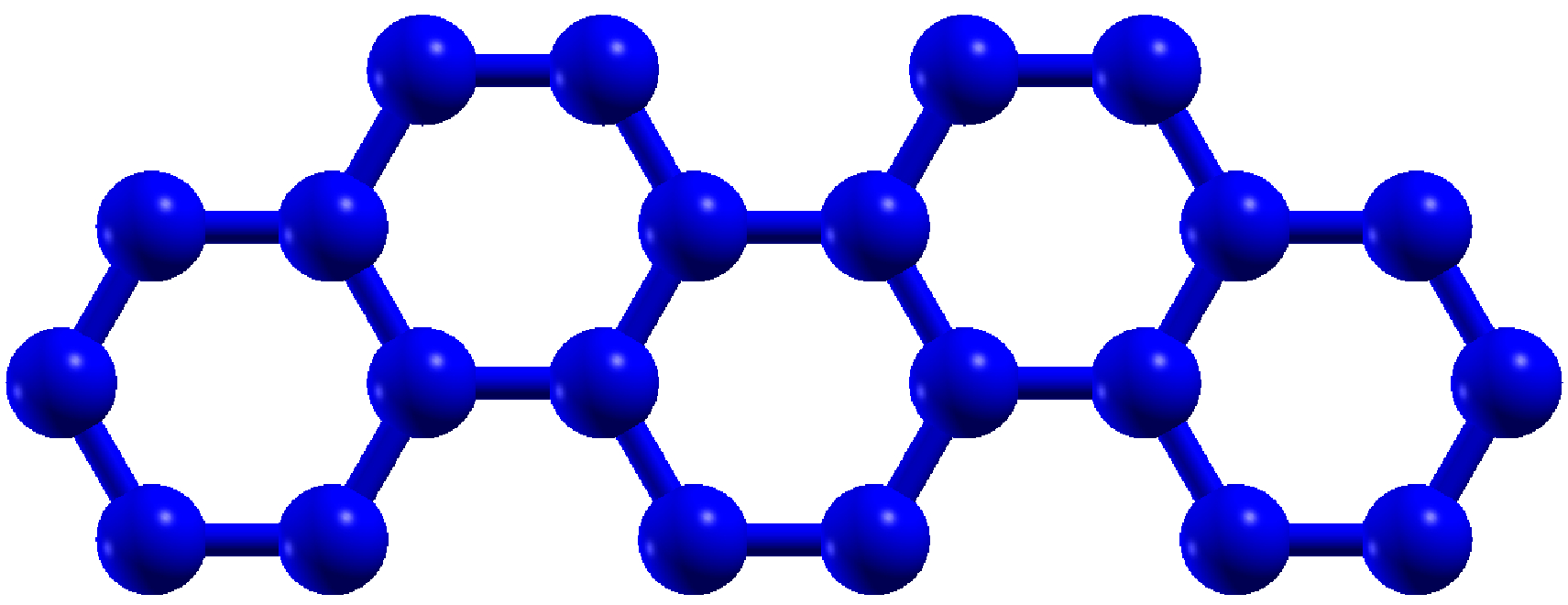}}

~~~~~~~\subfloat[Fulminene-($C_{2h}$) ]{\includegraphics[scale=0.145]{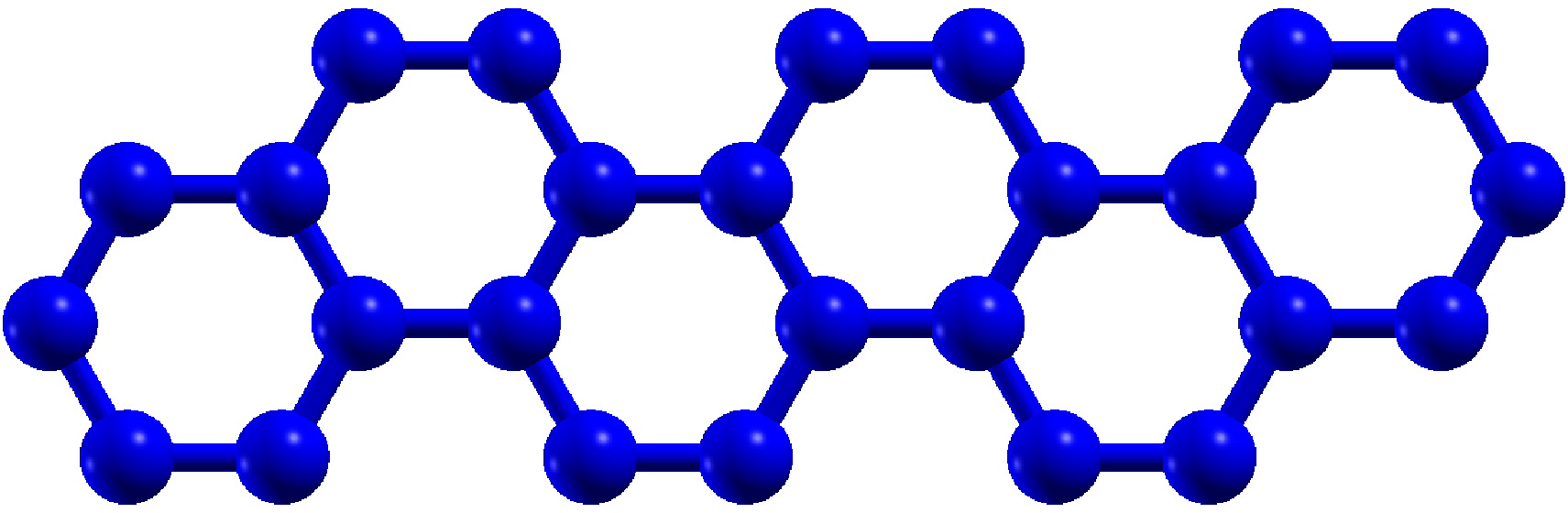}

}~~~~~~~~~~~~~~~~~\subfloat[{{[}7{]}Phenacene-($C_{2v}$) }]{\includegraphics[scale=0.18]{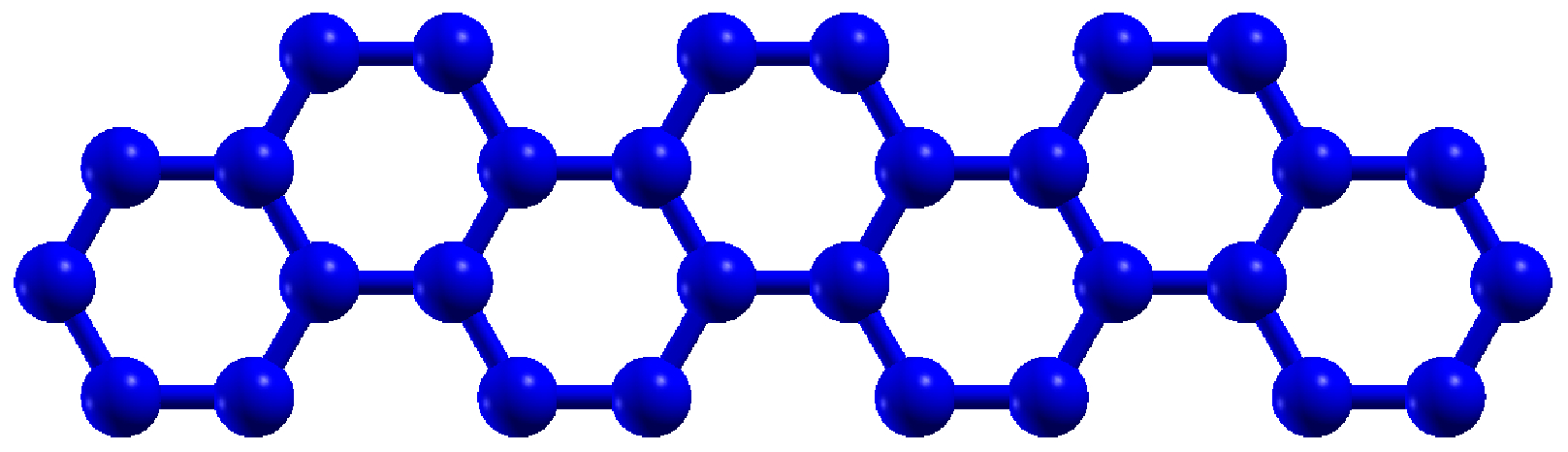}}

\subfloat[{{[}8{]}Phenacene-($C_{2h}$) }]{\includegraphics[scale=0.195]{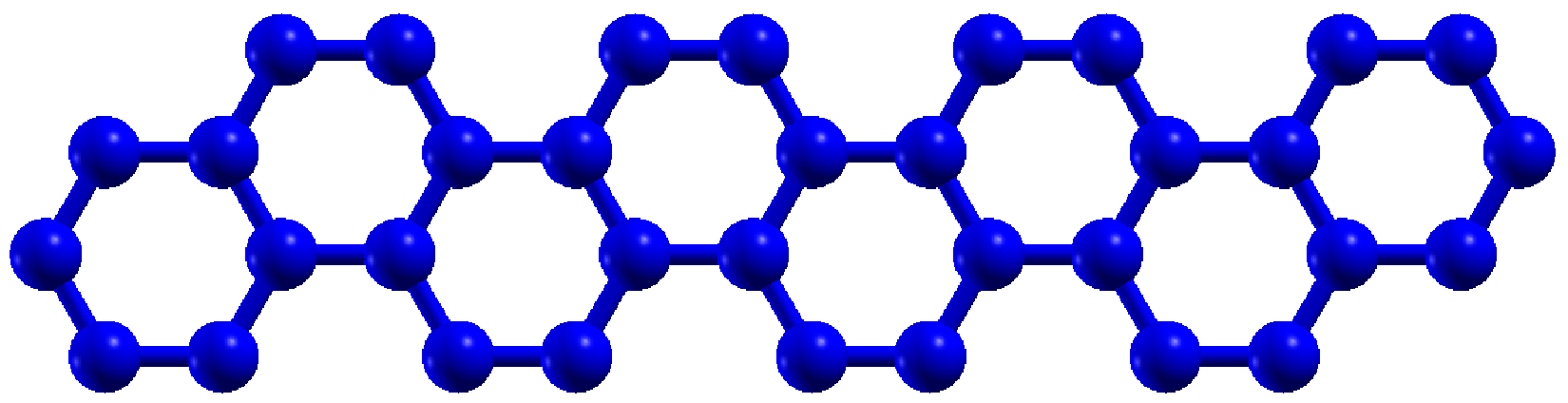}}\subfloat[{{[}9{]}Phenacene-($C_{2v}$) }]{\includegraphics[scale=0.215]{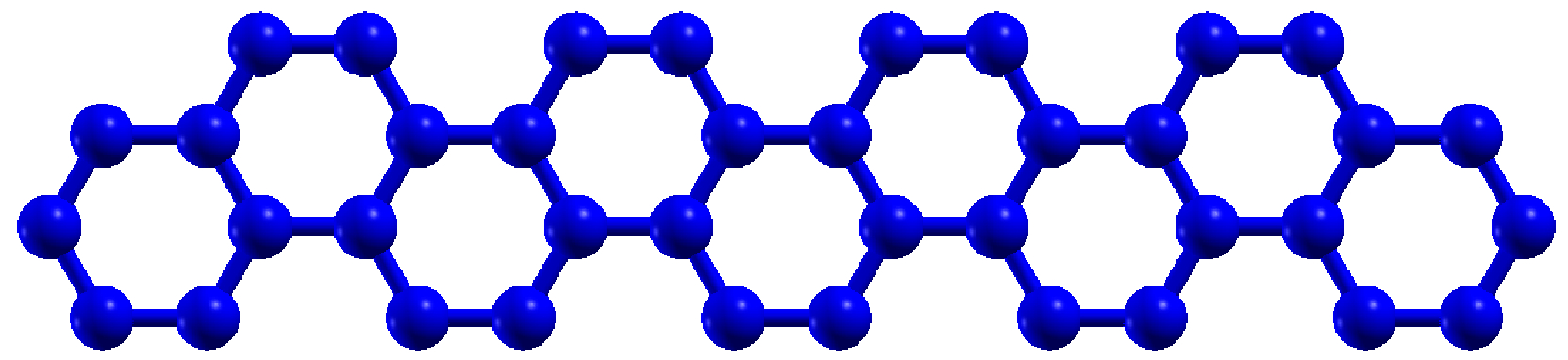}}\caption{\label{fig:Schematic-diagram-of PQDs}Schematic diagrams of phenacenes,
along with their point-group symmetries. }
\end{figure}
\par\end{flushleft}

\section{{\normalsize{}THEORETICAL METHODOLOGY}}

\label{sec:theory}

As described in the previous section, the molecules considered here
are $\pi$-conjugated systems, and, therefore, in this work we adopt
Pariser-Parr-Pople (PPP) model Hamiltonian,\citep{ppp-pople,ppp-pariser-parr}
reviewed in our earlier work\citep{springer-chapter} 

\begin{align}
H & \mbox{\ensuremath{=}\ensuremath{-}\ensuremath{\sum_{i,j,\sigma}t_{ij}\left(c_{i\sigma}^{\dagger}c_{j\sigma}+c_{j\sigma}^{\dagger}c_{i\sigma}\right)}\ensuremath{+}\ensuremath{U\sum_{i}n_{i\uparrow}n_{i\downarrow}}}\nonumber \\
\mbox{\mbox{\mbox{}}} & \mbox{+\ensuremath{\sum_{i<j}V_{ij}(n_{i}-1)(n_{j}-1),}}\label{eq:ppp}
\end{align}

where $c_{i\sigma}^{\dagger}($c$_{i\sigma})$ denotes creation (annihilation)
operators corresponding to an electron of spin $\sigma$ in a $p_{z}$
orbital, located on the $i$-th carbon atom, while the total number
of electrons on the atom is indicated by the number operator $n$$_{i}=\sum_{\sigma}c_{i\sigma}^{\dagger}c_{i\sigma}$.
In Eq. \ref{eq:ppp}, the first term denotes the one-electron hoppings
connecting $i$-th and $j$-th atoms, quantified by matrix elements
$t_{ij}$. It is assumed that the hopping connects only the nearest-neighbor
carbon atoms, with the matrix element $t_{0}=2.4$ eV, consistent
with our earlier calculations on conjugated polymers,\citep{PhysRevB.65.125204Shukla65,PhysRevB.69.165218Shukla69,PhysRevB.71.165204Priya_t0,:/content/aip/journal/jcp/131/1/10.1063/1.3159670Priyaanthracene,doi:10.1021/jp408535u,himanshu-triplet,sony-acene-lo}
polyaromatic hydrocarbons,\citep{doi:10.1021/jp410793rAryanpour,:/content/aip/journal/jcp/140/10/10.1063/1.4867363Aryanpour}
and graphene quantum dots.\citep{Tista1,Tista2} The remaining two
terms in Eq. \ref{eq:ppp} represent the electron-electron repulsion
terms, with the parameters $U$, and $V_{ij}$, denoting the on-site,
and the long-range Coulomb interactions, respectively. The distance-dependent
Coulomb parameters $V_{ij}$ are computed according to the Ohno relationship\citep{Theor.chim.act.2Ohno}

\begin{equation}
V_{ij}=U/\kappa_{i,j}(1+0.6117R_{i,j}^{2})^{\nicefrac{1}{2}},\label{eq:ohno}
\end{equation}

where $\kappa_{i,j}$ is the dielectric constant of the system, included
to take into account the screening effects, and $R_{i,j}$ is the
distance (in \AA) between the $i$th and $j$th carbon atoms. In
the present set of calculations we have used two sets of Coulomb parameters:
(a) the ``screened parameters''\citep{PhysRevB.55.1497Chandross}
with $U=8.0$ eV, $\kappa_{i,j}=2.0(i\neq j)$, and $\kappa_{i,i}=1.0$,
and (b) the ``standard parameters'' with $U=11.13$ eV and $\kappa_{i,j}=1.0$. 

The calculations are initiated by performing restricted Hartree-Fock
(RHF) calculations for the closed-shell singlet ground states of phenacenes
considered here, using a computer program developed in our group.\citep{Sony2010821}
The molecular orbitals (MOs) obtained from the RHF calculations form
a single-particle basis set used to transform the PPP Hamiltonian
from the site representation, to the MO representation. Subsequently,
correlated-electron calculations using the configuration interaction
(CI) approach are performed. The level of the CI calculations is decided
by the size of the molecule under consideration. For smaller molecules,
one can use full-CI (FCI) or quadruple-CI (QCI) approaches, however,
for the larger systems only the multi-reference singles-doubles configuration
interaction (MRSDCI) approach is tractable. In the MRSDCI calculatisons,
the CI expansion is generated by exciting up to two electrons, from
a chosen list of reference configurations, to the unoccupied MOs.\citep{peyerimhoff_energy_CI,buenker1978applicability}
The reference configurations included in the MRSDCI method depend
upon the states in consideration, which can be the ground state, or
optically excited states.\citep{PhysRevB.65.125204Shukla65,PhysRevB.69.165218Shukla69,PhysRevB.71.165204Priya_t0,:/content/aip/journal/jcp/131/1/10.1063/1.3159670Priyaanthracene,doi:10.1021/jp408535u,himanshu-triplet,sony-acene-lo,Tista1} 

Once the many-body wave functions and the energies of the ground and
the excited states are obtained from the CI calculations, we compute
the optical absorption cross-section $\sigma(\omega)$, according
to the formula 
\begin{equation}
\sigma(\omega)=4\pi\alpha\sum_{i}\frac{\omega_{i0}|\langle i|\boldsymbol{\hat{e.r}}|0\rangle|^{2}\gamma^{2}}{(\omega_{i0}-\omega)^{2}+\gamma^{2}}.\label{eq:sigma}
\end{equation}
In the equation above, $\boldsymbol{\hat{e}}$ represents the polarization
direction of the incident light, $\omega$ denotes its frequency,
$\boldsymbol{r}$ is the electronic position operator, indices $0$
and $i$ represent, respectively, the ground and excited states, $\omega_{i0}$
is the frequency difference between those states, $\alpha$ denotes
the fine structure constant, and $\gamma$ is the assumed universal
line width. The summation over $i$, in principle, is an infinite
sum, which, in practice, is restricted to those dipole-connected excited
states, whose excitation energies are within a certain cutoff, taken
to be 10 eV in these calculations.

\section{Results and Discussions }

\label{sec:results_PQDs}

In this section we present the calculated linear optical absorption
spectra and optical gaps of {[}$n${]}phenacenes, and compare our
results with the experiments, wherever available. Our calculations
were performed using both the tight-binding (TB) model, as well as
PPP model using the CI approach, and we find that our PPP-CI results
are in much better agreement with experiments.

\subsection{Tight-Binding Model Results}

Because the tight-binding (TB) model is an independent electron approach,
therefore, results obtained using it will help us understand the influence
of electron correlation effects, when compared with the results computed
using the PPP-model. We first present and discuss the optical absorption
spectra obtained using the TB-model, followed by a discussion of the
optical gaps. 

\subsubsection{Linear Optical absorption spectra}

\begin{figure}[H]
\begin{centering}
\includegraphics[scale=0.5]{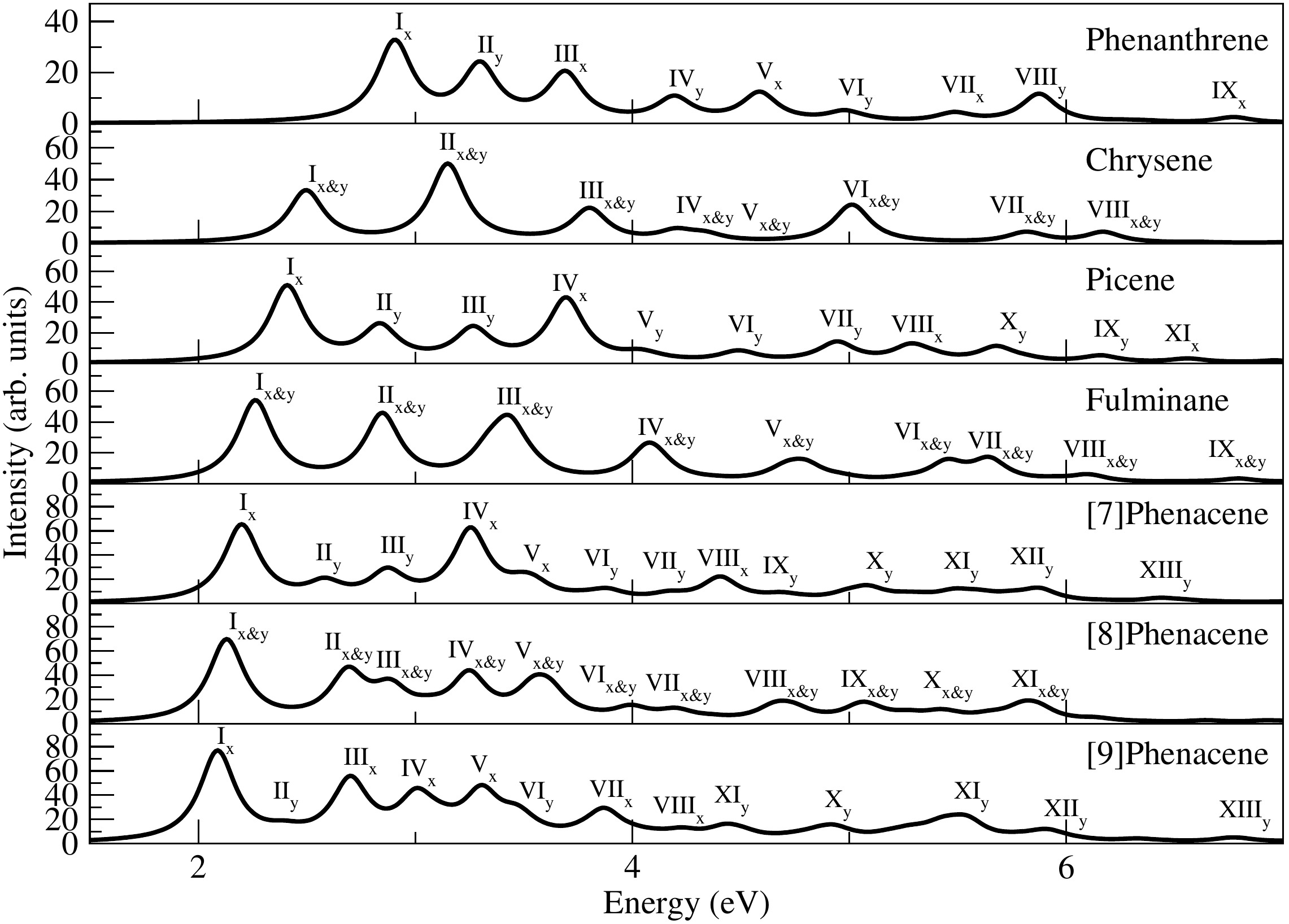}\protect
\par\end{centering}
\centering{}\caption{{\scriptsize{}\label{Computed-linear-opt_abs_ spectra_PQDS-TB}} Optical
absorption spectra of {[}$n${]}phenacenes ($n=3-9$), computed using
the TB model. The spectra have been broadened using a uniform line-width
of $0.1$ eV.}
\end{figure}

In Fig. \ref{Computed-linear-opt_abs_ spectra_PQDS-TB}, we present
the optical absorption spectra of {[}$n${]}phenacenes obtained using
the TB method. An examination of the spectra reveals the following
trends: (a) With the increasing lengths of the oligomers, absorption
spectra are red shifted, consistent with the phenomenon of the quantum
confinement effect. (b) The first peak for all the oligomers corresponds
to the excitation of an electron from HOMO ($H$) to LUMO ($L$).
It corresponds to transition to $1^{1}B_{1}$ state via absorption
of a photon polarized along the length ($x$-direction) of the $C_{2v}$
symmetric oligomers. For $C_{2h}$ symmetry, the first peak is due
to $1^{1}B_{u}$ state, reached via absorption of an $xy-$polarized
photon, with the $x-$component much stronger than the $y-$component.
(c) The maximum intensity peak is the first peak for all the oligomers,
except for chrysene for which the second peak is the most intense
one, and it is due to $\left|H\rightarrow L+1\right\rangle $$+c.c.$
excitation, where $c.c.$ denotes the corresponding charge-conjugated
configuration. 

\subsubsection{Optical Gap}

The locations of the first absorption peaks of {[}$n${]}phenancenes,
i.e. their optical gaps, computed using various approaches are presented
in Table \ref{tab:Comparison-of-optical-gap}, where they are also
compared to the experimental results. The following conclusions can
be drawn from this table: (a) Independent of the Hamiltonian, gaps
decrease with the increasing length of {[}$n${]}phenacene, (b) the
gaps obtained for {[}$n${]}phenacenes using the TB method are much
smaller compared to the corresponding experimental, as well as PPP-CI
values, (c) the gaps obtained using the PPP-CI method, employing screened
parameters, are in very good agreement with the experimental values. 

\begin{table}[H]
\begin{centering}
\caption{\label{tab:Comparison-of-optical-gap}Locations of the first absorption
peaks for increasing length of {[}$n${]}phenacenes, obtained using
the TB and the PPP-CI approaches. For the PPP-CI calculations, both
the screened (Scr.) and the standard (Std.) parameters are employed.}
\par\end{centering}
\centering{}%
\begin{tabular}{|c|c|c|c|c|c|}
\hline 
\multirow{3}{*}{System} & \multicolumn{5}{c|}{Optical gap (eV)}\tabularnewline
\cline{2-6} 
 & \multirow{2}{*}{TB Model} & \multicolumn{2}{c|}{PPP-CI} & \multirow{2}{*}{Experimental } & \multirow{2}{*}{Theory (other authors) }\tabularnewline
\cline{3-4} 
 &  & Scr. & Std. &  & \tabularnewline
\hline 
\multirow{4}{*}{Phenanthrene} & \multirow{4}{*}{2.90} & \multirow{4}{*}{4.31} & \multirow{4}{*}{4.26} & 4.09\citep{klevens1949spectral}, 4.17\citep{okamoto2013photochemical},  & 3.91\citep{parac2003tddft}, 4.19 \citep{malloci2011electronic}, \tabularnewline
 &  &  &  & 4.24\citep{clar1964polycyclic}, 4.24\citep{birks1970photophysics}, & 4.31\citep{parac2003tddft}, 4.34\citep{parac2003tddft}, \tabularnewline
 &  &  &  & 4.25\citep{salama1993neutral}, 4.36 \citep{halasinski2005investigation}, & 4.36 \citep{skancke1965semi}, 4.53\citep{malloci2004electronic}, \tabularnewline
 &  &  &  &  & 4.60\citep{hedges1968electronic}, 4.67 \citep{malloci2011electronic}, \tabularnewline
\hline 
\multirow{5}{*}{Chrysene} & \multirow{5}{*}{2.49} & \multirow{5}{*}{3.86} & \multirow{5}{*}{3.96} & 3.74\citep{klevens1949spectral}, 3.84\citep{okamoto2013photochemical}, & 3.40 \citep{parac2003tddft}, 3.43\citep{malloci2004electronic}, \tabularnewline
 &  &  &  &  3.87\citep{becker1963comprehensive}, 3.89\citep{clar1964polycyclic},  & 3.73\citep{malloci2011electronic}, 3.82\citep{parac2003tddft}, \tabularnewline
 &  &  &  & 3.89\citep{birks1970photophysics,mallory1997phenacenes},  & 3.90\citep{ham1956electronic}, 3.92\citep{parac2003tddft}, \tabularnewline
 &  &  &  &  & 4.13\citep{skancke1965semi}, 4.21\citep{malloci2011electronic}, \tabularnewline
 &  &  &  &  & 4.22\citep{hedges1968electronic} \tabularnewline
\hline 
\multirow{3}{*}{Picene} & \multirow{3}{*}{2.40} & \multirow{3}{*}{3.75} & \multirow{3}{*}{3.88} & 3.76\citep{okamoto2013photochemical,mallory1997phenacenes}, 3.77\citep{clar1964polycyclic}, & 3.32\citep{parac2003tddft}, 3.70 \citep{malloci2011electronic}, \tabularnewline
 &  &  &  & 3.80\citep{birks1970photophysics}, 3.82\citep{fanetti2012homo}, & 3.72\citep{parac2003tddft}, 3.83\citep{parac2003tddft}, \tabularnewline
 &  &  &  &  & 4.13\citep{hedges1968electronic}, 4.19\citep{malloci2011electronic}, \tabularnewline
\hline 
\multirow{2}{*}{Fulminene} & \multirow{2}{*}{2.26} & \multirow{2}{*}{3.34} & \multirow{2}{*}{3.52} & 3.14\citep{okamoto2014extended}, 3.24\citep{okamoto2013photochemical},  & \multirow{2}{*}{3.47\citep{malloci2011electronic}, 4\citep{malloci2011electronic}, }\tabularnewline
 &  &  &  & 3.76\citep{mallory1997phenacenes}, & \tabularnewline
\hline 
{[}7{]}Phenacene & 2.19 & 3.34 & 3.68 & 3.10\citep{okamoto2014extended}, 3.60\citep{mallory1997phenacenes},  & 3.50\citep{malloci2011electronic}, 4\citep{malloci2011electronic}, \tabularnewline
\hline 
{[}8{]}Phenacene & 2.13 & 3.11 & 3.41 & 3.08\citep{okamoto2014extended} & \textbf{---}\tabularnewline
\hline 
{[}9{]}Phenacene & 2.08 & 3.09 & 3.46 & 3.05\citep{shimo2016synthesis},  & \textbf{---}\tabularnewline
\hline 
\end{tabular}
\end{table}

\subsection{PPP Model Based Optical absorption Spectra}

In this section we present the results of our calculations of optical
absorption spectra of {[}$n${]}phenacenes, performed using the PPP-CI
approach. Before we discuss our results, we present the dimensions
of the CI matrices involved in the calculations, for various irreducible
representations of phenancenes. 

\subsubsection{Dimensions of the CI matrices }

Most accurate results within the CI approach are obtained when the
full-CI (FCI) calculations are performed, which involves distributing
all electrons, in all available molecule orbitals, in all possible
ways, consistent with the symmetries. Therefore, the size of the FCI
matrix increases exponentially with the increasing size of the molecule
involved, making it feasible only for small molecules. Thus, in this
work we were able to perform FCI calculations only on the smallest
oligomer, namely phenanthrene. For chrysene and picene we were able
to perform QCI calculations. For remaning oligomers, because of their
larger sizes, we employed the MRSDCI approach. Even within the truncated
CI approaches such as the QCI and the MRSDCI methods, larger-sized
CI expansions normally lead to more accurate results. Therefore, in
the section, to illustrate the accuracy of our calculations, we present
the dimensions of the MRSDCI matrices in Table \ref{size_of_matrix-phenacene}. 

\begin{table}[H]
\caption{\label{size_of_matrix-phenacene}Dimension of CI matrices ($N_{total}$)
involved in the calculations of the optical absorption spectra, for
different symmetry subspaces of {[}$n${]}phenacenes.}

\centering{}%
\begin{tabular}{|c|c|c|c|c|}
\hline 
\multirow{2}{*}{Molecules} & \multicolumn{4}{c|}{$N_{total}$}\tabularnewline
\cline{2-5} 
 & $^{1}A_{1}$  & $^{1}B_{1}$  & $^{1}A_{g}$  & $^{1}B_{u}$ \tabularnewline
\hline 
Phenanthrene & $1244504^{a}$ & $1239406^{a}$ & - & -\tabularnewline
\hline 
Chrysene & - & - & $386498^{b}$ & $670593^{b}$\tabularnewline
\hline 
Picene & $2003907^{b}$ & $3416371^{b}$ & - & -\tabularnewline
\hline 
\multirow{2}{*}{Fulminene} & - & - & $7877992^{b}$ & $1053746^{c}$\tabularnewline
\cline{2-5} 
 & - & - & $7877992^{b}$ & $1300041^{d}$\tabularnewline
\hline 
\multirow{2}{*}{{[}7{]}Phenacene} & $2303318^{c}$ & $5720562^{c}$ & - & -\tabularnewline
\cline{2-5} 
 & $3043013^{d}$ & $2645512^{d}$ & - & -\tabularnewline
\hline 
\multirow{2}{*}{{[}8{]}Phenacene} &  &  & $377903^{c}$ & $3300974^{c}$\tabularnewline
\cline{2-5} 
 &  &  & $628766^{b}$ & $5223159^{d}$\tabularnewline
\hline 
\multirow{2}{*}{{[}9{]}Phenacene} & $2617790^{c}$ & $3056846^{c}$ &  & \tabularnewline
\cline{2-5} 
 & $4672916^{d}$ & $5876058^{d}$ &  & \tabularnewline
\hline 
\multicolumn{5}{|c|}{FCI$^{a}$ method with screened and standard parameters. QCI$^{b}$
method with}\tabularnewline
\multicolumn{5}{|c|}{screened and standard parameters. MRSDCI$^{c}$ method with screened }\tabularnewline
\multicolumn{5}{|c|}{parameters. MRSDCI$^{d}$ method with standard parameters.}\tabularnewline
\hline 
\end{tabular}
\end{table}
It is obvious from the table that in various calculations $N_{total}$
ranges from $3.6\times10^{5}$ to $7.8\times10^{6}$. This implies
that our calculations employ large CI expansions, and, therefore,
should be fairly accurate, yielding reliable results.

\subsubsection{Optical absorption spectra }

\begin{figure}
\subfloat[screened parameters]{\begin{centering}
\includegraphics[scale=0.5]{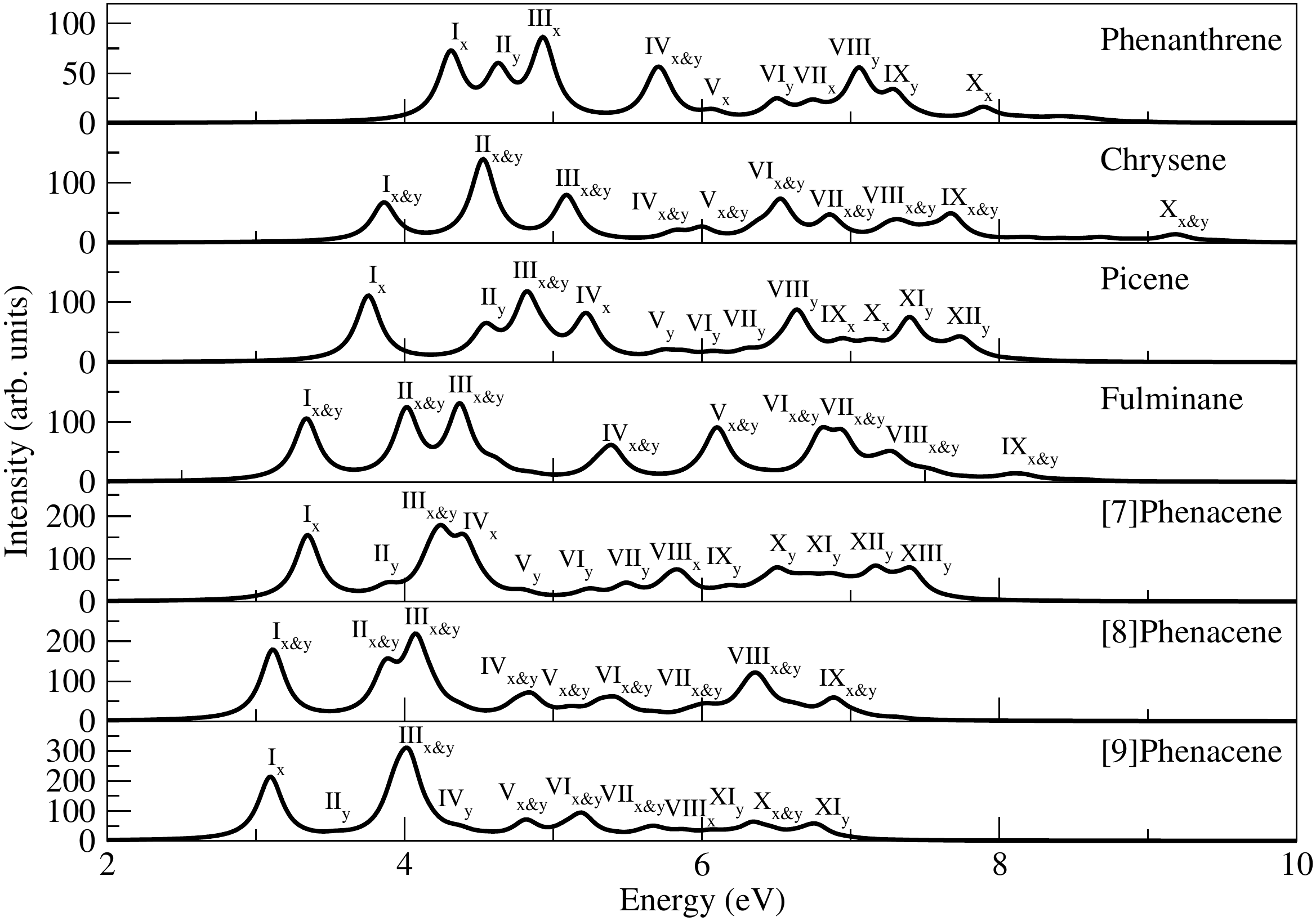}
\par\end{centering}

}

\subfloat[standard parameters]{\begin{centering}
\includegraphics[scale=0.5]{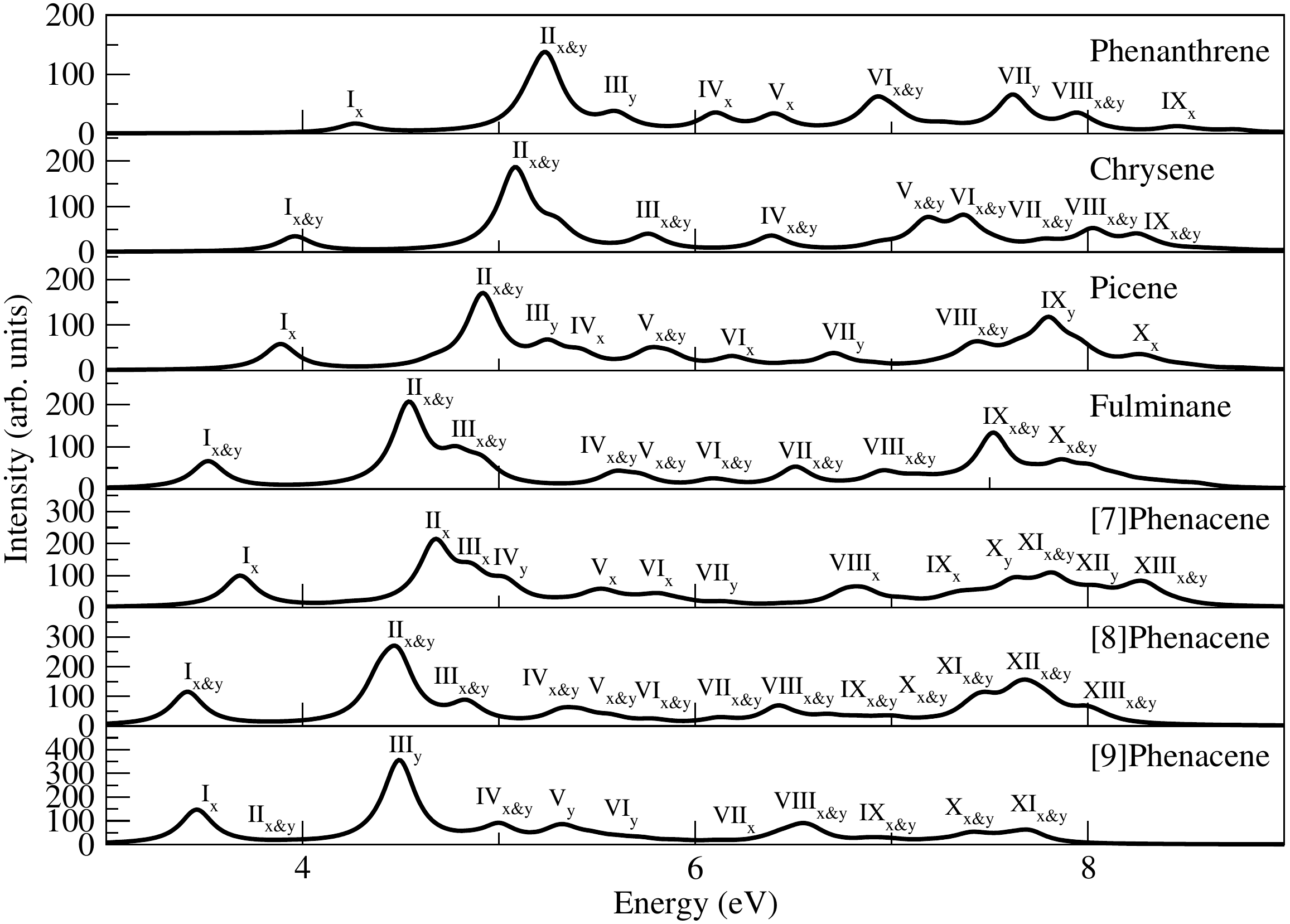}
\par\end{centering}

}

\caption{{\scriptsize{}\label{fig:Computed-linear-opt_abs_ spectra_PQDS-PPP-CI}}Calculated
optical absorption spectra of {[}$n${]}phenacenes, using the PPP-CI
method, by employing screened, as well as standard parameters. The
spectra have been broadened with a uniform line-width of $0.1$ eV.}

\end{figure}

First we discuss the general trends observed in the optical absorption
spectra of {[}$n${]}phenacenes calculated using the PPP-model, and
the MRSDCI approach, presented in Fig. \ref{fig:Computed-linear-opt_abs_ spectra_PQDS-PPP-CI}. 

Detailed information related to peaks contributing to the spectra,
such as the many-particle wave functions, energies, and transition
dipole moment etc. are presented in Tables S1-S14 of Supporting Information.
By carefully examining the optical absorption spectra (Fig. \ref{fig:Computed-linear-opt_abs_ spectra_PQDS-PPP-CI}),
we observed the following trends: (a) Similar to the case of TB model,
with the increasing length of {[}$n${]}phenacene, spectra is red-shifted
for the screened parameter calculations. For the standard parameters,
again the spectra are red-shifted, except for the case of {[}7{]}phenacene.
(b) the lowest energy absorption in {[}$n${]}phenacenes corresponds
to a dipole forbidden state of symmetry $^{1}A_{1}$ for $C_{2v}$
molecules, and $^{1}B_{u}$ for $C_{2h}$ molecules. The wave functions
of the corresponding state is dominated by single excitations $\left|H\rightarrow L+1\right\rangle $$+c.c.$
for oligomers up to {[}7{]}phenacene, and $\left|H\rightarrow L+2\right\rangle $$+c.c.$,
for longer oligomers. (c) In all the calculations, the first dipole-allowed
peak corresponding to the optical gap is not the most intense one
of the spectrum, in contradiction with the TB model results, and in
agreement with the experiments. In the screened parameter calculations,
the relative intensity of the first peak, as compared to the most
intense peak, is much larger than that in the standard parameter calculations.
In agreement with the results of TB model, the first peak for the
$C_{2v}$ symmetric molecules corresponds to $1^{1}B_{1}$ state,
and for $C_{2h}$ symmetric molecules to $1^{1}B_{u}$ state, and
in all the cases the dominating configuration in the many-particle
wave function of this state is $|H\rightarrow L\rangle$, single excitation.
(d) The maximum intensity peak occurs at higher energies, and the
wave functions of the two excited states contributing to it are dominated
by single excitations: (i) $|H-1\rightarrow L+1\rangle$ or $|H-2\rightarrow L+2\rangle$,
and (ii) $\left|H\rightarrow L+1\right\rangle $$+c.c.$ or $\left|H\rightarrow L+2\right\rangle $$+c.c.$,
which are the same excitations which contribute to the D.F. state.
Next we discuss the optical absorption spectra of the individual phenacenes
in detail. 

\subsubsection*{Phenanthrene}

Phenanthrene has $C_{2v}$ symmetry, and Klevens $et$ $al.$\citep{klevens1949spectral},
Okamoto $et$ $al.$\citep{okamoto2013photochemical}, Clar $et$
$al.$\citep{clar1964polycyclic}, Salama $et$ $al.$ \citep{salama1993neutral}
and Halasinski $et$ $al.$ \citep{halasinski2005investigation} have
reported the measurements of its absorption spectrum. In Fig. \ref{fig:Computed-linear-opt_abs_ spectra_PQDS-PPP-CI}
(a) and (b), we present our calculated spectra using the screened
and standard parameters, respectively, within the PPP-CI approach,
and in Table \ref{tab:comparison-Phenanthrene}, we have compared
our results on the locations of various peaks with the experiments,
and other theoretical results. Several experimentalist have measured
the first absorption peak in phenanthrene to be a very weak one, due
to a dipole forbidden (D.F.) state\citep{klevens1949spectral,okamoto2013photochemical,clar1964polycyclic,salama1993neutral}.
In particular Klevens $et$ $al.$\citep{klevens1949spectral}, Okomoto
$et$ $al.$\citep{okamoto2013photochemical}, Clar $et$ $al.$\citep{clar1964polycyclic}
and Salama $et$ $al.$\citep{salama1993neutral} measured its locations
at 3.50 eV, 3.57 eV, 3.59 eV and 3.61 eV, respectively. Our standard
parameter value of the D.F. state at 3.35 eV is closer to the experimental
values than the screened parameter value computed to be 2.96 eV. Both
our screened and standard parameter calculations predict this D.F.
state to be of $^{1}A_{1}$ symmetry, in agreement with the work of
Skancke \citep{skancke1965semi}. Our PPP-CI calculations predict
that the first dipole allowed state is not the most intense one, in
contradiction with the TB results. We would like to point out that
this PPP-CI result is in perfect agreement with the experimental measurement
of Klevens $et$ $al.$\citep{klevens1949spectral}, Okamoto $et$
$al.$\citep{okamoto2013photochemical} and Clar $et$ $al.$\citep{clar1964polycyclic}.
The location of the first dipole allowed peak, which is also the optical
gap, was calculated to be 4.31 eV with screened parameters, and $4.26$
eV with standard parameters. These values are in good agreement with
the experimental values which are measured in the range 4.09--4.36
eV (see Table \ref{tab:comparison-Phenanthrene}). In particular,
our results are in excellent agreement with the values 4.24 eV, 4.25
eV, and 4.36 eV, reported by Clar $et$ $al.$\citep{clar1964polycyclic},
Salama $et$ $al.$\citep{salama1993neutral}, and Halasinski $et$
$al.$ \citep{halasinski2005investigation}, respectively. 

As far as higher energy peaks are concerned, our screened parameter
spectra has a peak at 4.62 eV which is in excellent agreement with
the reported value 4.64 eV by Halasinski $et$ $al.$\citep{halasinski2005investigation}.
The next peak is the most intense (MI) peak in our calculated spectrum
for both screened as well as standard parameter calculations, and
it is located at 4.93 eV and 5.19 eV, respectively. Clar $et$ $al.$\citep{clar1964polycyclic}
experimentally measured the most intense peak at 4.93 eV which exactly
matches with our screened parameter value. The reported value of MI
peak by Klevens $et$ $al.$ \citep{klevens1949spectral} (4.88 eV)
and Salama $et$ $al.$ \citep{salama1993neutral} (4.80 eV) are also
very close with our obtained screened parameter value, while the MI
peak obtained using standard parameter is little bit on the higher
side compared to the experimental results. 

After that Halasinski $et$ $al$.\citep{halasinski2005investigation}
report a peak at 5.78 eV, while Clar $et$ $al.$\citep{clar1964polycyclic}
report one at 5.65 eV. Our screened parameter peak at 5.71 eV is in
excellent agreement with Halasinski $et$ $al$.\citep{halasinski2005investigation},
while the standard parameter peak at 5.59 eV, is in very good agreement
with the peak reported by Clar $et$ $al.$\citep{clar1964polycyclic}. 

Next experimental peak located at 6.62 eV, reported by Klevens $et$
$al.$\citep{klevens1949spectral}, is in good agreement with our
screened parameter peak computed at 6.74 eV. The highest measured
peak located at 6.99 eV, reported by Klevens $et$ $al.$\citep{klevens1949spectral}
is in excellent agreement with the peaks obtained both from screened
and standard parameter calculations at 7.05 eV and 6.96 eV, respectively.
Furthermore, we have computed several higher energy peaks as well,
for which no experimental results exist. We hope that in future measurements
of the absorption spectrum of phenanthrene, energy range beyond 7
eV will be explored. 

Dutta and Mazumdar\citep{tirtho-phenanthrene2014,tirtho-phenanthrene2016}
studied the ground state of metal-intercalated crystalline phenanthrene
using both \emph{ab initio} density functional theory (DFT), and PPP
model based approaches, with the aim of understanding the nature of
superconductivity in these materials. Parac $et$ $al.$ \citep{parac2003tddft}
have computed the absorption spectra of phenanthrene using time-dependent
DFT (TDDFT) and time-dependent PPP (TDPPP) method, while Malloci $et$
$al.$ \citep{malloci2011electronic} have computed the same using
DFT and TDDFT method. By using PPP model at the singles-CI level,
Skancke $et$ $al.$ and Hedges $et$ $al.$ have also computed the
absorption spectra of phenanthrene. We present the results of these
authors in Table \ref{tab:comparison-Phenanthrene}, from where it
is obvious that their calculated peak locations lie in a broad spectral
range. Given the fact that peaks measured by various experiments also
lie in a broad spectral range, the agreement between these theoretical
results and experiments is quite reasonable. The detailed wave function
analysis of all the excited states contributing to peaks in the computed
spectra of phenanthrene, is presented in Tables S1-S2 of Supporting
Information. 
\begin{center}
\textcolor{black}{\footnotesize{}}
\begin{table}[H]
\textcolor{black}{\footnotesize{}\caption{Comparison of computed peak locations in the spectra of phenanthrene
with the experimental, and other theoretical values; all energies
are in eV. Columns with headings Scr/Std contain results of calculations
performed using the screened/standard parameters in the PPP model.
MI denotes the peak of maximum intensity, OG denotes the peak corresponding
to the optical gap, and D.F. indicates the dipole forbidden sate.
\label{tab:comparison-Phenanthrene}}
}{\footnotesize\par}
\centering{}\textcolor{black}{}%
\begin{tabular}{|c|c|c|c|}
\hline 
\multirow{2}{*}{\textcolor{black}{Experimental }} & Theoretical & \multicolumn{2}{c|}{\textcolor{black}{This Work}}\tabularnewline
\cline{3-4} 
 & (other authors) & \textcolor{black}{Scr} & \textcolor{black}{Std}\tabularnewline
\hline 
\textcolor{black}{3.50\citep{klevens1949spectral}(D.F.), 3.57\citep{okamoto2013photochemical}(D.F.), } & 3.65$^{a}$/3.82$^{b}$/3.97$^{c}$(D.F.)\citep{parac2003tddft},  & 2.96 \textcolor{black}{(}$^{1}A_{1}$\textcolor{black}{)} & 3.35 \textcolor{black}{(}$^{1}A_{1}$\textcolor{black}{)}\tabularnewline
3.59\citep{clar1964polycyclic}(D.F.), 3.61\citep{salama1993neutral}(D.F.),  & 4.01\citep{skancke1965semi}(D.F.), 4.22\citep{hedges1968electronic}(D.F.) & \textcolor{black}{(D.F.)} & \textcolor{black}{(D.F.)}\tabularnewline
\hline 
4.09\citep{klevens1949spectral}(OG), 4.17\citep{okamoto2013photochemical}(OG),  & 3.91$^{a}$/4.21$^{c}$/4.34$^{b}$\citep{parac2003tddft}(OG),  & \multirow{2}{*}{4.31 ($^{1}B_{1}$)} & \multirow{2}{*}{4.26 ($^{1}B_{1}$)}\tabularnewline
 4.24\citep{birks1970photophysics,clar1964polycyclic}(OG), 4.25\citep{salama1993neutral}(OG), & 4.19$^{c}$/4.67$^{d}$ \citep{malloci2011electronic,malloci2004electronic}(OG), &  & \tabularnewline
4.36 \citep{halasinski2005investigation}(OG), & 4.36 \citep{skancke1965semi}(OG), 4.60\citep{hedges1968electronic}(OG), & (OG) & (OG)\tabularnewline
\hline 
4.34\citep{okamoto2013photochemical}, &  &  & \tabularnewline
\hline 
4.42\citep{salama1993neutral}, 4.44\citep{okamoto2013photochemical},  & \multirow{3}{*}{\textbf{---}} & \multirow{3}{*}{\textcolor{black}{$4.62$ (}$^{1}A_{1}$\textcolor{black}{)}} & \multirow{3}{*}{\textbf{---}}\tabularnewline
 4.53\citep{halasinski2005investigation}, 4.56\citep{salama1993neutral}, &  &  & \tabularnewline
4.64\citep{halasinski2005investigation}(MI), &  &  & \tabularnewline
\hline 
4.80\citep{salama1993neutral}(MI) , 4.93\citep{clar1964polycyclic}(MI),  & \multirow{3}{*}{\textbf{---}} & \multirow{2}{*}{\textcolor{black}{4.93 }($^{1}B_{1}$) } & \textcolor{black}{$5.19$ }\tabularnewline
4.88\citep{klevens1949spectral}(MI), 5.10\citep{halasinski2005investigation},  &  &  & ($^{1}A_{1}/{}^{1}B_{1}$)\tabularnewline
5.27\citep{halasinski2005investigation},  &  &  (MI) & (MI)\tabularnewline
\hline 
5.65\citep{clar1964polycyclic}, 5.78\citep{halasinski2005investigation}
\textcolor{black}{, } & 5.45\citep{hedges1968electronic}, 5.47\citep{malloci2004electronic}(MI),  & \textcolor{black}{$5.71$ } & \multirow{2}{*}{\textcolor{black}{$5.59$ }($^{1}A_{1}$)}\tabularnewline
\textcolor{black}{5.84}\citep{klevens1949spectral},  & 5.52\citep{skancke1965semi},5.89\citep{skancke1965semi}(MI),  & ($^{1}A_{1}/{}^{1}B_{1}$) & \tabularnewline
\hline 
\textbf{---} & 6.01\citep{malloci2004electronic}, 6.25\citep{skancke1965semi},  & \textcolor{black}{6.07 $(^{1}B_{1})$} & \textcolor{black}{6.10 $(^{1}B_{1})$}\tabularnewline
\hline 
\textbf{---} & 6.38\citep{malloci2004electronic}, & \textcolor{black}{6.49 }($^{1}A_{1}$) & \textcolor{black}{6.40 }($^{1}B_{1}$)\tabularnewline
\hline 
6.62\citep{klevens1949spectral} &  6.72\citep{malloci2004electronic}, 6.75\citep{malloci2004electronic}, & \textcolor{black}{6.74 }($^{1}B_{1}$) & ---\tabularnewline
\hline 
6.99\citep{klevens1949spectral} & 6.90\citep{skancke1965semi}, 7.39\citep{malloci2004electronic}, & 7.05 ($^{1}A_{1}$) & \textcolor{black}{6.96 }($^{1}A_{1}/{}^{1}B_{1}$)\tabularnewline
\hline 
\multicolumn{4}{|c|}{$^{a}$TDDFT(BP86) method, $^{b}$TDPPP method,}\tabularnewline
\multicolumn{4}{|c|}{$^{c}$TDDFT(B3LYP) method and $^{d}$DFT(Kohan-Sham) method}\tabularnewline
\hline 
\end{tabular}
\end{table}
{\footnotesize\par}
\par\end{center}

\subsubsection*{Chrysene}

Chrysene has $C_{2h}$ symmetry, and Okamoto $et$ $al.$\citep{okamoto2013photochemical},
Klevens $et$ $al.$\citep{klevens1949spectral}, Clar $et$ $al.$\citep{clar1964polycyclic}
and Becker $et$ $al.$ \citep{becker1963comprehensive} have reported
the measurements of its absorption spectrum. In Fig. \ref{fig:Computed-linear-opt_abs_ spectra_PQDS-PPP-CI}
(a) and (b), we present our calculated spectra using the screened
and standard parameters, respectively, within the PPP-CI approach,
and in Table \ref{tab:comparison-Chrysen}, we have compared the experimental
peak locations with our results, and those of other theoretical calculations.
The first peaks observed experimentally by Klevens $et$ $al.$\citep{klevens1949spectral},
Okomoto $et$ $al.$\citep{okamoto2013photochemical}, Becker $et$
$al.$ \citep{becker1963comprehensive} and Clar $et$ $al.$\citep{clar1964polycyclic}
are located at 3.40 eV, 3.42 eV, 3.43 eV, and 3.44 eV respectively,
and again correspond to a dipole forbidden (D.F.) state. Using the
standard parameters, we obtain the D.F. state at 3.33 eV, which agrees
well with the measured values. The screened parameter value computed
at 3.11 eV, is slightly lower as compared to the experimental values.
Both screened and standard parameter calculation predict the D.F.
state to be of $^{1}B_{u}$ symmetry, in agreement with the work of
Skancke \citep{skancke1965semi}.

The intensity of the first dipole allowed peak, corresponding to the
optical gap, was not found to be maximum in our calculated spectra.
This result is in perfect agreement with the measurements of Klevens
$et$ $al.$\citep{klevens1949spectral}, Okamoto $et$ $al.$\citep{okamoto2013photochemical},
Becker $et$ $al.$ \citep{becker1963comprehensive} and Clar $et$
$al.$\citep{clar1964polycyclic}, and in disagreement with the results
of the TB model. Our calculations predict this peak at 3.86 eV using
screened parameters, and 3.96 eV using standard parameters. As is
obvious from Table \ref{tab:comparison-Phenanthrene}, the experimental
values of the optical gap range from 3.74 eV to 3.89 eV, implying
that both the calculated values of optical gap are quite close to
the experimental values. In particular, we also note that the Okamoto
$et$ $al.$\citep{okamoto2013photochemical} and Becker $et$ $al.$
\citep{becker1963comprehensive} reported the values of optical gap
at 3.84 eV, and 3.87 eV, respectively, almost in perfect agreement
with our screened parameter value 3.86 eV. The standard parameter
value of 3.96 eV is slightly higher than the highest measured value
3.89 eV by Mallory $et$ $al.$\citep{mallory1997phenacenes} and
Clar $et$ $al.$ \citep{clar1964polycyclic}.

The second peak is the most intense (MI) peak in our calculated spectrum
for both the screened and the standard parameters, and is located
at 4.52 eV, and 5.08 eV, respectively. Okamoto $et$ $al.$\citep{okamoto2013photochemical}
experimentally measured the most intense peak at 4.54 eV which almost
exactly matches with our screened parameter value. The reported value
of MI peak by Klevens $et$ $al.$ \citep{klevens1949spectral} (4.61
eV), Becker $et$ $al.$ \citep{becker1963comprehensive} (4.63 eV),
and Clar $et$ $al.$\citep{clar1964polycyclic} (4.64 eV) are slightly
higher than the screened parameter value, while the standard parameter
result is significantly higher than the experimental values.

As far as higher energy peaks are concerned, Klevens $et$ $al.$
\citep{klevens1949spectral}, and Becker $et$ $al.$\citep{becker1963comprehensive}
found a peak at 5.13 eV, and, Clar $et$ $al.$\citep{clar1964polycyclic}
found one at 5.14 eV. In our computed spectra, we have a peak at 5.09
eV using screened parameters, and 5.08 eV using standard parameters,
both of which are in good agreement with experiments. After that,
in our standard parameter spectrum we have a peak at 5.76 eV, which
is in very good agreement with the 5.71 eV peak detected by Becker
$et$ $al.$\citep{becker1963comprehensive}, while the screened parameter
peak at 5.81 eV is somewhat higher as compared to the experiments.
Next experimental peak located at 6.36 eV, reported by Klevens $et$
$al.$\citep{klevens1949spectral}, is in perfect agreement with our
standard parameter peak computed at 6.38 eV. We have a screened parameter
peak at 6.52 eV which is in reasonable agreement with a peak at 6.43
eV, measured by Becker $et$ $al.$\citep{becker1963comprehensive}.
The highest measured peak located at 6.73 eV reported by Klevens $et$
$al.$\citep{klevens1949spectral} is at a slightly lower energy as
compared to the corresponding screened parameter peak at 6.86 eV.
Furthermore, we have computed several higher energy peaks as well,
for which no experimental results exist. We hope that in future measurements
of the absorption spectrum of chrysene, in higher energy range will
be probed.

First principles TDDFT method was employed to calculate the absorption
spectrum by Parac $et$ $al.$ \citep{parac2003tddft} and Malloci
$et$ $al.$ \citep{malloci2011electronic}. Malloci $et$ $al.$
\citep{malloci2011electronic} also computed the absorption spectrum
using the first principles DFT. Additionally, PPP model based calculations
were performed by Parac $et$ $al.$ \citep{parac2003tddft}, Skancke
$et$ $al.$\citep{skancke1965semi}, Ham $et$ $al.$ \citep{ham1956electronic}
and Hedges $et$ $al.$\citep{hedges1968electronic}. 

The predictions on the location of D.F. state by other authors are
in a broad energy range 3.48-4.13 eV\citep{parac2003tddft,ham1956electronic,hedges1968electronic,skancke1965semi},
while the experimental values are in a very narrow range 3.40-3.44
eV\citep{klevens1949spectral,okamoto2013photochemical,becker1963comprehensive,clar1964polycyclic}.
This means that most of the calculations of other authors overestimate
the experimental results. Regarding the optical gap, calculations
of other authors predict it in the range 3.40-4.22 eV, while the experimental
values are in the range 3.74-3.89 eV, implying that most other calculations
either underestimate or overestimate the experimental values. But
we note that the optical gaps computed using the TDDFT approach with
B3LYP functional by Parac $et$ $al.$ \citep{parac2003tddft}, and
Malloci $et$ $al.$ \citep{malloci2011electronic}, are in good agreement
with experiments. As far as the location of MI peak is concerned,
results of other authors are either below, or significantly above
the experimental value. The wave function of the excited states contributing
to peaks in the computed spectra of chrysene, are presented in Tables
S3-S4 of Supporting Information. 
\begin{center}
\textcolor{black}{\footnotesize{}}
\begin{table}[H]
\textcolor{black}{\footnotesize{}\caption{Comparison of computed peak locations in the spectra of chrysene with
the experimental values, and other theoretical values; all energies
are in eV. Rest of the information is same as in the caption of Table
\ref{tab:comparison-Phenanthrene}. \label{tab:comparison-Chrysen}}
}{\footnotesize\par}
\centering{}\textcolor{black}{}%
\begin{tabular}{|c|c|c|c|}
\hline 
\multirow{2}{*}{\textcolor{black}{Experimental }} & Theoretical  & \multicolumn{2}{c|}{\textcolor{black}{This Work}}\tabularnewline
\cline{3-4} 
 & (other authors) & \textcolor{black}{Scr} & \textcolor{black}{Std}\tabularnewline
\hline 
3.40\citep{klevens1949spectral}(D.F.), 3.42\citep{okamoto2013photochemical}(D.F.),  & 3.48$^{a}$/3.62$^{b}$/3.75$^{c}$\citep{parac2003tddft}(D.F.), & \multirow{2}{*}{3.11 \textcolor{black}{(}$^{1}B_{u}$\textcolor{black}{)}} & \multirow{2}{*}{3.33 \textcolor{black}{(}$^{1}B_{u}$\textcolor{black}{)}}\tabularnewline
3.43\citep{becker1963comprehensive}(D.F.), 3.44\citep{clar1964polycyclic}(D.F.) & 3.47\citep{ham1956electronic}(D.F.), 4.10\citep{hedges1968electronic}(D.F.) &  & \tabularnewline
 & 4.13\citep{skancke1965semi}(D.F.),  & \textcolor{black}{(D.F.)} & \textcolor{black}{(D.F.)}\tabularnewline
\hline 
3.74\citep{klevens1949spectral}(OG), 3.84\citep{okamoto2013photochemical}(OG),  & 3.40$^{a}$/3.92$^{b}$/3.82$^{c}$\citep{parac2003tddft}(OG), & \multirow{3}{*}{3.86 ($^{1}B_{u}$)} & \multirow{3}{*}{3.96 ($^{1}B_{u}$)}\tabularnewline
3.87\citep{becker1963comprehensive}(OG), & 3.73$^{c}$/4.21$^{d}$ \citep{malloci2011electronic,malloci2004electronic}(OG)  &  & \tabularnewline
3.89\citep{mallory1997phenacenes,clar1964polycyclic}(OG), & 3.90\citep{ham1956electronic}(OG), 4.13\citep{skancke1965semi}(OG), &  & \tabularnewline
 & 4.22\citep{hedges1968electronic}(OG) & (OG) & (OG)\tabularnewline
\hline 
4.00\citep{okamoto2013photochemical}, 4.17\citep{okamoto2013photochemical} & \textbf{---} & \textbf{---} & \textbf{---}\tabularnewline
\hline 
4.54\citep{okamoto2013photochemical}(MI),4.61\citep{klevens1949spectral}
(MI),  & \multirow{3}{*}{4.26\citep{malloci2004electronic,skancke1965semi} } & \multirow{2}{*}{4.52 ($^{1}B_{u}$) } & \multirow{3}{*}{\textbf{---}}\tabularnewline
4.63\citep{becker1963comprehensive}(MI), 4.64\citep{clar1964polycyclic}(MI),  &  &  & \tabularnewline
4.71 \citep{okamoto2013photochemical} &  & (MI) & \tabularnewline
\hline 
\multirow{3}{*}{5.13\citep{klevens1949spectral,becker1963comprehensive}, 5.14\citep{clar1964polycyclic}, } & 4.84\citep{malloci2004electronic}(MI), 4.93\citep{ham1956electronic}, & \multirow{3}{*}{5.09 ($^{1}B_{u}$)} & \multirow{2}{*}{5.08 ($^{1}B_{u}$) }\tabularnewline
 &  5.35\citep{ham1956electronic}, 5.39\citep{hedges1968electronic}, &  & \tabularnewline
 & 5.47\citep{hedges1968electronic} &  & (MI)\tabularnewline
\hline 
5.59\citep{becker1963comprehensive}, 5.65\citep{klevens1949spectral},  & 5.43\citep{malloci2004electronic}, 5.48\citep{skancke1965semi},  & \multirow{2}{*}{5.81 ($^{1}B_{u}$)} & \multirow{2}{*}{5.76 ($^{1}B_{u}$)}\tabularnewline
5.71\citep{becker1963comprehensive} &  5.75\citep{skancke1965semi} (MI), &  & \tabularnewline
\hline 
\textbf{---} & 6.13\citep{malloci2004electronic} & 6.00 ($^{1}B_{u}$) & \textbf{---}\tabularnewline
\hline 
6.36\citep{klevens1949spectral}, 6.43\citep{becker1963comprehensive} & 6.36\citep{skancke1965semi} & 6.52 ($^{1}B_{u}$) & 6.38 ($^{1}B_{u}$)\tabularnewline
\hline 
6.73\citep{klevens1949spectral} & 6.99\citep{malloci2004electronic} & 6.86 ($^{1}B_{u}$) & 7.17 ($^{1}B_{u}$)\tabularnewline
\hline 
\textbf{---} & 7.32\citep{skancke1965semi}, & 7.32 ($^{1}B_{u}$) & 7.37 ($^{1}B_{u}$)\tabularnewline
\hline 
\multicolumn{4}{|c|}{$^{a}$TDDFT(BP86) method, $^{b}$TDPPP method,}\tabularnewline
\multicolumn{4}{|c|}{$^{c}$TDDFT(B3LYP) method and $^{d}$DFT(Kohan-Sham) method}\tabularnewline
\hline 
\end{tabular}
\end{table}
{\footnotesize\par}
\par\end{center}

\subsubsection*{Picene}

Picene has $C_{2v}$ symmetry, and Okamoto $et$ $al.$\citep{okamoto2013photochemical},
Clar $et$ $al.$\citep{clar1964polycyclic} and Fanetti $et$ $al.$
\citep{fanetti2012homo} have reported the measurements of its absorption
spectrum. In Fig. \ref{fig:Computed-linear-opt_abs_ spectra_PQDS-PPP-CI}
(a) and (b), we present our calculated spectra using the screened
and standard parameters, respectively, within the PPP-CI approach,
and in Table \ref{tab:comparison-Picene}, we have compared the locations
of various peaks obtained in our calculations with the experimental
results, and other theoretical results. The first peak corresponding
to the dipole forbidden (D.F.) state was measured to be at 3.30 eV
by Okomoto $et$ $al.$,\citep{okamoto2013photochemical} and Clar
$et$ $al.$\citep{clar1964polycyclic}, while Fanetti $et$ $al.$
\citep{fanetti2012homo} measured it at 3.31 eV. Our standard parameter
calculation predicts the D.F. state at 3.33 eV, in excellent agreement
with the experiments, while the screened parameter value at 3.20 eV
is slightly lower than the experiments. Both sets of calculations
predict the D.F. state to be of $^{1}A_{1}$ symmetry.

The first dipole allowed state is computed to be of $^{1}B_{1}$ symmetry,
and leads to fairly intense absorption peaks located at 3.75 eV in
the screened parameter spectrum, and 3.88 eV in the standard parameter
spectrum. As it is obvious from Table \ref{tab:comparison-Phenanthrene},
the experimental values of the optical gap range from 3.76 eV to 3.82
eV. Thus, we find that both our screened and standard parameter of
optical gap are quite close to the range of experimental values. We
also note that the Okamoto $et$ $al.$\citep{okamoto2013photochemical}
and Mallory $et$ $al.$\citep{mallory1997phenacenes} reported the
value of optical gap at 3.76 eV, and the Clar $et$ $al.$\citep{clar1964polycyclic}
reported it at 3.77 eV, in excellent agreement with our screened parameter
value. While our standard parameter value 3.88 eV agrees well with
the optical gap value 3.82 eV, measured by Fanetti $et$ $al.$ \citep{fanetti2012homo}.
Furthermore, our calculation predict that this peak is not the most
intense one, in disagreement with the TB model results, and in complete
agreement with the experiments\citep{fanetti2012homo,clar1964polycyclic,okamoto2013photochemical}. 

As far as higher energy peaks are concerned, our screened parameter
spectrum has a peak at 4.54 eV which is in very good agreement with
a peak at 4.57 eV measured by Fanetti $et$ $al.$\citep{fanetti2012homo}.
The next peaks which are the most intense (MI) ones in our calculated
spectra using both screened and standard parameters, are located at
4.87 eV and 4.79 eV, respectively. Okamoto $et$ $al.$\citep{okamoto2013photochemical}
measured a peak at 4.75 eV, which is in very good agreement with the
location of our standard parameter peak. A peak measured at 4.85 eV
by Fanetti $et$ $al.$\citep{fanetti2012homo} is in excellent agreement
with the energy of our screened parameter peak.

Our calculated peaks at 5.22 eV (screened) and 5.24 eV (standard)
are the nearest peaks to the highest measured peak at 5.08 eV reported
by Fanetti $et$ $al.$ \citep{fanetti2012homo}. Additionally, we
have computed several higher energy peaks as well, for which no experimental
results exist. We hope that in future measurements of the absorption
spectrum of picene, energy range beyond 5 eV will be explored. 

The measured experimental value of D.F. state are in very tight energy
range 3.30-3.31 eV \citep{mallory1997phenacenes,fanetti2012homo,clar1964polycyclic,okamoto2013photochemical},
while the D.F. state calculated by other authors lie in a broad spectral
range 3.18-4.33 eV\citep{parac2003tddft,hedges1968electronic,malloci2011electronic}.
As far as optical gap is concerned, the computed values of other authors
are in the range of 3.32-4.13 eV, while the experimental values are
in the range 3.76-3.82 eV. Therefore, both for D.F. state and optical
gap several other calculations have either underestimated or overestimated
the data. But we note that Parac and Grimme \citep{parac2003tddft}
have obtained the optical gap value at 3.72 eV using TDDFT method
which is in good agreement with experimental value 3.76 eV. They have
also calculated the optical gap value using TDPPP method, which is
also in good agreement with the experimental results. The detailed
analysis of wave functions of the excited states contributing to peaks
in the calculated spectra of picene, is presented in Tables S5-S6
of Supporting Information. 
\begin{center}
\textcolor{black}{\footnotesize{}}
\begin{table}[H]
\textcolor{black}{\footnotesize{}\caption{Comparison of computed peak locations in the spectra of picene with
the experimental values, and other theoretical values; all energies
are in eV. Rest of the information is same as in the caption of Table
\ref{tab:comparison-Phenanthrene}. \label{tab:comparison-Picene}}
}{\footnotesize\par}
\centering{}\textcolor{black}{}%
\begin{tabular}{|c|c|c|c|}
\hline 
\multirow{2}{*}{\textcolor{black}{Experimental }} & Theoretical & \multicolumn{2}{c|}{\textcolor{black}{This Work}}\tabularnewline
\cline{3-4} 
 & (other authors) & \textcolor{black}{Scr} & \textcolor{black}{Std}\tabularnewline
\hline 
3.30\citep{okamoto2013photochemical,clar1964polycyclic}(D.F.),  & 3.18$^{a}$/3.49$^{b}$/3.56$^{c}$\citep{parac2003tddft}(D.F.),  & \multirow{2}{*}{3.20 \textcolor{black}{(}$^{1}A_{1}$\textcolor{black}{)(D.F.)}} & \multirow{2}{*}{3.33 \textcolor{black}{(}$^{1}A_{1}$\textcolor{black}{)(D.F.)}}\tabularnewline
3.31\citep{fanetti2012homo}(D.F.) & 4.33\citep{hedges1968electronic}(D.F.) &  & \tabularnewline
\hline 
3.76\citep{mallory1997phenacenes,okamoto2013photochemical}(OG),  & 3.32$^{a}$/3.83$^{b}$/3.72$^{c}$\citep{parac2003tddft}(OG),  & \multirow{3}{*}{3.75 ($^{1}B_{1}$)(OG)} & \multirow{3}{*}{3.88 ($^{1}B_{1}$)(OG)}\tabularnewline
3.77\citep{clar1964polycyclic}(OG), 3.82\citep{fanetti2012homo}(OG), & 3.70$^{c}$/4.19$^{d}$ \citep{malloci2011electronic}(OG), &  & \tabularnewline
 &  4.13\citep{hedges1968electronic}(OG), &  & \tabularnewline
\hline 
3.93\citep{okamoto2013photochemical}, 3.98\citep{fanetti2012homo},  & \multirow{2}{*}{\textbf{---}} & \multirow{2}{*}{\textbf{---}} & \multirow{2}{*}{\textbf{---}}\tabularnewline
 4.06\citep{okamoto2013photochemical}, 4.13\citep{fanetti2012homo},  &  &  & \tabularnewline
\hline 
4.26(MI)\citep{okamoto2013photochemical}, 4.32\citep{clar1964polycyclic}(MI),  & \multirow{3}{*}{\textbf{---}} & \multirow{3}{*}{4.54 ($^{1}A_{1}$)} & \multirow{3}{*}{\textbf{---}}\tabularnewline
4.39\citep{fanetti2012homo}(MI), 4.45\citep{okamoto2013photochemical}, &  &  & \tabularnewline
4.57\citep{fanetti2012homo}, &  &  & \tabularnewline
\hline 
4.72\citep{fanetti2012homo}, 4.75\citep{okamoto2013photochemical},  & \multirow{2}{*}{\textbf{---}} & 4.87(MI) & 4.79(MI) \tabularnewline
4.85\citep{fanetti2012homo} , &  &  ($^{1}A_{1}/{}^{1}B_{1}$) & ($^{1}A_{1}/{}^{1}B_{1}$)\tabularnewline
\hline 
5.08\citep{fanetti2012homo} & 5.13\citep{hedges1968electronic}, 5.22\citep{hedges1968electronic} & 5.22 ($^{1}B_{1}$)  & 5.24 ($^{1}A_{1}$)\tabularnewline
\hline 
\multicolumn{4}{|c|}{$^{a}$TDDFT(BP86) method, $^{b}$TDPPP method,}\tabularnewline
\multicolumn{4}{|c|}{$^{c}$TDDFT(B3LYP) method and $^{d}$DFT(Kohan-Sham) method}\tabularnewline
\hline 
\end{tabular}
\end{table}
{\footnotesize\par}
\par\end{center}

\subsubsection*{Fulminene}

Fulminene has $C_{2h}$ symmetry, and Okamoto $et$ $al.$\citep{okamoto2013photochemical,okamoto2014extended}
and Mallory $et$ $al.$ \citep{mallory1997phenacenes} have measured
the absorption in fulminene. In Fig. \ref{fig:Computed-linear-opt_abs_ spectra_PQDS-PPP-CI}
(a) and (b), we present our calculated spectra using the screened
and standard parameters, respectively, within the PPP-CI approach,
and in Table \ref{tab:comparison-fulminene}, we have compared the
experimental results, and theoretical results of other authors, with
our calculations. The detailed wave functions analysis of all the
excited states contributing to peaks in the computed spectra of fulminene,
is presented in Tables S7-S8 of Supporting Information. 

The first peak observed experimentally by Okomoto $et$ $al.$\citep{okamoto2013photochemical}
is located at 3.24 eV, and it corresponds to a dipole forbidden (D.F.)
state. Both our standard parameter and screened parameter calculations
predict the symmetry of this state to be $^{1}B_{u}$, and located
at 3.07 eV, and 2.86 eV, respectively. This implies that our calculated
locations of the D.F. state are lower than the experimental value,
with the screened parameter value being significnatly lower. 

The first dipole allowed peak, corresponding to the optical gap, was
measured to be at 3.14 eV for the thin film sample by Okamoto $et$
$al.$\citep{okamoto2014extended}, and 3.65 eV, and 3.66 eV for the
solution sample, by Okamoto $et$ $al.$\citep{okamoto2014extended},
and Mallory $et$ $al.$ \citep{mallory1997phenacenes}, respectively.
Our screened parameter calculations predict the optical gap to be
3.34 eV, which is closer to the measured value of thin film sample,
while our standard parameter value at 3.52 eV is closer to the solution
based sample. This is understandable on physical grounds because,
in thin films, electron correlations may be getting screened due to
presence of other molecules, an effect screened parameters may be
mimicking. We also note that intensity of the first dipole allowed
peak is not maximum when compared to other peaks, in agreement with
the measurements of the Okamoto $et$ $al.$\citep{okamoto2014extended}. 

As far as higher energy features are concerned, our screened parameter
spectrum has a peak at 4.01 eV, which is in good agreement with the
peaks measured at 3.95 eV\citep{okamoto2013photochemical}, and 4.08
eV\citep{okamoto2014extended}, in solution, and thin film, based
samples, respectively. The next peak is the most intense (MI) one
in our calculated spectrum for the both screened and standard parameters,
located at 4.37 eV, and 4.54 eV, respectively. In solution based spectrum
the most intense peak lies at 4.17 eV\citep{okamoto2013photochemical}
which is closer to the screened parameter value, than the standard
one. Furthermore, we have computed several higher energy peaks as
well, for which no experimental results exist. We hope that in future
measurements of the absorption spectrum of fulminene, the energy range
beyond 5 eV will be explored. 

The only other calculation on fulminene is by Malloci $et$ $al.$
\citep{malloci2011electronic}, who reported the values of optical
gap at 3.47 eV (TDDFT approach), and 4.00 eV (Kohn-Sham). The former
value is within the range of experimental measurements, while the
latter is well above it. 
\begin{center}
\textcolor{black}{\footnotesize{}}
\begin{table}[H]
\textcolor{black}{\footnotesize{}\caption{Comparison of computed peak locations in the spectra of fulminene
with the experimental values, and other theoretical values; all energies
are in eV. Rest of the information is same as in the caption of Table
\ref{tab:comparison-Phenanthrene}. \label{tab:comparison-fulminene}}
}{\footnotesize\par}
\centering{}\textcolor{black}{}%
\begin{tabular}{|c|c|c|c|}
\hline 
\multirow{2}{*}{\textcolor{black}{Experimental }} & Theoretical & \multicolumn{2}{c|}{\textcolor{black}{This Work}}\tabularnewline
\cline{3-4} 
 & (other authors) & \textcolor{black}{Scr} & \textcolor{black}{Std}\tabularnewline
\hline 
3.24 \citep{okamoto2013photochemical}(D.F.) & \textbf{---} & 2.86 \textcolor{black}{(}$^{1}B_{u}$\textcolor{black}{)(D.F.)} & 3.07 \textcolor{black}{(}$^{1}B_{u}$\textcolor{black}{)(D.F.)}\tabularnewline
\hline 
3.14\citep{okamoto2014extended}(OG), 3.65\citep{okamoto2013photochemical}(OG), & \multirow{2}{*}{3.47$^{c}$/ 4.00$^{d}$\citep{malloci2011electronic}(OG)} & \multirow{2}{*}{3.34 ($^{1}B_{u}$)(OG)} & \multirow{2}{*}{3.52 ($^{1}B_{u}$)(OG)}\tabularnewline
3.66\citep{mallory1997phenacenes} (OG), &  &  & \tabularnewline
\hline 
3.29\citep{okamoto2014extended}, 3.44\citep{okamoto2014extended}, &  &  & \tabularnewline
\hline 
3.90\citep{okamoto2014extended}, 3.80\citep{okamoto2013photochemical},  & \multirow{2}{*}{\textbf{---}} & \multirow{2}{*}{4.01 ($^{1}B_{u}$)} & \multirow{2}{*}{\textbf{---}}\tabularnewline
3.95\citep{okamoto2013photochemical}, 4.08\citep{okamoto2014extended} &  &  & \tabularnewline
\hline 
4.17\citep{okamoto2013photochemical}(MI), 4.36\citep{okamoto2013photochemical},
4.66\citep{okamoto2013photochemical} & \textbf{---} & 4.37 ($^{1}B_{u}$)(MI) & 4.54 ($^{1}B_{u}$)(MI)\tabularnewline
\hline 
\multicolumn{4}{|c|}{$^{c}$TDDFT(B3LYP) method and $^{d}$DFT(Kohan-Sham) method}\tabularnewline
\hline 
\end{tabular}
\end{table}
{\footnotesize\par}
\par\end{center}

\subsubsection*{{[}7{]}Phenacene}

{[}7{]}phenacene has $C_{2v}$ symmetry, and Okamoto $et$ $al.$\citep{okamoto2014extended},
and Mallory $et$ $al.$ \citep{mallory1997phenacenes} have reported
the measurements of its absorption spectrum. In Fig. \ref{fig:Computed-linear-opt_abs_ spectra_PQDS-PPP-CI}
(a) and (b), we present our calculated spectra using the screened
and standard parameters, respectively, within the PPP-CI approach,
and in Table \ref{tab:comparison-=00005B7=00005DPhenacene}, we have
compared the experimental results, and theoretical results of other
authors, with our calculations. Analysis of the calculated CI wave
functions of the excited states contributing to the absorption spectra,
is presented in Tables S9-S10 of Supporting Information. 

Our calculations predict the dipole forbidden state to be of $^{1}A_{1}$
symmetry, located at 2.96 eV(screened parameters) and 3.31 eV (standard
parameters). Because no prior experimental measurements of D.F. state
are available for {[}7{]}phenacene, our results could be tested in
future experiments.

On comparing the relative intensity of first dipole allowed peak,
corresponding to the optical gap, we find that it is not of maximum
intensity, in agreement with the measurements\citep{okamoto2014extended,mallory1997phenacenes}.
Our calculations predicts this peak at 3.34 eV (screened parameters),
and at 3.68 eV (standard parameters). We note that the Okamoto $et$
$al.$\citep{okamoto2014extended} reported the value of optical gap
at 3.10 eV, which is closer to our screened parameter value, while
our standard parameter value is in very good agreement with 3.60 eV,
measured by Mallory $et$ $al.$ \citep{mallory1997phenacenes}.

As far as higher energy peaks are concerned, our screened parameter
spectra has a peak at 3.88 eV which is in excellent agreement with
the measured values 3.87 eV\citep{mallory1997phenacenes}, and 3.90
eV\citep{okamoto2014extended}. The next peak is the most intense
(MI) peak in our calculated spectrum located at 4.20 eV (screened
parameters) and 4.68 eV (standard parameters). Okamoto $et$ $al.$\citep{okamoto2014extended}
and Mallory $et$ $al.$ \citep{mallory1997phenacenes} experimentally
measured the most intense peak at 4.08 eV which  is a little lower
than our screened parameter value. Furthermore, we have computed several
higher energy peaks as well, for which no experimental results exist.
We hope that in future measurements of the absorption spectrum of
{[}7{]}phenacene, the energy range beyond 5 eV will be probed. 

The only other calculation on {[}7{]}phenacene is by Malloci $et$
$al.$ \citep{malloci2011electronic}, who reported the values of
optical gap at 3.50 eV (TDDFT approach), and 4.00 eV (Kohn-Sham).
The former value is within the range of experimental measurements,
while the latter is well above it. 
\begin{center}
\textcolor{black}{\footnotesize{}}
\begin{table}[H]
\textcolor{black}{\footnotesize{}\caption{Comparison of computed peak locations in the spectra of {[}7{]}phenacene
with the experimental values , and other theoretical values; all energies
are in eV. Rest of the information is same as in the caption of Table
\ref{tab:comparison-Phenanthrene}. \label{tab:comparison-=00005B7=00005DPhenacene}}
}{\footnotesize\par}
\centering{}\textcolor{black}{}%
\begin{tabular}{|c|c|c|c|}
\hline 
\multirow{2}{*}{\textcolor{black}{Experimental }} & Theoretical & \multicolumn{2}{c|}{\textcolor{black}{This Work}}\tabularnewline
\cline{3-4} 
 & (other authors) & \textcolor{black}{Scr} & \textcolor{black}{Std}\tabularnewline
\hline 
\textbf{---} & \textbf{---} & 2.96 \textcolor{black}{(}$^{1}A_{1}$\textcolor{black}{)(D.F.)} & 3.31 \textcolor{black}{(}$^{1}A_{1}$\textcolor{black}{)(D.F.)}\tabularnewline
\hline 
3.10\citep{okamoto2014extended}(OG), 3.29\citep{okamoto2014extended}, & \multirow{2}{*}{3.50$^{c}$/4.00$^{d,}$\citep{malloci2011electronic}(OG), } & 3.34 ($^{1}B_{1}$)  & 3.68 ($^{1}B_{1}$) \tabularnewline
3.44\citep{okamoto2014extended}, 3.60\citep{mallory1997phenacenes}(OG),  &  & (OG) & (OG)\tabularnewline
\hline 
3.87\citep{mallory1997phenacenes}, 3.90\citep{okamoto2014extended} & \textbf{---} & 3.88 ($^{1}A_{1}$)  & \textbf{---}\tabularnewline
\hline 
4.08\citep{okamoto2014extended,mallory1997phenacenes} (MI) & \textbf{---} & 4.20 ($^{1}A_{1}$/$^{1}B_{1}$) (MI) & 4.68 ($^{1}B_{1}$) (MI) \tabularnewline
\hline 
\multicolumn{4}{|c|}{$^{c}$TDDFT(B3LYP) method; $^{d}$DFT(Kohan-Sham) method}\tabularnewline
\hline 
\end{tabular}
\end{table}
{\footnotesize\par}
\par\end{center}

\subsubsection*{{[}8{]}Phenacene}

{[}8{]}phenacene has $C_{2h}$ symmetry, and Okamoto $et$ $al.$\citep{okamoto2014extended}
have reported the measurement its absorption spectrum. In Fig. \ref{fig:Computed-linear-opt_abs_ spectra_PQDS-PPP-CI}
(a) and (b), we present our calculated spectra using the screened
and standard parameters, respectively, within the PPP-CI approach,
and in Table \ref{tab:comparison-=00005B8=00005DPhenacene}, we have
compared the experimental results, with our calculations. In this
molecule we have calculated a dipole forbidden (D.F.) state of $^{1}B_{u}$
symmetry, located at 2.84 eV with screened parameters, and 3.13 eV
with standard parameters. However, we are unable to compare our results
with the experiments, because no measurements of this state have been
performed so far.

In our calculated spectra, the first dipole allowed peak corresponding
to the optical gap is not the most intense peak of the spectrum, in
full agreement with the experimental measurements\citep{okamoto2014extended}.
The calculated locations of this peak is 3.11 eV using screened parameters,
and 3.41 eV using standard parameters. We find that our screened parameter
results are in excellent agreement with the experimentally measured
value of 3.08 eV\citep{okamoto2014extended}.

As far as higher energy peaks are concerned, our screened parameter
spectrum has a peak at 3.87 eV which is a bit higher than the measured
peak at 3.64 eV\citep{okamoto2014extended}. The next peak is the
most intense (MI) peak in our calculated spectrum for the both screened
as well as standard parameters, and is located at 4.08 eV, and 4.48
eV, respectively. Okamoto $et$ $al.$\citep{okamoto2014extended}
experimentally measured the most intense peak at 4.00 eV which  is
in very good agreement with our screened parameter value. Furthermore,
we have computed several higher energy peaks as well, for which no
experimental results exist. We hope that in future measurements of
the absorption spectrum of {[}8{]}phenacene, energy range beyond 4
eV will be explored. Detailed information about the wave functions
of the excited states contributing to peaks in the computed spectra,
can be obtained in Tables S11-S12 of Supporting Information. 
\begin{center}
\textcolor{black}{\footnotesize{}}
\begin{table}[H]
\textcolor{black}{\footnotesize{}\caption{Comparison of computed peak locations in the spectra of {[}8{]}phenacene
with the experimental values; all energies are in eV. Rest of the
information is same as in the caption of Table \ref{tab:comparison-Phenanthrene}.
\label{tab:comparison-=00005B8=00005DPhenacene}}
}{\footnotesize\par}
\centering{}\textcolor{black}{}%
\begin{tabular}{|c|c|c|}
\hline 
\multirow{2}{*}{\textcolor{black}{Experimental }} & \multicolumn{2}{c|}{\textcolor{black}{This Work}}\tabularnewline
\cline{2-3} 
 & \textcolor{black}{Scr} & \textcolor{black}{Std}\tabularnewline
\hline 
\textbf{---} & 2.84 \textcolor{black}{(}$^{1}B_{u}$\textcolor{black}{)(D.F.)} & 3.13 \textcolor{black}{(}$^{1}B_{u}$\textcolor{black}{)(D.F.)}\tabularnewline
\hline 
3.08\citep{okamoto2014extended}(OG), 3.26\citep{okamoto2014extended},
3.44\citep{okamoto2014extended} & 3.11($^{1}B_{u}$)(OG)  & 3.41 ($^{1}B_{u}$)(OG)\tabularnewline
\hline 
3.64\citep{okamoto2014extended} & 3.87 ($^{1}B_{u}$) & \textbf{---}\tabularnewline
\hline 
4.00\citep{okamoto2014extended} (MI) & 4.08 ($^{1}B_{u}$)(MI)  & 4.48 ($^{1}B_{u}$)(MI)\tabularnewline
\hline 
\end{tabular}
\end{table}
{\footnotesize\par}
\par\end{center}

\subsubsection*{{[}9{]}Phenacene}

{[}9{]}phenacene has $C_{2v}$ symmetry, and Shimo $et$ $al.$\citep{shimo2016synthesis}
have reported the measurement of its absorption spectrum. In Fig.
\ref{fig:Computed-linear-opt_abs_ spectra_PQDS-PPP-CI} (a) and (b),
we present our calculated spectra using the screened and standard
parameters, respectively, within the PPP-CI approach, and in Table
\ref{tab:comparison-=00005B9=00005DPhenacene}, we have compared the
experimental results, with our calculations. Additionally, detailed
information about the excited states contributing to peaks in the
computed spectra is presented in Tables S13-S14 of Supporting Information.

Our calculations locate a dipole forbidden (D.F.) state of $^{1}A_{1}$
symmetry, at 2.92 eV, with screened parameters, and at 3.29 eV with
standard parameters. However, we are unable to compare our results
with the experiments, because no measurements of this state have been
performed so far.

In this molecule as well, our calculations predict that the first
dipole-allowed peak corresponding to the optical gap, is not the most
intense peak of the spectrum, in full agreement with the experiment,\citep{shimo2016synthesis}
and in disagreement with the results of the TB model calculations.
As far as the value of optical gap is concerned, our calculations
predict it to be 3.09 eV obtained using the screened parameters, and
3.46 eV using the standard parameters. We note that our screened parameter
value is in very good agreement with the value 3.05 measured by Shimo
$et$ $al.$\citep{shimo2016synthesis}. 

As far as higher energy peaks are concerned, our screened parameter
spectrum has a peak at 3.54 eV, which is somewhat higher than 3.33
eV measured by Shimo $et$ $al.$\citep{shimo2016synthesis}. The
most intense (MI) peaks in our calculated spectra for screened as
well as standard parameters are located at 3.98 eV, and 5.0 eV, respectively.
In the experimental spectrum of Shimo $et$ $al.$\citep{shimo2016synthesis},
the intensity appears to increase monotonically in the region 3.60-4.00
eV, beyond which no measurements exist. This implies that the maximum
intensity peak is at an energy higher than 4.00 eV, which we hope
will be confirmed in future measurements. Furthermore, we have computed
several higher energy peaks as well, which we hope will also be verified
in future measurements beyond 4 eV. We are unable to compare our results
with calculations of other authors, because our calculations appear
to be the first ones on this molecule.
\begin{center}
\textcolor{black}{\footnotesize{}}
\begin{table}[H]
\textcolor{black}{\footnotesize{}\caption{Comparison of computed peak locations in the spectra of {[}9{]}phenacene
with the experimental values; all energies are in eV. Rest of the
information is same as in the caption of Table \ref{tab:comparison-Phenanthrene}.
\label{tab:comparison-=00005B9=00005DPhenacene}}
}{\footnotesize\par}
\centering{}\textcolor{black}{}%
\begin{tabular}{|c|c|c|}
\hline 
\multirow{2}{*}{\textcolor{black}{Experimental }} & \multicolumn{2}{c|}{\textcolor{black}{This Work}}\tabularnewline
\cline{2-3} 
 & \textcolor{black}{Scr} & \textcolor{black}{Std}\tabularnewline
\hline 
\textbf{---} & 2.92 \textcolor{black}{(}$^{1}A_{1}$\textcolor{black}{)(D.F.)} & 3.29 \textcolor{black}{(}$^{1}A_{1}$\textcolor{black}{)(D.F.)}\tabularnewline
\hline 
3.05\citep{shimo2016synthesis}(OG), 3.22\citep{shimo2016synthesis} & 3.09 ($^{1}B_{1}$)(OG)  & 3.46 ($^{1}B_{1}$)(OG) \tabularnewline
\hline 
3.33\citep{shimo2016synthesis} & 3.54 ($^{1}A_{1}$)  & \textbf{---}\tabularnewline
\hline 
\textbf{---} & 3.98($^{1}A_{1}$/$^{1}B_{1}$) (MI) & \textbf{---}\tabularnewline
\hline 
\textbf{---} & 4.38 ($^{1}A_{1}$) & 4.21 ($^{1}A_{1}$/$^{1}B_{1}$) \tabularnewline
\hline 
\textbf{---} & 4.8 ($^{1}A_{1}$/$^{1}B_{1}$) & \textbf{---}\tabularnewline
\hline 
\textbf{---} & 5.13 ($^{1}A_{1}$/$^{1}B_{1}$) & 5.0 ($^{1}A_{1}$)(MI)\tabularnewline
\hline 
\end{tabular}
\end{table}
{\footnotesize\par}
\par\end{center}

\subsection{Comparison between Phenacenes and Polyacenes }

As mentioned in the Introduction section, phenacenes and polyacenes
are isomers, $i.e.$ they have same chemical formula but different
structural arrangement. In polyacenes, benzene rings are fused in
a straight line and they belong to $D_{2h}$ point group. While, in
phenacenes, benzene rings are fused in a zig-zag manner, leading either
to $C_{2v}$ or $C_{2h}$ symmetry. In an earlier work in our group,
Sony $et$ $al$.\citep{sony-acene-lo} computed the absorption spectra
of oligoacenes ranging from naphthalene to heptacene, and, later on
Chakraborty $et$ $al.$\citep{doi:10.1021/jp408535u} extended the
work till decacene. In Table \ref{tab: comparison_polyacene_phenaacene}
we compare our calculated optical gaps for isomers containing 3 to
9 benzene rings, and we find that irrespective of parameters used,
the optical gaps of {[}$n${]}phenacenes are always larger than those
of acene-$n$. We also note the relative difference in the optical
gaps of two set of compounds increases with the increasing conjugation
length. These facts are also verified in the optical absorption experiments
on phenacenes cited in the present work, as well those on acenes reviewed
in our earlier work.\citep{sony-acene-lo} Additionally, Roth \emph{et
al}.\citep{Roth2013} performed a comparative study of singlet states
in two of the smallest phenacenes (phenanthrene and chrysene), and
acenes (anthracene and tetracene) in the crystalline phase, using
the electron-energy-loss spectroscopy (EELS), and concluded that absorptions
occur in acenes at much lower energies as compared to corresponding
phenacenes. This, combined with our theoretical calculations, suggests
that the lowest singlet excitations in both acene and phenacene molecular
crystals are intramolecular in nature. Thus, experimental and theoretical
evidence suggests that as far as optoelectronic device applications
are concerned, phenacenes will be useful in higher frequency range,
as compared to oligoacenes. 

\begin{table}[H]
\caption{\label{tab: comparison_polyacene_phenaacene}Comparison of the optical
gaps of phenacenes and oligoacenes, computed using the PPP-CI approach. }

\centering{}%
\begin{tabular}{|c|cc|c|c|c|}
\hline 
Phenacenes & \multicolumn{2}{c|}{Optical } & Oligoacenes & \multicolumn{2}{c|}{Optical }\tabularnewline
(This work) & \multicolumn{2}{c|}{gap (eV)} & (Sony $et$ $al$.\citep{sony-acene-lo} and Chakraborty $et$ $al.$\citep{doi:10.1021/jp408535u}) & \multicolumn{2}{c|}{gap (eV)}\tabularnewline
\cline{2-3} \cline{5-6} 
 & Scr. & Std. &  & Scr. & Std.\tabularnewline
\hline 
Phenanthrene & 4.31 & 4.26 & Anthracene\citep{sony-acene-lo} & 3.55 & 3.66\tabularnewline
\hline 
Chrysene & 3.86 & 3.96 & Tetracene\citep{sony-acene-lo} & 2.97 & 3.16\tabularnewline
\hline 
Picene & 3.75 & 3.88 & Pentacene\citep{sony-acene-lo} & 2.65 & 2.86\tabularnewline
\hline 
Fulminene & 3.34 & 3.52 & Hexacene\citep{sony-acene-lo} & 2.38 & 2.71\tabularnewline
\hline 
{[}7{]}Phenacene & 3.34 & 3.68 & Heptacene\citep{sony-acene-lo} & 2.24 & 2.63\tabularnewline
\hline 
{[}8{]}Phenacene & 3.11 & 3.41 & Octacene\citep{doi:10.1021/jp408535u} & 1.49 & 2.24\tabularnewline
\hline 
{[}9{]}Phenacene & 3.09 & 3.46 & Nonacene\citep{doi:10.1021/jp408535u} & 1.46 & 1.82\tabularnewline
\hline 
\end{tabular}
\end{table}

\section{Summary and Conclusions \label{sec:conclusions_PQDS}}

In this paper, we presented the results of our calculations of optical
absorption spectra of {[}$n${]}phenacenes, with $n=$3-9. The calculation
were performed using both the tight-binding, and PPP models, and for
the case of PPP model, electron correlation effects were taken into
account within the configuration-interaction approach. Our calculations
reveal that the inclusion of electron correlation effect is very important
for the correct qualitative and quantitative description of optical
properties of these materials. For example, optical gaps predicted
by TB model are much smaller than their experimental values, and the
predictions of our PPP-CI calculations. We find our PPP-CI values
of the optical gaps are generally in very good agreement with the
experimentally values. Furthermore, the TB model predicts for all
the molecules that the first dipole allowed peak corresponding the
optical gap is the most intense peak of the spectrum, in complete
disagreement with the experiments, as well as results of our PPP-CI
calculations. Moreover, predictions of our PPP-CI calculations on
absorption peaks higher than the optical gap are also in very good
agreement with the experiments. 

We also compared the calculated optical gaps of {[}$n${]}phenacene
with their isomeric oligoacenes, and noted that gaps of {[}$n${]}phenacenes
are significantly larger. This is in agreement not only with numerous
optical absorption experiments performed on these molecules, but also
with a comparative EELS study of crystalline phenanthrene, chryesene,
anthracene, and tetracene.\citep{Roth2013} This further validates
our theory, and also confirms that the lowest optical excitations
in these materials are intramolecular excitons. Furthermore, this
suggests that {[}$n${]}phenacenes can have optoelectronic applications
in the higher energy range. 

In this paper, we have confined ourselves to the study of the optical
properties of {[}$n${]}phenacenes for their ground states, i.e.,
in the singlet manifold. However, in these materials, triplet states,
and their optics, are also very interesting, from the point of view
of light harvesting through the route of singlet fission, which we
aim to study in future. We also plan to explore the non-linear optical
processes in phenacenes such as two-photon absorption, and third-harmonic
generation, in future works.
\begin{acknowledgments}
This research was supported in part by Department of Science and Technology,
Government of India, under project no. SB/S2/CMP-066/2013. 
\end{acknowledgments}

\bibliographystyle{apsrev4-1}
\bibliography{phenacene}

\begin{thebibliography}{50}%
\makeatletter
\providecommand \@ifxundefined [1]{%
 \@ifx{#1\undefined}
}%
\providecommand \@ifnum [1]{%
 \ifnum #1\expandafter \@firstoftwo
 \else \expandafter \@secondoftwo
 \fi
}%
\providecommand \@ifx [1]{%
 \ifx #1\expandafter \@firstoftwo
 \else \expandafter \@secondoftwo
 \fi
}%
\providecommand \natexlab [1]{#1}%
\providecommand \enquote  [1]{``#1''}%
\providecommand \bibnamefont  [1]{#1}%
\providecommand \bibfnamefont [1]{#1}%
\providecommand \citenamefont [1]{#1}%
\providecommand \href@noop [0]{\@secondoftwo}%
\providecommand \href [0]{\begingroup \@sanitize@url \@href}%
\providecommand \@href[1]{\@@startlink{#1}\@@href}%
\providecommand \@@href[1]{\endgroup#1\@@endlink}%
\providecommand \@sanitize@url [0]{\catcode `\\12\catcode `\$12\catcode
  `\&12\catcode `\#12\catcode `\^12\catcode `\_12\catcode `\%12\relax}%
\providecommand \@@startlink[1]{}%
\providecommand \@@endlink[0]{}%
\providecommand \url  [0]{\begingroup\@sanitize@url \@url }%
\providecommand \@url [1]{\endgroup\@href {#1}{\urlprefix }}%
\providecommand \urlprefix  [0]{URL }%
\providecommand \Eprint [0]{\href }%
\providecommand \doibase [0]{http://dx.doi.org/}%
\providecommand \selectlanguage [0]{\@gobble}%
\providecommand \bibinfo  [0]{\@secondoftwo}%
\providecommand \bibfield  [0]{\@secondoftwo}%
\providecommand \translation [1]{[#1]}%
\providecommand \BibitemOpen [0]{}%
\providecommand \bibitemStop [0]{}%
\providecommand \bibitemNoStop [0]{.\EOS\space}%
\providecommand \EOS [0]{\spacefactor3000\relax}%
\providecommand \BibitemShut  [1]{\csname bibitem#1\endcsname}%
\let\auto@bib@innerbib\@empty
\bibitem [{\citenamefont {Clar}\ and\ \citenamefont
  {Schoental}(1964)}]{clar1964polycyclic}%
  \BibitemOpen
  \bibfield  {author} {\bibinfo {author} {\bibfnamefont {E.}~\bibnamefont
  {Clar}}\ and\ \bibinfo {author} {\bibfnamefont {R.}~\bibnamefont
  {Schoental}},\ }\href@noop {} {\emph {\bibinfo {title} {Polycyclic
  hydrocarbons}}},\ Vol.~\bibinfo {volume} {1}\ (\bibinfo  {publisher}
  {Springer},\ \bibinfo {year} {1964})\BibitemShut {NoStop}%
\bibitem [{\citenamefont {Dwek}\ \emph {et~al.}(1997)\citenamefont {Dwek},
  \citenamefont {Arendt}, \citenamefont {Fixsen}, \citenamefont {Sodroski},
  \citenamefont {Odegard}, \citenamefont {Weiland}, \citenamefont {Reach},
  \citenamefont {Hauser}, \citenamefont {Kelsall}, \citenamefont {Moseley}
  \emph {et~al.}}]{dwek1997detection}%
  \BibitemOpen
  \bibfield  {author} {\bibinfo {author} {\bibfnamefont {E.}~\bibnamefont
  {Dwek}}, \bibinfo {author} {\bibfnamefont {R.}~\bibnamefont {Arendt}},
  \bibinfo {author} {\bibfnamefont {D.}~\bibnamefont {Fixsen}}, \bibinfo
  {author} {\bibfnamefont {T.}~\bibnamefont {Sodroski}}, \bibinfo {author}
  {\bibfnamefont {N.}~\bibnamefont {Odegard}}, \bibinfo {author} {\bibfnamefont
  {J.}~\bibnamefont {Weiland}}, \bibinfo {author} {\bibfnamefont
  {W.}~\bibnamefont {Reach}}, \bibinfo {author} {\bibfnamefont
  {M.}~\bibnamefont {Hauser}}, \bibinfo {author} {\bibfnamefont
  {T.}~\bibnamefont {Kelsall}}, \bibinfo {author} {\bibfnamefont
  {S.}~\bibnamefont {Moseley}},  \emph {et~al.},\ }\href@noop {} {\bibfield
  {journal} {\bibinfo  {journal} {The Astrophysical Journal}\ }\textbf
  {\bibinfo {volume} {475}},\ \bibinfo {pages} {565} (\bibinfo {year}
  {1997})}\BibitemShut {NoStop}%
\bibitem [{\citenamefont {Bostr{\"o}m}\ \emph {et~al.}(2002)\citenamefont
  {Bostr{\"o}m}, \citenamefont {Gerde}, \citenamefont {Hanberg}, \citenamefont
  {Jernstr{\"o}m}, \citenamefont {Johansson}, \citenamefont {Kyrklund},
  \citenamefont {Rannug}, \citenamefont {T{\"o}rnqvist}, \citenamefont
  {Victorin},\ and\ \citenamefont {Westerholm}}]{bostrom2002cancer}%
  \BibitemOpen
  \bibfield  {author} {\bibinfo {author} {\bibfnamefont {C.-E.}\ \bibnamefont
  {Bostr{\"o}m}}, \bibinfo {author} {\bibfnamefont {P.}~\bibnamefont {Gerde}},
  \bibinfo {author} {\bibfnamefont {A.}~\bibnamefont {Hanberg}}, \bibinfo
  {author} {\bibfnamefont {B.}~\bibnamefont {Jernstr{\"o}m}}, \bibinfo {author}
  {\bibfnamefont {C.}~\bibnamefont {Johansson}}, \bibinfo {author}
  {\bibfnamefont {T.}~\bibnamefont {Kyrklund}}, \bibinfo {author}
  {\bibfnamefont {A.}~\bibnamefont {Rannug}}, \bibinfo {author} {\bibfnamefont
  {M.}~\bibnamefont {T{\"o}rnqvist}}, \bibinfo {author} {\bibfnamefont
  {K.}~\bibnamefont {Victorin}}, \ and\ \bibinfo {author} {\bibfnamefont
  {R.}~\bibnamefont {Westerholm}},\ }\href@noop {} {\bibfield  {journal}
  {\bibinfo  {journal} {Environmental health perspectives}\ }\textbf {\bibinfo
  {volume} {110}},\ \bibinfo {pages} {451} (\bibinfo {year}
  {2002})}\BibitemShut {NoStop}%
\bibitem [{\citenamefont {Lawal}(2017)}]{Abdulazeez2017}%
  \BibitemOpen
  \bibfield  {author} {\bibinfo {author} {\bibfnamefont {A.~T.}\ \bibnamefont
  {Lawal}},\ }\href {\doibase 10.1080/23311843.2017.1339841} {\bibfield
  {journal} {\bibinfo  {journal} {Cogent Environmental Science}\ }\textbf
  {\bibinfo {volume} {3}},\ \bibinfo {pages} {1339841} (\bibinfo {year}
  {2017})},\ \Eprint
  {http://arxiv.org/abs/https://www.tandfonline.com/doi/pdf/10.1080/23311843.2017.1339841}
  {https://www.tandfonline.com/doi/pdf/10.1080/23311843.2017.1339841}
  \BibitemShut {NoStop}%
\bibitem [{\citenamefont {Okamoto}\ \emph {et~al.}(2014)\citenamefont
  {Okamoto}, \citenamefont {Eguchi}, \citenamefont {Hamao}, \citenamefont
  {Goto}, \citenamefont {Gotoh}, \citenamefont {Sakai}, \citenamefont {Izumi},
  \citenamefont {Takaguchi}, \citenamefont {Gohda},\ and\ \citenamefont
  {Kubozono}}]{okamoto2014extended}%
  \BibitemOpen
  \bibfield  {author} {\bibinfo {author} {\bibfnamefont {H.}~\bibnamefont
  {Okamoto}}, \bibinfo {author} {\bibfnamefont {R.}~\bibnamefont {Eguchi}},
  \bibinfo {author} {\bibfnamefont {S.}~\bibnamefont {Hamao}}, \bibinfo
  {author} {\bibfnamefont {H.}~\bibnamefont {Goto}}, \bibinfo {author}
  {\bibfnamefont {K.}~\bibnamefont {Gotoh}}, \bibinfo {author} {\bibfnamefont
  {Y.}~\bibnamefont {Sakai}}, \bibinfo {author} {\bibfnamefont
  {M.}~\bibnamefont {Izumi}}, \bibinfo {author} {\bibfnamefont
  {Y.}~\bibnamefont {Takaguchi}}, \bibinfo {author} {\bibfnamefont
  {S.}~\bibnamefont {Gohda}}, \ and\ \bibinfo {author} {\bibfnamefont
  {Y.}~\bibnamefont {Kubozono}},\ }\href@noop {} {\bibfield  {journal}
  {\bibinfo  {journal} {Scientific reports}\ }\textbf {\bibinfo {volume} {4}},\
  \bibinfo {pages} {5330} (\bibinfo {year} {2014})}\BibitemShut {NoStop}%
\bibitem [{\citenamefont {Witte}\ and\ \citenamefont
  {W{\"o}ll}(2004)}]{witte2004growth}%
  \BibitemOpen
  \bibfield  {author} {\bibinfo {author} {\bibfnamefont {G.}~\bibnamefont
  {Witte}}\ and\ \bibinfo {author} {\bibfnamefont {C.}~\bibnamefont
  {W{\"o}ll}},\ }\href@noop {} {\bibfield  {journal} {\bibinfo  {journal}
  {Journal of Materials Research}\ }\textbf {\bibinfo {volume} {19}},\ \bibinfo
  {pages} {1889} (\bibinfo {year} {2004})}\BibitemShut {NoStop}%
\bibitem [{\citenamefont {Cicoira}\ and\ \citenamefont
  {Santato}(2007)}]{cicoira2007organic}%
  \BibitemOpen
  \bibfield  {author} {\bibinfo {author} {\bibfnamefont {F.}~\bibnamefont
  {Cicoira}}\ and\ \bibinfo {author} {\bibfnamefont {C.}~\bibnamefont
  {Santato}},\ }\href@noop {} {\bibfield  {journal} {\bibinfo  {journal}
  {Advanced Functional Materials}\ }\textbf {\bibinfo {volume} {17}},\ \bibinfo
  {pages} {3421} (\bibinfo {year} {2007})}\BibitemShut {NoStop}%
\bibitem [{\citenamefont {Raghunath}\ \emph {et~al.}(2006)\citenamefont
  {Raghunath}, \citenamefont {Reddy}, \citenamefont {Gouri}, \citenamefont
  {Bhanuprakash},\ and\ \citenamefont {Rao}}]{raghunath2006electronic}%
  \BibitemOpen
  \bibfield  {author} {\bibinfo {author} {\bibfnamefont {P.}~\bibnamefont
  {Raghunath}}, \bibinfo {author} {\bibfnamefont {M.~A.}\ \bibnamefont
  {Reddy}}, \bibinfo {author} {\bibfnamefont {C.}~\bibnamefont {Gouri}},
  \bibinfo {author} {\bibfnamefont {K.}~\bibnamefont {Bhanuprakash}}, \ and\
  \bibinfo {author} {\bibfnamefont {V.~J.}\ \bibnamefont {Rao}},\ }\href@noop
  {} {\bibfield  {journal} {\bibinfo  {journal} {The Journal of Physical
  Chemistry A}\ }\textbf {\bibinfo {volume} {110}},\ \bibinfo {pages} {1152}
  (\bibinfo {year} {2006})}\BibitemShut {NoStop}%
\bibitem [{\citenamefont {Yamashita}(2009)}]{phenacene-review-yamashita}%
  \BibitemOpen
  \bibfield  {author} {\bibinfo {author} {\bibfnamefont {Y.}~\bibnamefont
  {Yamashita}},\ }\href {\doibase 10.1088/1468-6996/10/2/024313} {\bibfield
  {journal} {\bibinfo  {journal} {Science and Technology of Advanced
  Materials}\ }\textbf {\bibinfo {volume} {10}},\ \bibinfo {pages} {024313}
  (\bibinfo {year} {2009})},\ \Eprint
  {http://arxiv.org/abs/https://doi.org/10.1088/1468-6996/10/2/024313}
  {https://doi.org/10.1088/1468-6996/10/2/024313} \BibitemShut {NoStop}%
\bibitem [{\citenamefont {Kubozono}\ \emph {et~al.}(2014)\citenamefont
  {Kubozono}, \citenamefont {He}, \citenamefont {Hamao}, \citenamefont
  {Teranishi}, \citenamefont {Goto}, \citenamefont {Eguchi}, \citenamefont
  {Kambe}, \citenamefont {Gohda},\ and\ \citenamefont
  {Nishihara}}]{kubozono2014transistor}%
  \BibitemOpen
  \bibfield  {author} {\bibinfo {author} {\bibfnamefont {Y.}~\bibnamefont
  {Kubozono}}, \bibinfo {author} {\bibfnamefont {X.}~\bibnamefont {He}},
  \bibinfo {author} {\bibfnamefont {S.}~\bibnamefont {Hamao}}, \bibinfo
  {author} {\bibfnamefont {K.}~\bibnamefont {Teranishi}}, \bibinfo {author}
  {\bibfnamefont {H.}~\bibnamefont {Goto}}, \bibinfo {author} {\bibfnamefont
  {R.}~\bibnamefont {Eguchi}}, \bibinfo {author} {\bibfnamefont
  {T.}~\bibnamefont {Kambe}}, \bibinfo {author} {\bibfnamefont
  {S.}~\bibnamefont {Gohda}}, \ and\ \bibinfo {author} {\bibfnamefont
  {Y.}~\bibnamefont {Nishihara}},\ }\href@noop {} {\bibfield  {journal}
  {\bibinfo  {journal} {European Journal of Inorganic Chemistry}\ }\textbf
  {\bibinfo {volume} {2014}},\ \bibinfo {pages} {3806} (\bibinfo {year}
  {2014})}\BibitemShut {NoStop}%
\bibitem [{\citenamefont {Shimo}\ \emph
  {et~al.}(2016{\natexlab{a}})\citenamefont {Shimo}, \citenamefont {Mikami},
  \citenamefont {Hamao}, \citenamefont {Goto}, \citenamefont {Okamoto},
  \citenamefont {Eguchi}, \citenamefont {Gohda}, \citenamefont {Hayashi},\ and\
  \citenamefont {Kubozono}}]{shimo2016synthesis}%
  \BibitemOpen
  \bibfield  {author} {\bibinfo {author} {\bibfnamefont {Y.}~\bibnamefont
  {Shimo}}, \bibinfo {author} {\bibfnamefont {T.}~\bibnamefont {Mikami}},
  \bibinfo {author} {\bibfnamefont {S.}~\bibnamefont {Hamao}}, \bibinfo
  {author} {\bibfnamefont {H.}~\bibnamefont {Goto}}, \bibinfo {author}
  {\bibfnamefont {H.}~\bibnamefont {Okamoto}}, \bibinfo {author} {\bibfnamefont
  {R.}~\bibnamefont {Eguchi}}, \bibinfo {author} {\bibfnamefont
  {S.}~\bibnamefont {Gohda}}, \bibinfo {author} {\bibfnamefont
  {Y.}~\bibnamefont {Hayashi}}, \ and\ \bibinfo {author} {\bibfnamefont
  {Y.}~\bibnamefont {Kubozono}},\ }\href@noop {} {\bibfield  {journal}
  {\bibinfo  {journal} {Scientific reports}\ }\textbf {\bibinfo {volume} {6}},\
  \bibinfo {pages} {21008} (\bibinfo {year} {2016}{\natexlab{a}})}\BibitemShut
  {NoStop}%
\bibitem [{\citenamefont {Okamoto}\ \emph {et~al.}(2013)\citenamefont
  {Okamoto}, \citenamefont {Yamaji}, \citenamefont {Gohda}, \citenamefont
  {Sato}, \citenamefont {Sugino},\ and\ \citenamefont
  {Satake}}]{okamoto2013photochemical}%
  \BibitemOpen
  \bibfield  {author} {\bibinfo {author} {\bibfnamefont {H.}~\bibnamefont
  {Okamoto}}, \bibinfo {author} {\bibfnamefont {M.}~\bibnamefont {Yamaji}},
  \bibinfo {author} {\bibfnamefont {S.}~\bibnamefont {Gohda}}, \bibinfo
  {author} {\bibfnamefont {K.}~\bibnamefont {Sato}}, \bibinfo {author}
  {\bibfnamefont {H.}~\bibnamefont {Sugino}}, \ and\ \bibinfo {author}
  {\bibfnamefont {K.}~\bibnamefont {Satake}},\ }\href@noop {} {\bibfield
  {journal} {\bibinfo  {journal} {Research on Chemical Intermediates}\ }\textbf
  {\bibinfo {volume} {39}},\ \bibinfo {pages} {147} (\bibinfo {year}
  {2013})}\BibitemShut {NoStop}%
\bibitem [{\citenamefont {Komura}\ \emph {et~al.}(2012)\citenamefont {Komura},
  \citenamefont {Goto}, \citenamefont {He}, \citenamefont {Mitamura},
  \citenamefont {Eguchi}, \citenamefont {Kaji}, \citenamefont {Okamoto},
  \citenamefont {Sugawara}, \citenamefont {Gohda}, \citenamefont {Sato} \emph
  {et~al.}}]{komura2012characteristics}%
  \BibitemOpen
  \bibfield  {author} {\bibinfo {author} {\bibfnamefont {N.}~\bibnamefont
  {Komura}}, \bibinfo {author} {\bibfnamefont {H.}~\bibnamefont {Goto}},
  \bibinfo {author} {\bibfnamefont {X.}~\bibnamefont {He}}, \bibinfo {author}
  {\bibfnamefont {H.}~\bibnamefont {Mitamura}}, \bibinfo {author}
  {\bibfnamefont {R.}~\bibnamefont {Eguchi}}, \bibinfo {author} {\bibfnamefont
  {Y.}~\bibnamefont {Kaji}}, \bibinfo {author} {\bibfnamefont {H.}~\bibnamefont
  {Okamoto}}, \bibinfo {author} {\bibfnamefont {Y.}~\bibnamefont {Sugawara}},
  \bibinfo {author} {\bibfnamefont {S.}~\bibnamefont {Gohda}}, \bibinfo
  {author} {\bibfnamefont {K.}~\bibnamefont {Sato}},  \emph {et~al.},\
  }\href@noop {} {\bibfield  {journal} {\bibinfo  {journal} {Applied Physics
  Letters}\ }\textbf {\bibinfo {volume} {101}},\ \bibinfo {pages} {083301}
  (\bibinfo {year} {2012})}\BibitemShut {NoStop}%
\bibitem [{\citenamefont {Shimo}\ \emph
  {et~al.}(2016{\natexlab{b}})\citenamefont {Shimo}, \citenamefont {Mikami},
  \citenamefont {Hamao}, \citenamefont {Goto}, \citenamefont {Okamoto},
  \citenamefont {Eguchi}, \citenamefont {Gohda}, \citenamefont {Hayashi},\ and\
  \citenamefont {Kubozono}}]{9phenacene-fet}%
  \BibitemOpen
  \bibfield  {author} {\bibinfo {author} {\bibfnamefont {Y.}~\bibnamefont
  {Shimo}}, \bibinfo {author} {\bibfnamefont {T.}~\bibnamefont {Mikami}},
  \bibinfo {author} {\bibfnamefont {S.}~\bibnamefont {Hamao}}, \bibinfo
  {author} {\bibfnamefont {H.}~\bibnamefont {Goto}}, \bibinfo {author}
  {\bibfnamefont {H.}~\bibnamefont {Okamoto}}, \bibinfo {author} {\bibfnamefont
  {R.}~\bibnamefont {Eguchi}}, \bibinfo {author} {\bibfnamefont
  {S.}~\bibnamefont {Gohda}}, \bibinfo {author} {\bibfnamefont
  {Y.}~\bibnamefont {Hayashi}}, \ and\ \bibinfo {author} {\bibfnamefont
  {Y.}~\bibnamefont {Kubozono}},\ }\href {\doibase 10.1038/srep21008}
  {\bibfield  {journal} {\bibinfo  {journal} {Scientific Reports}\ }\textbf
  {\bibinfo {volume} {6}},\ \bibinfo {pages} {21008} (\bibinfo {year}
  {2016}{\natexlab{b}})},\ \Eprint
  {http://arxiv.org/abs/https://doi.org/10.1038/srep21008}
  {https://doi.org/10.1038/srep21008} \BibitemShut {NoStop}%
\bibitem [{\citenamefont {Mitsuhashi}\ \emph {et~al.}(2010)\citenamefont
  {Mitsuhashi}, \citenamefont {Suzuki}, \citenamefont {Yamanari}, \citenamefont
  {Mitamura}, \citenamefont {Takashi~Kambe}, \citenamefont {Okamoto},
  \citenamefont {Fujiwara}, \citenamefont {Yamaji}, \citenamefont {Kawasaki},
  \citenamefont {Maniwa},\ and\ \citenamefont
  {Kubozono}}]{picene-superconductivity}%
  \BibitemOpen
  \bibfield  {author} {\bibinfo {author} {\bibfnamefont {R.}~\bibnamefont
  {Mitsuhashi}}, \bibinfo {author} {\bibfnamefont {Y.}~\bibnamefont {Suzuki}},
  \bibinfo {author} {\bibfnamefont {Y.}~\bibnamefont {Yamanari}}, \bibinfo
  {author} {\bibfnamefont {H.}~\bibnamefont {Mitamura}}, \bibinfo {author}
  {\bibfnamefont {N.~I.}\ \bibnamefont {Takashi~Kambe}}, \bibinfo {author}
  {\bibfnamefont {H.}~\bibnamefont {Okamoto}}, \bibinfo {author} {\bibfnamefont
  {A.}~\bibnamefont {Fujiwara}}, \bibinfo {author} {\bibfnamefont
  {M.}~\bibnamefont {Yamaji}}, \bibinfo {author} {\bibfnamefont
  {N.}~\bibnamefont {Kawasaki}}, \bibinfo {author} {\bibfnamefont
  {Y.}~\bibnamefont {Maniwa}}, \ and\ \bibinfo {author} {\bibfnamefont
  {Y.}~\bibnamefont {Kubozono}},\ }\href {\doibase 10.1038/nature08859}
  {\bibfield  {journal} {\bibinfo  {journal} {Nature}\ }\textbf {\bibinfo
  {volume} {464}},\ \bibinfo {pages} {76} (\bibinfo {year} {2010})},\ \Eprint
  {http://arxiv.org/abs/https://www.nature.com/articles/nature08859}
  {https://www.nature.com/articles/nature08859} \BibitemShut {NoStop}%
\bibitem [{\citenamefont {Gundra}\ and\ \citenamefont
  {Shukla}(2013)}]{springer-chapter}%
  \BibitemOpen
  \bibfield  {author} {\bibinfo {author} {\bibfnamefont {K.}~\bibnamefont
  {Gundra}}\ and\ \bibinfo {author} {\bibfnamefont {A.}~\bibnamefont
  {Shukla}},\ }\enquote {\bibinfo {title} {A pariser--parr--pople model
  hamiltonian-based approach to the electronic structure and optical properties
  of graphene nanostructures},}\ in\ \href {\doibase
  10.1007/978-94-007-6413-2_6} {\emph {\bibinfo {booktitle} {Topological
  Modelling of Nanostructures and Extended Systems}}},\ \bibinfo {editor}
  {edited by\ \bibinfo {editor} {\bibfnamefont {A.~R.}\ \bibnamefont
  {Ashrafi}}, \bibinfo {editor} {\bibfnamefont {F.}~\bibnamefont {Cataldo}},
  \bibinfo {editor} {\bibfnamefont {A.}~\bibnamefont {Iranmanesh}}, \ and\
  \bibinfo {editor} {\bibfnamefont {O.}~\bibnamefont {Ori}}}\ (\bibinfo
  {publisher} {Springer Netherlands},\ \bibinfo {address} {Dordrecht},\
  \bibinfo {year} {2013})\ pp.\ \bibinfo {pages} {199--227}\BibitemShut
  {NoStop}%
\bibitem [{\citenamefont {Sony}\ and\ \citenamefont
  {Shukla}(2007)}]{sony-acene-lo}%
  \BibitemOpen
  \bibfield  {author} {\bibinfo {author} {\bibfnamefont {P.}~\bibnamefont
  {Sony}}\ and\ \bibinfo {author} {\bibfnamefont {A.}~\bibnamefont {Shukla}},\
  }\href {\doibase 10.1103/PhysRevB.75.155208} {\bibfield  {journal} {\bibinfo
  {journal} {Phys. Rev. B}\ }\textbf {\bibinfo {volume} {75}},\ \bibinfo
  {pages} {155208} (\bibinfo {year} {2007})}\BibitemShut {NoStop}%
\bibitem [{\citenamefont {Pople}(1953)}]{ppp-pople}%
  \BibitemOpen
  \bibfield  {author} {\bibinfo {author} {\bibfnamefont {J.~A.}\ \bibnamefont
  {Pople}},\ }\href {\doibase 10.1039/TF9534901375} {\bibfield  {journal}
  {\bibinfo  {journal} {Trans. Faraday Soc.}\ }\textbf {\bibinfo {volume}
  {49}},\ \bibinfo {pages} {1375} (\bibinfo {year} {1953})}\BibitemShut
  {NoStop}%
\bibitem [{\citenamefont {Pariser}\ and\ \citenamefont
  {Parr}(1953)}]{ppp-pariser-parr}%
  \BibitemOpen
  \bibfield  {author} {\bibinfo {author} {\bibfnamefont {R.}~\bibnamefont
  {Pariser}}\ and\ \bibinfo {author} {\bibfnamefont {R.~G.}\ \bibnamefont
  {Parr}},\ }\href {\doibase http://dx.doi.org/10.1063/1.1699030} {\bibfield
  {journal} {\bibinfo  {journal} {J. Chem. Phys.}\ }\textbf {\bibinfo {volume}
  {21}},\ \bibinfo {pages} {767} (\bibinfo {year} {1953})}\BibitemShut
  {NoStop}%
\bibitem [{\citenamefont {Shukla}(2002)}]{PhysRevB.65.125204Shukla65}%
  \BibitemOpen
  \bibfield  {author} {\bibinfo {author} {\bibfnamefont {A.}~\bibnamefont
  {Shukla}},\ }\href {\doibase 10.1103/PhysRevB.65.125204} {\bibfield
  {journal} {\bibinfo  {journal} {Phys. Rev. B}\ }\textbf {\bibinfo {volume}
  {65}},\ \bibinfo {pages} {125204} (\bibinfo {year} {2002})}\BibitemShut
  {NoStop}%
\bibitem [{\citenamefont {Shukla}(2004)}]{PhysRevB.69.165218Shukla69}%
  \BibitemOpen
  \bibfield  {author} {\bibinfo {author} {\bibfnamefont {A.}~\bibnamefont
  {Shukla}},\ }\href {\doibase 10.1103/PhysRevB.69.165218} {\bibfield
  {journal} {\bibinfo  {journal} {Phys. Rev. B}\ }\textbf {\bibinfo {volume}
  {69}},\ \bibinfo {pages} {165218} (\bibinfo {year} {2004})}\BibitemShut
  {NoStop}%
\bibitem [{\citenamefont {Sony}\ and\ \citenamefont
  {Shukla}(2005)}]{PhysRevB.71.165204Priya_t0}%
  \BibitemOpen
  \bibfield  {author} {\bibinfo {author} {\bibfnamefont {P.}~\bibnamefont
  {Sony}}\ and\ \bibinfo {author} {\bibfnamefont {A.}~\bibnamefont {Shukla}},\
  }\href {\doibase 10.1103/PhysRevB.71.165204} {\bibfield  {journal} {\bibinfo
  {journal} {Phys. Rev. B}\ }\textbf {\bibinfo {volume} {71}},\ \bibinfo
  {pages} {165204} (\bibinfo {year} {2005})}\BibitemShut {NoStop}%
\bibitem [{\citenamefont {Sony}\ and\ \citenamefont
  {Shukla}(2009)}]{:/content/aip/journal/jcp/131/1/10.1063/1.3159670Priyaanthracene}%
  \BibitemOpen
  \bibfield  {author} {\bibinfo {author} {\bibfnamefont {P.}~\bibnamefont
  {Sony}}\ and\ \bibinfo {author} {\bibfnamefont {A.}~\bibnamefont {Shukla}},\
  }\href {\doibase http://dx.doi.org/10.1063/1.3159670} {\bibfield  {journal}
  {\bibinfo  {journal} {The Journal of Chemical Physics}\ }\textbf {\bibinfo
  {volume} {131}},\ \bibinfo {pages} {014302} (\bibinfo {year}
  {2009})}\BibitemShut {NoStop}%
\bibitem [{\citenamefont {Chakraborty}\ and\ \citenamefont
  {Shukla}(2013)}]{doi:10.1021/jp408535u}%
  \BibitemOpen
  \bibfield  {author} {\bibinfo {author} {\bibfnamefont {H.}~\bibnamefont
  {Chakraborty}}\ and\ \bibinfo {author} {\bibfnamefont {A.}~\bibnamefont
  {Shukla}},\ }\href {\doibase 10.1021/jp408535u} {\bibfield  {journal}
  {\bibinfo  {journal} {The Journal of Physical Chemistry A}\ }\textbf
  {\bibinfo {volume} {117}},\ \bibinfo {pages} {14220} (\bibinfo {year}
  {2013})}\BibitemShut {NoStop}%
\bibitem [{\citenamefont {Chakraborty}\ and\ \citenamefont
  {Shukla}(2014)}]{himanshu-triplet}%
  \BibitemOpen
  \bibfield  {author} {\bibinfo {author} {\bibfnamefont {H.}~\bibnamefont
  {Chakraborty}}\ and\ \bibinfo {author} {\bibfnamefont {A.}~\bibnamefont
  {Shukla}},\ }\href {\doibase http://dx.doi.org/10.1063/1.4897955} {\bibfield
  {journal} {\bibinfo  {journal} {The Journal of Chemical Physics}\ }\textbf
  {\bibinfo {volume} {141}},\ \bibinfo {pages} {164301} (\bibinfo {year}
  {2014})}\BibitemShut {NoStop}%
\bibitem [{\citenamefont {Aryanpour}\ \emph
  {et~al.}(2014{\natexlab{a}})\citenamefont {Aryanpour}, \citenamefont
  {Roberts}, \citenamefont {Sandhu}, \citenamefont {Rathore}, \citenamefont
  {Shukla},\ and\ \citenamefont {Mazumdar}}]{doi:10.1021/jp410793rAryanpour}%
  \BibitemOpen
  \bibfield  {author} {\bibinfo {author} {\bibfnamefont {K.}~\bibnamefont
  {Aryanpour}}, \bibinfo {author} {\bibfnamefont {A.}~\bibnamefont {Roberts}},
  \bibinfo {author} {\bibfnamefont {A.}~\bibnamefont {Sandhu}}, \bibinfo
  {author} {\bibfnamefont {R.}~\bibnamefont {Rathore}}, \bibinfo {author}
  {\bibfnamefont {A.}~\bibnamefont {Shukla}}, \ and\ \bibinfo {author}
  {\bibfnamefont {S.}~\bibnamefont {Mazumdar}},\ }\href {\doibase
  10.1021/jp410793r} {\bibfield  {journal} {\bibinfo  {journal} {The Journal of
  Physical Chemistry C}\ }\textbf {\bibinfo {volume} {118}},\ \bibinfo {pages}
  {3331} (\bibinfo {year} {2014}{\natexlab{a}})}\BibitemShut {NoStop}%
\bibitem [{\citenamefont {Aryanpour}\ \emph
  {et~al.}(2014{\natexlab{b}})\citenamefont {Aryanpour}, \citenamefont
  {Shukla},\ and\ \citenamefont
  {Mazumdar}}]{:/content/aip/journal/jcp/140/10/10.1063/1.4867363Aryanpour}%
  \BibitemOpen
  \bibfield  {author} {\bibinfo {author} {\bibfnamefont {K.}~\bibnamefont
  {Aryanpour}}, \bibinfo {author} {\bibfnamefont {A.}~\bibnamefont {Shukla}}, \
  and\ \bibinfo {author} {\bibfnamefont {S.}~\bibnamefont {Mazumdar}},\ }\href
  {\doibase http://dx.doi.org/10.1063/1.4867363} {\bibfield  {journal}
  {\bibinfo  {journal} {The Journal of Chemical Physics}\ }\textbf {\bibinfo
  {volume} {140}},\ \bibinfo {pages} {104301} (\bibinfo {year}
  {2014}{\natexlab{b}})}\BibitemShut {NoStop}%
\bibitem [{\citenamefont {Basak}\ \emph {et~al.}(2015)\citenamefont {Basak},
  \citenamefont {Chakraborty},\ and\ \citenamefont {Shukla}}]{Tista1}%
  \BibitemOpen
  \bibfield  {author} {\bibinfo {author} {\bibfnamefont {T.}~\bibnamefont
  {Basak}}, \bibinfo {author} {\bibfnamefont {H.}~\bibnamefont {Chakraborty}},
  \ and\ \bibinfo {author} {\bibfnamefont {A.}~\bibnamefont {Shukla}},\ }\href
  {\doibase 10.1103/PhysRevB.92.205404} {\bibfield  {journal} {\bibinfo
  {journal} {Phys. Rev. B}\ }\textbf {\bibinfo {volume} {92}},\ \bibinfo
  {pages} {205404} (\bibinfo {year} {2015})}\BibitemShut {NoStop}%
\bibitem [{\citenamefont {Basak}\ and\ \citenamefont {Shukla}(2016)}]{Tista2}%
  \BibitemOpen
  \bibfield  {author} {\bibinfo {author} {\bibfnamefont {T.}~\bibnamefont
  {Basak}}\ and\ \bibinfo {author} {\bibfnamefont {A.}~\bibnamefont {Shukla}},\
  }\href {\doibase 10.1103/PhysRevB.93.235432} {\bibfield  {journal} {\bibinfo
  {journal} {Phys. Rev. B}\ }\textbf {\bibinfo {volume} {93}},\ \bibinfo
  {pages} {235432} (\bibinfo {year} {2016})}\BibitemShut {NoStop}%
\bibitem [{\citenamefont {Ohno}(1964)}]{Theor.chim.act.2Ohno}%
  \BibitemOpen
  \bibfield  {author} {\bibinfo {author} {\bibfnamefont {K.}~\bibnamefont
  {Ohno}},\ }\href {\doibase 10.1007/BF00528281} {\bibfield  {journal}
  {\bibinfo  {journal} {Theoretica chimica acta}\ }\textbf {\bibinfo {volume}
  {2}},\ \bibinfo {pages} {219} (\bibinfo {year} {1964})}\BibitemShut {NoStop}%
\bibitem [{\citenamefont {Chandross}\ and\ \citenamefont
  {Mazumdar}(1997)}]{PhysRevB.55.1497Chandross}%
  \BibitemOpen
  \bibfield  {author} {\bibinfo {author} {\bibfnamefont {M.}~\bibnamefont
  {Chandross}}\ and\ \bibinfo {author} {\bibfnamefont {S.}~\bibnamefont
  {Mazumdar}},\ }\href {\doibase 10.1103/PhysRevB.55.1497} {\bibfield
  {journal} {\bibinfo  {journal} {Phys. Rev. B}\ }\textbf {\bibinfo {volume}
  {55}},\ \bibinfo {pages} {1497} (\bibinfo {year} {1997})}\BibitemShut
  {NoStop}%
\bibitem [{\citenamefont {Sony}\ and\ \citenamefont
  {Shukla}(2010)}]{Sony2010821}%
  \BibitemOpen
  \bibfield  {author} {\bibinfo {author} {\bibfnamefont {P.}~\bibnamefont
  {Sony}}\ and\ \bibinfo {author} {\bibfnamefont {A.}~\bibnamefont {Shukla}},\
  }\href {\doibase http://dx.doi.org/10.1016/j.cpc.2009.12.015} {\bibfield
  {journal} {\bibinfo  {journal} {Computer Physics Communications}\ }\textbf
  {\bibinfo {volume} {181}},\ \bibinfo {pages} {821 } (\bibinfo {year}
  {2010})}\BibitemShut {NoStop}%
\bibitem [{\citenamefont {Buenker}\ and\ \citenamefont
  {Peyerimhoff}(1974)}]{peyerimhoff_energy_CI}%
  \BibitemOpen
  \bibfield  {author} {\bibinfo {author} {\bibfnamefont {R.}~\bibnamefont
  {Buenker}}\ and\ \bibinfo {author} {\bibfnamefont {S.}~\bibnamefont
  {Peyerimhoff}},\ }\href {\doibase 10.1007/PL00020553} {\bibfield  {journal}
  {\bibinfo  {journal} {Theor. Chim. Acta}\ }\textbf {\bibinfo {volume} {35}},\
  \bibinfo {pages} {33} (\bibinfo {year} {1974})}\BibitemShut {NoStop}%
\bibitem [{\citenamefont {Buenker}\ \emph {et~al.}(1978)\citenamefont
  {Buenker}, \citenamefont {Peyerimhoff},\ and\ \citenamefont
  {Butscher}}]{buenker1978applicability}%
  \BibitemOpen
  \bibfield  {author} {\bibinfo {author} {\bibfnamefont {R.~J.}\ \bibnamefont
  {Buenker}}, \bibinfo {author} {\bibfnamefont {S.~D.}\ \bibnamefont
  {Peyerimhoff}}, \ and\ \bibinfo {author} {\bibfnamefont {W.}~\bibnamefont
  {Butscher}},\ }\href@noop {} {\bibfield  {journal} {\bibinfo  {journal}
  {Molecular Physics}\ }\textbf {\bibinfo {volume} {35}},\ \bibinfo {pages}
  {771} (\bibinfo {year} {1978})}\BibitemShut {NoStop}%
\bibitem [{\citenamefont {Klevens}\ and\ \citenamefont
  {Platt}(1949)}]{klevens1949spectral}%
  \BibitemOpen
  \bibfield  {author} {\bibinfo {author} {\bibfnamefont {H.}~\bibnamefont
  {Klevens}}\ and\ \bibinfo {author} {\bibfnamefont {J.}~\bibnamefont
  {Platt}},\ }\href@noop {} {\bibfield  {journal} {\bibinfo  {journal} {The
  Journal of Chemical Physics}\ }\textbf {\bibinfo {volume} {17}},\ \bibinfo
  {pages} {470} (\bibinfo {year} {1949})}\BibitemShut {NoStop}%
\bibitem [{\citenamefont {Parac}\ and\ \citenamefont
  {Grimme}(2003)}]{parac2003tddft}%
  \BibitemOpen
  \bibfield  {author} {\bibinfo {author} {\bibfnamefont {M.}~\bibnamefont
  {Parac}}\ and\ \bibinfo {author} {\bibfnamefont {S.}~\bibnamefont {Grimme}},\
  }\href@noop {} {\bibfield  {journal} {\bibinfo  {journal} {Chemical physics}\
  }\textbf {\bibinfo {volume} {292}},\ \bibinfo {pages} {11} (\bibinfo {year}
  {2003})}\BibitemShut {NoStop}%
\bibitem [{\citenamefont {Malloci}\ \emph {et~al.}(2011)\citenamefont
  {Malloci}, \citenamefont {Cappellini}, \citenamefont {Mulas},\ and\
  \citenamefont {Mattoni}}]{malloci2011electronic}%
  \BibitemOpen
  \bibfield  {author} {\bibinfo {author} {\bibfnamefont {G.}~\bibnamefont
  {Malloci}}, \bibinfo {author} {\bibfnamefont {G.}~\bibnamefont {Cappellini}},
  \bibinfo {author} {\bibfnamefont {G.}~\bibnamefont {Mulas}}, \ and\ \bibinfo
  {author} {\bibfnamefont {A.}~\bibnamefont {Mattoni}},\ }\href@noop {}
  {\bibfield  {journal} {\bibinfo  {journal} {Chemical Physics}\ }\textbf
  {\bibinfo {volume} {384}},\ \bibinfo {pages} {19} (\bibinfo {year}
  {2011})}\BibitemShut {NoStop}%
\bibitem [{\citenamefont {Birks}(1970)}]{birks1970photophysics}%
  \BibitemOpen
  \bibfield  {author} {\bibinfo {author} {\bibfnamefont {J.~B.}\ \bibnamefont
  {Birks}},\ }\href@noop {} {\emph {\bibinfo {title} {Photophysics of aromatic
  molecules}}}\ (\bibinfo {year} {1970})\BibitemShut {NoStop}%
\bibitem [{\citenamefont {Salama}\ and\ \citenamefont
  {Allamandola}(1993)}]{salama1993neutral}%
  \BibitemOpen
  \bibfield  {author} {\bibinfo {author} {\bibfnamefont {F.}~\bibnamefont
  {Salama}}\ and\ \bibinfo {author} {\bibfnamefont {L.~J.}\ \bibnamefont
  {Allamandola}},\ }\href@noop {} {\bibfield  {journal} {\bibinfo  {journal}
  {Journal of the Chemical lectronic absorption spectra of PAHsSociety, Faraday
  Transactions}\ }\textbf {\bibinfo {volume} {89}},\ \bibinfo {pages} {2277}
  (\bibinfo {year} {1993})}\BibitemShut {NoStop}%
\bibitem [{\citenamefont {Halasinski}\ \emph {et~al.}(2005)\citenamefont
  {Halasinski}, \citenamefont {Salama},\ and\ \citenamefont
  {Allamandola}}]{halasinski2005investigation}%
  \BibitemOpen
  \bibfield  {author} {\bibinfo {author} {\bibfnamefont {T.}~\bibnamefont
  {Halasinski}}, \bibinfo {author} {\bibfnamefont {F.}~\bibnamefont {Salama}},
  \ and\ \bibinfo {author} {\bibfnamefont {L.}~\bibnamefont {Allamandola}},\
  }\href@noop {} {\bibfield  {journal} {\bibinfo  {journal} {The Astrophysical
  Journal}\ }\textbf {\bibinfo {volume} {628}},\ \bibinfo {pages} {555}
  (\bibinfo {year} {2005})}\BibitemShut {NoStop}%
\bibitem [{\citenamefont {Skancke}(1965)}]{skancke1965semi}%
  \BibitemOpen
  \bibfield  {author} {\bibinfo {author} {\bibfnamefont {P.}~\bibnamefont
  {Skancke}},\ }\href@noop {} {\bibfield  {journal} {\bibinfo  {journal} {ACTA
  CHEMICA SCANDINAVICA}\ }\textbf {\bibinfo {volume} {19}},\ \bibinfo {pages}
  {401} (\bibinfo {year} {1965})}\BibitemShut {NoStop}%
\bibitem [{\citenamefont {Malloci}\ \emph {et~al.}(2004)\citenamefont
  {Malloci}, \citenamefont {Mulas},\ and\ \citenamefont
  {Joblin}}]{malloci2004electronic}%
  \BibitemOpen
  \bibfield  {author} {\bibinfo {author} {\bibfnamefont {G.}~\bibnamefont
  {Malloci}}, \bibinfo {author} {\bibfnamefont {G.}~\bibnamefont {Mulas}}, \
  and\ \bibinfo {author} {\bibfnamefont {C.}~\bibnamefont {Joblin}},\
  }\href@noop {} {\bibfield  {journal} {\bibinfo  {journal} {Astronomy \&
  Astrophysics}\ }\textbf {\bibinfo {volume} {426}},\ \bibinfo {pages} {105}
  (\bibinfo {year} {2004})}\BibitemShut {NoStop}%
\bibitem [{\citenamefont {Hedges}\ and\ \citenamefont
  {Phillips}(1968)}]{hedges1968electronic}%
  \BibitemOpen
  \bibfield  {author} {\bibinfo {author} {\bibfnamefont {R.}~\bibnamefont
  {Hedges}}\ and\ \bibinfo {author} {\bibfnamefont {L.}~\bibnamefont
  {Phillips}},\ }\href@noop {} {\bibfield  {journal} {\bibinfo  {journal}
  {Theoretical Chemistry Accounts: Theory, Computation, and Modeling
  (Theoretica Chimica Acta)}\ }\textbf {\bibinfo {volume} {10}},\ \bibinfo
  {pages} {73} (\bibinfo {year} {1968})}\BibitemShut {NoStop}%
\bibitem [{\citenamefont {Becker}\ \emph {et~al.}(1963)\citenamefont {Becker},
  \citenamefont {Singh},\ and\ \citenamefont
  {Jackson}}]{becker1963comprehensive}%
  \BibitemOpen
  \bibfield  {author} {\bibinfo {author} {\bibfnamefont {R.~S.}\ \bibnamefont
  {Becker}}, \bibinfo {author} {\bibfnamefont {I.~S.}\ \bibnamefont {Singh}}, \
  and\ \bibinfo {author} {\bibfnamefont {E.~A.}\ \bibnamefont {Jackson}},\
  }\href@noop {} {\bibfield  {journal} {\bibinfo  {journal} {The Journal of
  Chemical Physics}\ }\textbf {\bibinfo {volume} {38}},\ \bibinfo {pages}
  {2144} (\bibinfo {year} {1963})}\BibitemShut {NoStop}%
\bibitem [{\citenamefont {Mallory}\ \emph {et~al.}(1997)\citenamefont
  {Mallory}, \citenamefont {Butler}, \citenamefont {Evans}, \citenamefont
  {Brondyke}, \citenamefont {Mallory}, \citenamefont {Yang},\ and\
  \citenamefont {Ellenstein}}]{mallory1997phenacenes}%
  \BibitemOpen
  \bibfield  {author} {\bibinfo {author} {\bibfnamefont {F.~B.}\ \bibnamefont
  {Mallory}}, \bibinfo {author} {\bibfnamefont {K.~E.}\ \bibnamefont {Butler}},
  \bibinfo {author} {\bibfnamefont {A.~C.}\ \bibnamefont {Evans}}, \bibinfo
  {author} {\bibfnamefont {E.~J.}\ \bibnamefont {Brondyke}}, \bibinfo {author}
  {\bibfnamefont {C.~W.}\ \bibnamefont {Mallory}}, \bibinfo {author}
  {\bibfnamefont {C.}~\bibnamefont {Yang}}, \ and\ \bibinfo {author}
  {\bibfnamefont {A.}~\bibnamefont {Ellenstein}},\ }\href@noop {} {\bibfield
  {journal} {\bibinfo  {journal} {Journal of the American Chemical Society}\
  }\textbf {\bibinfo {volume} {119}},\ \bibinfo {pages} {2119} (\bibinfo {year}
  {1997})}\BibitemShut {NoStop}%
\bibitem [{\citenamefont {Ham}\ and\ \citenamefont
  {Ruedenberg}(1956)}]{ham1956electronic}%
  \BibitemOpen
  \bibfield  {author} {\bibinfo {author} {\bibfnamefont {N.~S.}\ \bibnamefont
  {Ham}}\ and\ \bibinfo {author} {\bibfnamefont {K.}~\bibnamefont
  {Ruedenberg}},\ }\href@noop {} {\bibfield  {journal} {\bibinfo  {journal}
  {The Journal of Chemical Physics}\ }\textbf {\bibinfo {volume} {25}},\
  \bibinfo {pages} {13} (\bibinfo {year} {1956})}\BibitemShut {NoStop}%
\bibitem [{\citenamefont {Fanetti}\ \emph {et~al.}(2012)\citenamefont
  {Fanetti}, \citenamefont {Citroni}, \citenamefont {Bini}, \citenamefont
  {Malavasi}, \citenamefont {Artioli},\ and\ \citenamefont
  {Postorino}}]{fanetti2012homo}%
  \BibitemOpen
  \bibfield  {author} {\bibinfo {author} {\bibfnamefont {S.}~\bibnamefont
  {Fanetti}}, \bibinfo {author} {\bibfnamefont {M.}~\bibnamefont {Citroni}},
  \bibinfo {author} {\bibfnamefont {R.}~\bibnamefont {Bini}}, \bibinfo {author}
  {\bibfnamefont {L.}~\bibnamefont {Malavasi}}, \bibinfo {author}
  {\bibfnamefont {G.~A.}\ \bibnamefont {Artioli}}, \ and\ \bibinfo {author}
  {\bibfnamefont {P.}~\bibnamefont {Postorino}},\ }\href@noop {} {\bibfield
  {journal} {\bibinfo  {journal} {The Journal of chemical physics}\ }\textbf
  {\bibinfo {volume} {137}},\ \bibinfo {pages} {224506} (\bibinfo {year}
  {2012})}\BibitemShut {NoStop}%
\bibitem [{\citenamefont {Dutta}\ and\ \citenamefont
  {Mazumdar}(2014)}]{tirtho-phenanthrene2014}%
  \BibitemOpen
  \bibfield  {author} {\bibinfo {author} {\bibfnamefont {T.}~\bibnamefont
  {Dutta}}\ and\ \bibinfo {author} {\bibfnamefont {S.}~\bibnamefont
  {Mazumdar}},\ }\href {\doibase 10.1103/PhysRevB.89.245129} {\bibfield
  {journal} {\bibinfo  {journal} {Phys. Rev. B}\ }\textbf {\bibinfo {volume}
  {89}},\ \bibinfo {pages} {245129} (\bibinfo {year} {2014})}\BibitemShut
  {NoStop}%
\bibitem [{\citenamefont {Dutta}\ and\ \citenamefont
  {Mazumdar}(2016)}]{tirtho-phenanthrene2016}%
  \BibitemOpen
  \bibfield  {author} {\bibinfo {author} {\bibfnamefont {T.}~\bibnamefont
  {Dutta}}\ and\ \bibinfo {author} {\bibfnamefont {S.}~\bibnamefont
  {Mazumdar}},\ }\href {https://arxiv.org/abs/1607.03198} {\bibfield  {journal}
  {\bibinfo  {journal} {arXiv:1607.03198}\ } (\bibinfo {year}
  {2016})}\BibitemShut {NoStop}%
\bibitem [{\citenamefont {Roth}\ \emph {et~al.}(2013)\citenamefont {Roth},
  \citenamefont {Mahns}, \citenamefont {Hampel}, \citenamefont {Nohr},
  \citenamefont {Berger}, \citenamefont {B{\"u}chner},\ and\ \citenamefont
  {Knupfer}}]{Roth2013}%
  \BibitemOpen
  \bibfield  {author} {\bibinfo {author} {\bibfnamefont {F.}~\bibnamefont
  {Roth}}, \bibinfo {author} {\bibfnamefont {B.}~\bibnamefont {Mahns}},
  \bibinfo {author} {\bibfnamefont {S.}~\bibnamefont {Hampel}}, \bibinfo
  {author} {\bibfnamefont {M.}~\bibnamefont {Nohr}}, \bibinfo {author}
  {\bibfnamefont {H.}~\bibnamefont {Berger}}, \bibinfo {author} {\bibfnamefont
  {B.}~\bibnamefont {B{\"u}chner}}, \ and\ \bibinfo {author} {\bibfnamefont
  {M.}~\bibnamefont {Knupfer}},\ }\href {\doibase 10.1140/epjb/e2012-30592-1}
  {\bibfield  {journal} {\bibinfo  {journal} {The European Physical Journal B}\
  }\textbf {\bibinfo {volume} {86}},\ \bibinfo {pages} {66} (\bibinfo {year}
  {2013})}\BibitemShut {NoStop}%
\end{thebibliography}%

\end{document}


\title{Supporting Information for A Pariser-Parr-Pople Model Based Study
of Optoelectronic Properties of Phenacenes}
\author{{\normalsize{}Deepak Kumar Rai}}
\email{dkriitb@gmail.com}

\author{Alok Shukla}
\email{shukla@phy.iitb.ac.in}

\affiliation{Department of Physics, Indian Institute of Technology Bombay, Powai,
Mumbai 400076, India}

\maketitle
The following tables contain the excitation energies, dominant many-body
wave functions, and transition dipole matrix elements of excited states
with respect to the ground state $1^{1}$A$_{g}$ for $C_{2h}$, and
$1^{1}A_{1}$ for $C_{2v}$ molecules. The symbols $H$ and $L$ represent
the highest occupied molecular orbital (HOMO), and the lowest unoccupied
molecular orbital (LUMO), respectively. Many-electron wave functions
of various excited states are written as linear combinations of configurations,
and the magnitudes of their coefficients are included in parentheses
next to them. The configurations are expressed as single-, double-,...,
excitations with respect to the closed-shell restricted-Hartree-Fock
(RHF) reference state, and  the configurations are denoted using arrows
from the occupied to unoccupied (virtual) orbitals, representing a
given $n$-particle excitation. The charge conjugate of a given configuration
is abbreviated as '$c.c.$', while the sign (+/-) preceding '$c.c.$'
indicates that the two coefficients have (same/opposite) signs. The
symbol D.F. denotes a dipole-forbidden state.

\begin{table}[H]
\protect\caption{\label{wave_analysis_14atoms_scr_PQD}Excited states giving rise to
peaks in the singlet linear absorption spectrum of phenanthrene, computed
employing the FCI approach along with the screened parameters in the
PPP model Hamiltonian. }

\selectlanguage{american}%
\begin{tabular}{ccccc}
\hline 
\selectlanguage{english}%
Peak\selectlanguage{american}%
 & \selectlanguage{english}%
State \selectlanguage{american}%
 & \selectlanguage{english}%
E ($eV$)\selectlanguage{american}%
 & \selectlanguage{english}%
Transition \selectlanguage{american}%
 & \selectlanguage{english}%
Dominant Contributing \selectlanguage{american}%
\tabularnewline
\selectlanguage{english}%
\selectlanguage{american}%
 & \selectlanguage{english}%
\selectlanguage{american}%
 & \selectlanguage{english}%
\selectlanguage{american}%
 & \selectlanguage{english}%
Dipole (\AA)\selectlanguage{american}%
 & \selectlanguage{english}%
Configurations\selectlanguage{american}%
\tabularnewline
\hline 
\selectlanguage{english}%
D.F.\selectlanguage{american}%
 & \selectlanguage{english}%
$2^{1}A_{1}$\selectlanguage{american}%
 & \selectlanguage{english}%
2.96\selectlanguage{american}%
 & \selectlanguage{english}%
0\selectlanguage{american}%
 & \selectlanguage{english}%
$\left|H\rightarrow L+1\right\rangle $$+c.c.$ (0.5564)\selectlanguage{american}%
\tabularnewline
\selectlanguage{english}%
\selectlanguage{american}%
 & \selectlanguage{english}%
\selectlanguage{american}%
 & \selectlanguage{english}%
\selectlanguage{american}%
 & \selectlanguage{english}%
\selectlanguage{american}%
 & \selectlanguage{english}%
$\left|H-4\rightarrow L\right\rangle $$-c.c.$(0.1076)\selectlanguage{american}%
\tabularnewline
$I_{x}$ & $1$$^{1}B_{1}$ & $4.31$ & $1.2393$ & \selectlanguage{english}%
$\left|H\rightarrow L\right\rangle $ $(0.8858)$\selectlanguage{american}%
\tabularnewline
\selectlanguage{english}%
\selectlanguage{american}%
 & \selectlanguage{english}%
\selectlanguage{american}%
 & \selectlanguage{english}%
\selectlanguage{american}%
 & \selectlanguage{english}%
\selectlanguage{american}%
 & \selectlanguage{english}%
$\left|H\rightarrow L;H-3\rightarrow L+3;H-1\rightarrow L+1\right\rangle $
$(0.0974)$\selectlanguage{american}%
\tabularnewline
\hline 
\end{tabular}\selectlanguage{english}%
\end{table}
\begin{table}[H]
\protect\caption{\label{wave_analysis_14atoms_std_PQD}Excited states giving rise to
peaks in the singlet linear absorption spectrum of phenanthrene, computed
employing the FCI approach along with the standard parameters in the
PPP model Hamiltonian. }

\selectlanguage{american}%
\begin{tabular}{ccccc}
\hline 
\selectlanguage{english}%
Peak\selectlanguage{american}%
 & \selectlanguage{english}%
State \selectlanguage{american}%
 & \selectlanguage{english}%
E ($eV$)\selectlanguage{american}%
 & \selectlanguage{english}%
Transition \selectlanguage{american}%
 & \selectlanguage{english}%
Dominant Contributing \selectlanguage{american}%
\tabularnewline
\selectlanguage{english}%
\selectlanguage{american}%
 & \selectlanguage{english}%
\selectlanguage{american}%
 & \selectlanguage{english}%
\selectlanguage{american}%
 & \selectlanguage{english}%
Dipole (\AA)\selectlanguage{american}%
 & \selectlanguage{english}%
Configurations\selectlanguage{american}%
\tabularnewline
\hline 
\selectlanguage{english}%
D.F.\selectlanguage{american}%
 & \selectlanguage{english}%
$2^{1}A_{1}$\selectlanguage{american}%
 & \selectlanguage{english}%
3.35\selectlanguage{american}%
 & \selectlanguage{english}%
0\selectlanguage{american}%
 & \selectlanguage{english}%
$\left|H\rightarrow L+1\right\rangle $$-c.c.$ (0.5841)\selectlanguage{american}%
\tabularnewline
\selectlanguage{english}%
\selectlanguage{american}%
 & \selectlanguage{english}%
\selectlanguage{american}%
 & \selectlanguage{english}%
\selectlanguage{american}%
 & \selectlanguage{english}%
\selectlanguage{american}%
 & \selectlanguage{english}%
$\left|H-4\rightarrow L\right\rangle $$-c.c.$(0.1411)\selectlanguage{american}%
\tabularnewline
$I_{x}$ & $1$$^{1}B_{1}$ & $4.26$ & $0.5925$ & \selectlanguage{english}%
$\left|H\rightarrow L\right\rangle $ $(0.7891)$\selectlanguage{american}%
\tabularnewline
\selectlanguage{english}%
\selectlanguage{american}%
 & \selectlanguage{english}%
\selectlanguage{american}%
 & \selectlanguage{english}%
\selectlanguage{american}%
 & \selectlanguage{english}%
\selectlanguage{american}%
 & \selectlanguage{english}%
$\left|H-1\rightarrow L+1\right\rangle $ $(0.4482)$\selectlanguage{american}%
\tabularnewline
\hline 
\end{tabular}\selectlanguage{english}%
\end{table}
\begin{table}[H]
\protect\caption{\label{wave_analysis_18atoms_scr_PQD}Excited states giving rise to
peaks in the singlet linear absorption spectrum of chrysene, computed
employing the QCI approach along with the screened parameters in the
PPP model Hamiltonian. }

\selectlanguage{american}%
\begin{tabular}{ccccc}
\hline 
\selectlanguage{english}%
Peak\selectlanguage{american}%
 & \selectlanguage{english}%
State \selectlanguage{american}%
 & \selectlanguage{english}%
E ($eV$)\selectlanguage{american}%
 & \selectlanguage{english}%
Transition \selectlanguage{american}%
 & \selectlanguage{english}%
Dominant Contributing \selectlanguage{american}%
\tabularnewline
\selectlanguage{english}%
\selectlanguage{american}%
 & \selectlanguage{english}%
\selectlanguage{american}%
 & \selectlanguage{english}%
\selectlanguage{american}%
 & \selectlanguage{english}%
Dipole (\AA)\selectlanguage{american}%
 & \selectlanguage{english}%
Configurations\selectlanguage{american}%
\tabularnewline
\hline 
D.F. & $1$$^{1}B_{u}$ & 3.11 & 0 & \selectlanguage{english}%
$\left|H\rightarrow L+1\right\rangle $$-c.c.$$(0.5658)$\selectlanguage{american}%
\tabularnewline
\selectlanguage{english}%
\selectlanguage{american}%
 & \selectlanguage{english}%
\selectlanguage{american}%
 & \selectlanguage{english}%
\selectlanguage{american}%
 & \selectlanguage{english}%
\selectlanguage{american}%
 & \selectlanguage{english}%
$\left|H-3\rightarrow L+2\right\rangle $$-c.c.$$(0.1241)$\selectlanguage{american}%
\tabularnewline
$I_{xy}$ & $2$$^{1}B_{u}$ & $3.86$ & $1.2820$ & \selectlanguage{english}%
$\left|H\rightarrow L\right\rangle $ $(0.8696)$\selectlanguage{american}%
\tabularnewline
\selectlanguage{english}%
\selectlanguage{american}%
 & \selectlanguage{english}%
\selectlanguage{american}%
 & \selectlanguage{english}%
\selectlanguage{american}%
 & \selectlanguage{english}%
\selectlanguage{american}%
 & \selectlanguage{english}%
$\left|H-1\rightarrow L+3;H\rightarrow L;H-3\rightarrow L+1\right\rangle $
$(0.0766)$\selectlanguage{american}%
\tabularnewline
$II_{xy}$ & $3$$^{1}B_{u}$ & $4.52$ & $1.7260$ & \selectlanguage{english}%
$\left|H-1\rightarrow L\right\rangle $$+c.c.$$(0.6034)$\selectlanguage{american}%
\tabularnewline
\hline 
\end{tabular}\selectlanguage{english}%
\end{table}
\begin{table}[H]
\protect\caption{\label{wave_analysis_18atoms_std_PQD}Excited states giving rise to
peaks in the singlet linear absorption spectrum of chrysene, computed
employing the QCI approach along with the standard parameters in the
PPP model Hamiltonian. }

\selectlanguage{american}%
\begin{tabular}{ccccc}
\hline 
\selectlanguage{english}%
Peak\selectlanguage{american}%
 & \selectlanguage{english}%
State \selectlanguage{american}%
 & \selectlanguage{english}%
E ($eV$)\selectlanguage{american}%
 & \selectlanguage{english}%
Transition \selectlanguage{american}%
 & \selectlanguage{english}%
Dominant Contributing \selectlanguage{american}%
\tabularnewline
\selectlanguage{english}%
\selectlanguage{american}%
 & \selectlanguage{english}%
\selectlanguage{american}%
 & \selectlanguage{english}%
\selectlanguage{american}%
 & \selectlanguage{english}%
Dipole (\AA)\selectlanguage{american}%
 & \selectlanguage{english}%
Configurations\selectlanguage{american}%
\tabularnewline
\hline 
D.F. & $1$$^{1}B_{u}$ & 3.33 & 0 & \selectlanguage{english}%
$\left|H\rightarrow L+1\right\rangle $$+c.c.$$(0.5783)$\selectlanguage{american}%
\tabularnewline
\selectlanguage{english}%
\selectlanguage{american}%
 & \selectlanguage{english}%
\selectlanguage{american}%
 & \selectlanguage{english}%
\selectlanguage{american}%
 & \selectlanguage{english}%
\selectlanguage{american}%
 & \selectlanguage{english}%
$\left|H-3\rightarrow L+2\right\rangle $$+c.c.$$(0.1720)$\selectlanguage{american}%
\tabularnewline
$I_{xy}$ & $2$$^{1}B_{u}$ & $3.96$ & $0.9099$ & \selectlanguage{english}%
$\left|H\rightarrow L\right\rangle $ $(0.8292)$\selectlanguage{american}%
\tabularnewline
\selectlanguage{english}%
\selectlanguage{american}%
 & \selectlanguage{english}%
\selectlanguage{american}%
 & \selectlanguage{english}%
\selectlanguage{american}%
 & \selectlanguage{english}%
\selectlanguage{american}%
 & \selectlanguage{english}%
$\left|H-1\rightarrow L+1\right\rangle $ $(0.3102)$\selectlanguage{american}%
\tabularnewline
$II_{xy}$ & $4$$^{1}B_{u}$ & $5.08$ & $1.8681$ & \selectlanguage{english}%
$\left|H\rightarrow L+1\right\rangle $$-c.c.$$(0.5097)$\selectlanguage{american}%
\tabularnewline
\hline 
\end{tabular}\selectlanguage{english}%
\end{table}
\begin{table}[H]
\protect\caption{\label{wave_analysis_22atoms_scr_PQD}Excited states giving rise to
peaks in the singlet linear absorption spectrum of picene, computed
employing the QCI approach along with the screened parameters in the
PPP model Hamiltonian. }

\selectlanguage{american}%
\begin{tabular}{ccccc}
\hline 
\selectlanguage{english}%
Peak\selectlanguage{american}%
 & \selectlanguage{english}%
State \selectlanguage{american}%
 & \selectlanguage{english}%
E ($eV$)\selectlanguage{american}%
 & \selectlanguage{english}%
Transition \selectlanguage{american}%
 & \selectlanguage{english}%
Dominant Contributing \selectlanguage{american}%
\tabularnewline
\selectlanguage{english}%
\selectlanguage{american}%
 & \selectlanguage{english}%
\selectlanguage{american}%
 & \selectlanguage{english}%
\selectlanguage{american}%
 & \selectlanguage{english}%
Dipole (\AA)\selectlanguage{american}%
 & \selectlanguage{english}%
Configurations\selectlanguage{american}%
\tabularnewline
\hline 
\selectlanguage{english}%
D.F.\selectlanguage{american}%
 & \selectlanguage{english}%
$2^{1}A_{1}$\selectlanguage{american}%
 & \selectlanguage{english}%
3.20\selectlanguage{american}%
 & \selectlanguage{english}%
0\selectlanguage{american}%
 & \selectlanguage{english}%
$\left|H\rightarrow L+1\right\rangle $$-c.c.$ (0.5072)\selectlanguage{american}%
\tabularnewline
\selectlanguage{english}%
\selectlanguage{american}%
 & \selectlanguage{english}%
\selectlanguage{american}%
 & \selectlanguage{english}%
\selectlanguage{american}%
 & \selectlanguage{english}%
\selectlanguage{american}%
 & \selectlanguage{english}%
$\left|H\rightarrow L+2\right\rangle $$-c.c.$(0.2177)\selectlanguage{american}%
\tabularnewline
$I_{x}$ & $1$$^{1}B_{1}$ & $3.75$ & $1.7007$ & \selectlanguage{english}%
$\left|H\rightarrow L\right\rangle $ $(0.8521)$\selectlanguage{american}%
\tabularnewline
\selectlanguage{english}%
\selectlanguage{american}%
 & \selectlanguage{english}%
\selectlanguage{american}%
 & \selectlanguage{english}%
\selectlanguage{american}%
 & \selectlanguage{english}%
\selectlanguage{american}%
 & \selectlanguage{english}%
$\left|H\rightarrow L;H-3\rightarrow L+1;H-1\rightarrow L+3\right\rangle $
$(0.0651)$\selectlanguage{american}%
\tabularnewline
\hline 
\end{tabular}\selectlanguage{english}%
\end{table}
\begin{table}[H]
\protect\caption{\label{wave_analysis_22atoms_std_PQD}Excited states giving rise to
peaks in the singlet linear absorption spectrum of picene, computed
employing the QCI approach along with the standard parameters in the
PPP model Hamiltonian. }

\selectlanguage{american}%
\begin{tabular}{ccccc}
\hline 
\selectlanguage{english}%
Peak\selectlanguage{american}%
 & \selectlanguage{english}%
State \selectlanguage{american}%
 & \selectlanguage{english}%
E ($eV$)\selectlanguage{american}%
 & \selectlanguage{english}%
Transition \selectlanguage{american}%
 & \selectlanguage{english}%
Dominant Contributing \selectlanguage{american}%
\tabularnewline
\selectlanguage{english}%
\selectlanguage{american}%
 & \selectlanguage{english}%
\selectlanguage{american}%
 & \selectlanguage{english}%
\selectlanguage{american}%
 & \selectlanguage{english}%
Dipole (\AA)\selectlanguage{american}%
 & \selectlanguage{english}%
Configurations\selectlanguage{american}%
\tabularnewline
\hline 
\selectlanguage{english}%
D.F.\selectlanguage{american}%
 & \selectlanguage{english}%
$2^{1}A_{1}$\selectlanguage{american}%
 & \selectlanguage{english}%
3.33\selectlanguage{american}%
 & \selectlanguage{english}%
0\selectlanguage{american}%
 & \selectlanguage{english}%
$\left|H\rightarrow L+1\right\rangle $$+c.c.$ (0.5280)\selectlanguage{american}%
\tabularnewline
\selectlanguage{english}%
\selectlanguage{american}%
 & \selectlanguage{english}%
\selectlanguage{american}%
 & \selectlanguage{english}%
\selectlanguage{american}%
 & \selectlanguage{english}%
\selectlanguage{american}%
 & \selectlanguage{english}%
$\left|H\rightarrow L+2\right\rangle $$-c.c.$(0.1878)\selectlanguage{american}%
\tabularnewline
$I_{x}$ & $1$$^{1}B_{1}$ & $3.88$ & $1.1982$ & \selectlanguage{english}%
$\left|H\rightarrow L\right\rangle $ $(0.8100)$\selectlanguage{american}%
\tabularnewline
\selectlanguage{english}%
\selectlanguage{american}%
 & \selectlanguage{english}%
\selectlanguage{american}%
 & \selectlanguage{english}%
\selectlanguage{american}%
 & \selectlanguage{english}%
\selectlanguage{american}%
 & \selectlanguage{english}%
$\left|H-1\rightarrow L+1\right\rangle $ $(0.2910)$\selectlanguage{american}%
\tabularnewline
\hline 
\end{tabular}\selectlanguage{english}%
\end{table}
\begin{table}[H]
\protect\caption{\label{wave_analysis_26atoms_scr_PQD}Excited states giving rise to
peaks in the singlet linear absorption spectrum of fulminene, computed
employing the MRSDCI approach along with the screened parameters in
the PPP model Hamiltonian. }

\selectlanguage{american}%
\begin{tabular}{ccccc}
\hline 
\selectlanguage{english}%
Peak\selectlanguage{american}%
 & \selectlanguage{english}%
State \selectlanguage{american}%
 & \selectlanguage{english}%
E ($eV$)\selectlanguage{american}%
 & \selectlanguage{english}%
Transition \selectlanguage{american}%
 & \selectlanguage{english}%
Dominant Contributing \selectlanguage{american}%
\tabularnewline
\selectlanguage{english}%
\selectlanguage{american}%
 & \selectlanguage{english}%
\selectlanguage{american}%
 & \selectlanguage{english}%
\selectlanguage{american}%
 & \selectlanguage{english}%
Dipole (\AA)\selectlanguage{american}%
 & \selectlanguage{english}%
Configurations\selectlanguage{american}%
\tabularnewline
\hline 
D.F. & $1$$^{1}B_{u}$ & 2.86 & 0 & \selectlanguage{english}%
$\left|H\rightarrow L+1\right\rangle $$-c.c.$$(0.5642)$\selectlanguage{american}%
\tabularnewline
\selectlanguage{english}%
\selectlanguage{american}%
 & \selectlanguage{english}%
\selectlanguage{american}%
 & \selectlanguage{english}%
\selectlanguage{american}%
 & \selectlanguage{english}%
\selectlanguage{american}%
 & \selectlanguage{english}%
$\left|H-3\rightarrow L+2\right\rangle $$-c.c.$$(0.2027)$\selectlanguage{american}%
\tabularnewline
$I_{xy}$ & $2$$^{1}B_{u}$ & $3.34$ & $1.7446$ & \selectlanguage{english}%
$\left|H\rightarrow L\right\rangle $ $(0.8729)$\selectlanguage{american}%
\tabularnewline
\selectlanguage{english}%
\selectlanguage{american}%
 & \selectlanguage{english}%
\selectlanguage{american}%
 & \selectlanguage{english}%
\selectlanguage{american}%
 & \selectlanguage{english}%
\selectlanguage{american}%
 & \selectlanguage{english}%
$\left|H\rightarrow L+1\right\rangle $$+c.c.$ $(0.0530)$\selectlanguage{american}%
\tabularnewline
$II_{xy}$ & $4$$^{1}B_{u}$ & $4.01$ & $1.6754$ & \selectlanguage{english}%
$\left|H\rightarrow L+1\right\rangle $$+c.c.$ $(0.5903)$\selectlanguage{american}%
\tabularnewline
\hline 
\end{tabular}\selectlanguage{english}%
\end{table}
\begin{table}[H]
\protect\caption{\label{wave_analysis_26atoms_std_PQD}Excited states giving rise to
peaks in the singlet linear absorption spectrum of fulminene, computed
employing the MRSDCI approach along with the standard parameters in
the PPP model Hamiltonian. }

\selectlanguage{american}%
\begin{tabular}{ccccc}
\hline 
\selectlanguage{english}%
Peak\selectlanguage{american}%
 & \selectlanguage{english}%
State \selectlanguage{american}%
 & \selectlanguage{english}%
E ($eV$)\selectlanguage{american}%
 & \selectlanguage{english}%
Transition \selectlanguage{american}%
 & \selectlanguage{english}%
Dominant Contributing \selectlanguage{american}%
\tabularnewline
\selectlanguage{english}%
\selectlanguage{american}%
 & \selectlanguage{english}%
\selectlanguage{american}%
 & \selectlanguage{english}%
\selectlanguage{american}%
 & \selectlanguage{english}%
Dipole (\AA)\selectlanguage{american}%
 & \selectlanguage{english}%
Configurations\selectlanguage{american}%
\tabularnewline
\hline 
D.F. & $1$$^{1}B_{u}$ & \selectlanguage{english}%
3.07\selectlanguage{american}%
 & 0 & \selectlanguage{english}%
$\left|H\rightarrow L+1\right\rangle $$+c.c.$$(0.5583)$\selectlanguage{american}%
\tabularnewline
\selectlanguage{english}%
\selectlanguage{american}%
 & \selectlanguage{english}%
\selectlanguage{american}%
 & \selectlanguage{english}%
\selectlanguage{american}%
 & \selectlanguage{english}%
\selectlanguage{american}%
 & \selectlanguage{english}%
$\left|H-3\rightarrow L+2\right\rangle $$-c.c.$$(0.1933)$\selectlanguage{american}%
\tabularnewline
$I_{xy}$ & $2$$^{1}B_{u}$ & $3.52$ & $1.3412$ & \selectlanguage{english}%
$\left|H\rightarrow L\right\rangle $ $(0.8026)$\selectlanguage{american}%
\tabularnewline
\selectlanguage{english}%
\selectlanguage{american}%
 & \selectlanguage{english}%
\selectlanguage{american}%
 & \selectlanguage{english}%
\selectlanguage{american}%
 & \selectlanguage{english}%
\selectlanguage{american}%
 & \selectlanguage{english}%
$\left|H-1\rightarrow L+1\right\rangle $ $(0.2694)$\selectlanguage{american}%
\tabularnewline
$II_{xy}$ & $5$$^{1}B_{u}$ & $4.54$ & $2.0679$ & \selectlanguage{english}%
$\left|H-1\rightarrow L+1\right\rangle $ $(0.6036)$\selectlanguage{american}%
\tabularnewline
\hline 
\end{tabular}\selectlanguage{english}%
\end{table}
\begin{table}[H]
\protect\caption{\label{wave_analysis_30atoms_scr_PQD}Excited states giving rise to
peaks in the singlet linear absorption spectrum of {[}7{]}phenacene,
computed employing the MRSDCI approach along with the screened parameters
in the PPP model Hamiltonian. }

\selectlanguage{american}%
\begin{tabular}{ccccc}
\hline 
\selectlanguage{english}%
Peak\selectlanguage{american}%
 & \selectlanguage{english}%
State \selectlanguage{american}%
 & \selectlanguage{english}%
E ($eV$)\selectlanguage{american}%
 & \selectlanguage{english}%
Transition \selectlanguage{american}%
 & \selectlanguage{english}%
Dominant Contributing \selectlanguage{american}%
\tabularnewline
\selectlanguage{english}%
\selectlanguage{american}%
 & \selectlanguage{english}%
\selectlanguage{american}%
 & \selectlanguage{english}%
\selectlanguage{american}%
 & \selectlanguage{english}%
Dipole (\AA)\selectlanguage{american}%
 & \selectlanguage{english}%
Configurations\selectlanguage{american}%
\tabularnewline
\hline 
\selectlanguage{english}%
D.F.\selectlanguage{american}%
 & \selectlanguage{english}%
$2^{1}A_{1}$\selectlanguage{american}%
 & \selectlanguage{english}%
2.96\selectlanguage{american}%
 & \selectlanguage{english}%
0\selectlanguage{american}%
 & \selectlanguage{english}%
$\left|H\rightarrow L+1\right\rangle $$+c.c.$ (0.4042)\selectlanguage{american}%
\tabularnewline
\selectlanguage{english}%
\selectlanguage{american}%
 & \selectlanguage{english}%
\selectlanguage{american}%
 & \selectlanguage{english}%
\selectlanguage{american}%
 & \selectlanguage{english}%
\selectlanguage{american}%
 & \selectlanguage{english}%
$\left|H\rightarrow L+2\right\rangle $$-c.c.$(0.3904)\selectlanguage{american}%
\tabularnewline
$I_{x}$ & $1$$^{1}B_{1}$ & $3.34$ & $2.1261$ & \selectlanguage{english}%
$\left|H\rightarrow L\right\rangle $ $(0.8681)$\selectlanguage{american}%
\tabularnewline
\selectlanguage{english}%
\selectlanguage{american}%
 & \selectlanguage{english}%
\selectlanguage{american}%
 & \selectlanguage{english}%
\selectlanguage{american}%
 & \selectlanguage{english}%
\selectlanguage{american}%
 & \selectlanguage{english}%
$\left|H-2\rightarrow L+1\right\rangle $ $-c.c.$$(0.0451)$\selectlanguage{american}%
\tabularnewline
\hline 
\end{tabular}\selectlanguage{english}%
\end{table}
\begin{table}[H]
\protect\caption{\label{wave_analysis_30atoms_std_PQD}Excited states giving rise to
peaks in the singlet linear absorption spectrum of {[}7{]}phenacene,
computed employing the MRSDCI approach along with the standard parameters
in the PPP model Hamiltonian. }

\selectlanguage{american}%
\begin{tabular}{ccccc}
\hline 
\selectlanguage{english}%
Peak\selectlanguage{american}%
 & \selectlanguage{english}%
State \selectlanguage{american}%
 & \selectlanguage{english}%
E ($eV$)\selectlanguage{american}%
 & \selectlanguage{english}%
Transition \selectlanguage{american}%
 & \selectlanguage{english}%
Dominant Contributing \selectlanguage{american}%
\tabularnewline
\selectlanguage{english}%
\selectlanguage{american}%
 & \selectlanguage{english}%
\selectlanguage{american}%
 & \selectlanguage{english}%
\selectlanguage{american}%
 & \selectlanguage{english}%
Dipole (\AA)\selectlanguage{american}%
 & \selectlanguage{english}%
Configurations\selectlanguage{american}%
\tabularnewline
\hline 
\selectlanguage{english}%
D.F.\selectlanguage{american}%
 & \selectlanguage{english}%
$2^{1}A_{1}$\selectlanguage{american}%
 & \selectlanguage{english}%
3.31\selectlanguage{american}%
 & \selectlanguage{english}%
0\selectlanguage{american}%
 & \selectlanguage{english}%
$\left|H\rightarrow L+1\right\rangle $$-c.c.$ (0.4320)\selectlanguage{american}%
\tabularnewline
\selectlanguage{english}%
\selectlanguage{american}%
 & \selectlanguage{english}%
\selectlanguage{american}%
 & \selectlanguage{english}%
\selectlanguage{american}%
 & \selectlanguage{english}%
\selectlanguage{american}%
 & \selectlanguage{english}%
$\left|H\rightarrow L+2\right\rangle $$-c.c.$(0.3321)\selectlanguage{american}%
\tabularnewline
$I_{x}$ & $1$$^{1}B_{1}$ & $3.68$ & $1.6182$ & \selectlanguage{english}%
$\left|H\rightarrow L\right\rangle $ $(0.7928)$\selectlanguage{american}%
\tabularnewline
\selectlanguage{english}%
\selectlanguage{american}%
 & \selectlanguage{english}%
\selectlanguage{american}%
 & \selectlanguage{english}%
\selectlanguage{american}%
 & \selectlanguage{english}%
\selectlanguage{american}%
 & \selectlanguage{english}%
$\left|H-1\rightarrow L+1\right\rangle $ $(0.2598)$\selectlanguage{american}%
\tabularnewline
\hline 
\end{tabular}\selectlanguage{english}%
\end{table}
\begin{table}[H]
\protect\caption{\label{wave_analysis_34atoms_scr_PQD}Excited states giving rise to
peaks in the singlet linear absorption spectrum of {[}8{]}phenacene,
computed employing the MRSDCI approach along with the screened parameters
in the PPP model Hamiltonian. }

\selectlanguage{american}%
\begin{tabular}{ccccc}
\hline 
\selectlanguage{english}%
Peak\selectlanguage{american}%
 & \selectlanguage{english}%
State \selectlanguage{american}%
 & \selectlanguage{english}%
E ($eV$)\selectlanguage{american}%
 & \selectlanguage{english}%
Transition \selectlanguage{american}%
 & \selectlanguage{english}%
Dominant Contributing \selectlanguage{american}%
\tabularnewline
\selectlanguage{english}%
\selectlanguage{american}%
 & \selectlanguage{english}%
\selectlanguage{american}%
 & \selectlanguage{english}%
\selectlanguage{american}%
 & \selectlanguage{english}%
Dipole (\AA)\selectlanguage{american}%
 & \selectlanguage{english}%
Configurations\selectlanguage{american}%
\tabularnewline
\hline 
D.F. & $1$$^{1}B_{u}$ & \selectlanguage{english}%
2.84\selectlanguage{american}%
 & 0 & \selectlanguage{english}%
$\left|H-2\rightarrow L\right\rangle $$+c.c.$$(0.5513)$\selectlanguage{american}%
\tabularnewline
\selectlanguage{english}%
\selectlanguage{american}%
 & \selectlanguage{english}%
\selectlanguage{american}%
 & \selectlanguage{english}%
\selectlanguage{american}%
 & \selectlanguage{english}%
\selectlanguage{american}%
 & \selectlanguage{english}%
$\left|H-1\rightarrow L+4\right\rangle $$+c.c.$$(0.1647)$\selectlanguage{american}%
\tabularnewline
$I_{xy}$ & $2$$^{1}B_{u}$ & $3.11$ & $2.3654$ & \selectlanguage{english}%
$\left|H\rightarrow L\right\rangle $ $(0.8595)$\selectlanguage{american}%
\tabularnewline
\selectlanguage{english}%
\selectlanguage{american}%
 & \selectlanguage{english}%
\selectlanguage{american}%
 & \selectlanguage{english}%
\selectlanguage{american}%
 & \selectlanguage{english}%
\selectlanguage{american}%
 & \selectlanguage{english}%
$\left|H-1\rightarrow L+1\right\rangle $$(0.0786)$\selectlanguage{american}%
\tabularnewline
$II_{xy}$ & $5$$^{1}B_{u}$ & $3.87$ & $1.7035$ & \selectlanguage{english}%
$\left|H-2\rightarrow L\right\rangle $$-c.c.$ $(0.5887)$\selectlanguage{american}%
\tabularnewline
\hline 
\end{tabular}\selectlanguage{english}%
\end{table}
\begin{table}[H]
\protect\caption{\label{wave_analysis_34atoms_std_PQD}Excited states giving rise to
peaks in the singlet linear absorption spectrum of {[}8{]}phenacene,
computed employing the MRSDCI approach along with the standard parameters
in the PPP model Hamiltonian. }

\selectlanguage{american}%
\begin{tabular}{ccccc}
\hline 
\selectlanguage{english}%
Peak\selectlanguage{american}%
 & \selectlanguage{english}%
State \selectlanguage{american}%
 & \selectlanguage{english}%
E ($eV$)\selectlanguage{american}%
 & \selectlanguage{english}%
Transition \selectlanguage{american}%
 & \selectlanguage{english}%
Dominant Contributing \selectlanguage{american}%
\tabularnewline
\selectlanguage{english}%
\selectlanguage{american}%
 & \selectlanguage{english}%
\selectlanguage{american}%
 & \selectlanguage{english}%
\selectlanguage{american}%
 & \selectlanguage{english}%
Dipole (\AA)\selectlanguage{american}%
 & \selectlanguage{english}%
Configurations\selectlanguage{american}%
\tabularnewline
\hline 
D.F. & $1$$^{1}B_{u}$ & \selectlanguage{english}%
3.13\selectlanguage{american}%
 & 0 & \selectlanguage{english}%
$\left|H-2\rightarrow L\right\rangle $$+c.c.$$(0.5307)$\selectlanguage{american}%
\tabularnewline
\selectlanguage{english}%
\selectlanguage{american}%
 & \selectlanguage{english}%
\selectlanguage{american}%
 & \selectlanguage{english}%
\selectlanguage{american}%
 & \selectlanguage{english}%
\selectlanguage{american}%
 & \selectlanguage{english}%
$\left|H-1\rightarrow L+4\right\rangle $$-c.c.$$(0.2166)$\selectlanguage{american}%
\tabularnewline
$I_{xy}$ & $2$$^{1}B_{u}$ & $3.41$ & $1.8097$ & \selectlanguage{english}%
$\left|H\rightarrow L\right\rangle $ $(0.7761)$\selectlanguage{american}%
\tabularnewline
\selectlanguage{english}%
\selectlanguage{american}%
 & \selectlanguage{english}%
\selectlanguage{american}%
 & \selectlanguage{english}%
\selectlanguage{american}%
 & \selectlanguage{english}%
\selectlanguage{american}%
 & \selectlanguage{english}%
$\left|H-1\rightarrow L+1\right\rangle $ $(0.2679)$\selectlanguage{american}%
\tabularnewline
$II_{xy}$ & $6$$^{1}B_{u}$ & $4.48$ & $2.1087$ & \selectlanguage{english}%
$\left|H-2\rightarrow L+2\right\rangle $ $(0.5421)$\selectlanguage{american}%
\tabularnewline
\hline 
\end{tabular}\selectlanguage{english}%
\end{table}
\begin{table}[H]
\protect\caption{\label{wave_analysis_38atoms_scr_PQD}Excited states giving rise to
peaks in the singlet linear absorption spectrum of {[}9{]}phenacene,
computed employing the MRSDCI approach along with the screened parameters
in the PPP model Hamiltonian. }

\selectlanguage{american}%
\begin{tabular}{ccccc}
\hline 
\selectlanguage{english}%
Peak\selectlanguage{american}%
 & \selectlanguage{english}%
State \selectlanguage{american}%
 & \selectlanguage{english}%
E ($eV$)\selectlanguage{american}%
 & \selectlanguage{english}%
Transition \selectlanguage{american}%
 & \selectlanguage{english}%
Dominant Contributing \selectlanguage{american}%
\tabularnewline
\selectlanguage{english}%
\selectlanguage{american}%
 & \selectlanguage{english}%
\selectlanguage{american}%
 & \selectlanguage{english}%
\selectlanguage{american}%
 & \selectlanguage{english}%
Dipole (\AA)\selectlanguage{american}%
 & \selectlanguage{english}%
Configurations\selectlanguage{american}%
\tabularnewline
\hline 
\selectlanguage{english}%
D.F.\selectlanguage{american}%
 & \selectlanguage{english}%
$2^{1}A_{1}$\selectlanguage{american}%
 & \selectlanguage{english}%
2.92\selectlanguage{american}%
 & \selectlanguage{english}%
0\selectlanguage{american}%
 & \selectlanguage{english}%
$\left|H-2\rightarrow L\right\rangle $$+c.c.$ (0.5012) \selectlanguage{american}%
\tabularnewline
\selectlanguage{english}%
\selectlanguage{american}%
 & \selectlanguage{english}%
\selectlanguage{american}%
 & \selectlanguage{english}%
\selectlanguage{american}%
 & \selectlanguage{english}%
\selectlanguage{american}%
 & \selectlanguage{english}%
$\left|H-1\rightarrow L\right\rangle $$-c.c.$(0.2363)\selectlanguage{american}%
\tabularnewline
$I_{x}$ & $1$$^{1}B_{1}$ & $3.09$ & $2.5945$ & \selectlanguage{english}%
$\left|H\rightarrow L\right\rangle $ $(0.8565)$\selectlanguage{american}%
\tabularnewline
\selectlanguage{english}%
\selectlanguage{american}%
 & \selectlanguage{english}%
\selectlanguage{american}%
 & \selectlanguage{english}%
\selectlanguage{american}%
 & \selectlanguage{english}%
\selectlanguage{american}%
 & \selectlanguage{english}%
$\left|H-1\rightarrow L+1\right\rangle $ $(0.1066)$\selectlanguage{american}%
\tabularnewline
\hline 
\end{tabular}\selectlanguage{english}%
\end{table}
\begin{table}[H]
\protect\caption{\label{wave_analysis_38atoms_std_PQD}Excited states giving rise to
peaks in the singlet linear absorption spectrum of {[}9{]}phenacene,
computed employing the MRSDCI approach along with the standard parameters
in the PPP model Hamiltonian. }

\selectlanguage{american}%
\begin{tabular}{ccccc}
\hline 
\selectlanguage{english}%
Peak\selectlanguage{american}%
 & \selectlanguage{english}%
State \selectlanguage{american}%
 & \selectlanguage{english}%
E ($eV$)\selectlanguage{american}%
 & \selectlanguage{english}%
Transition \selectlanguage{american}%
 & \selectlanguage{english}%
Dominant Contributing \selectlanguage{american}%
\tabularnewline
\selectlanguage{english}%
\selectlanguage{american}%
 & \selectlanguage{english}%
\selectlanguage{american}%
 & \selectlanguage{english}%
\selectlanguage{american}%
 & \selectlanguage{english}%
Dipole (\AA)\selectlanguage{american}%
 & \selectlanguage{english}%
Configurations\selectlanguage{american}%
\tabularnewline
\hline 
\selectlanguage{english}%
D.F.\selectlanguage{american}%
 & \selectlanguage{english}%
$2^{1}A_{1}$\selectlanguage{american}%
 & \selectlanguage{english}%
3.29\selectlanguage{american}%
 & \selectlanguage{english}%
0\selectlanguage{american}%
 & \selectlanguage{english}%
$\left|H-2\rightarrow L\right\rangle $$+c.c.$ (0.4580)\selectlanguage{american}%
\tabularnewline
\selectlanguage{english}%
\selectlanguage{american}%
 & \selectlanguage{english}%
\selectlanguage{american}%
 & \selectlanguage{english}%
\selectlanguage{american}%
 & \selectlanguage{english}%
\selectlanguage{american}%
 & \selectlanguage{english}%
$\left|H-1\rightarrow L\right\rangle $$+c.c.$(0.2550)\selectlanguage{american}%
\tabularnewline
$I_{x}$ & $1$$^{1}B_{1}$ & $3.46$ & $2.0333$ & \selectlanguage{english}%
$\left|H\rightarrow L\right\rangle $ $(0.7612)$\selectlanguage{american}%
\tabularnewline
\selectlanguage{english}%
\selectlanguage{american}%
 & \selectlanguage{english}%
\selectlanguage{american}%
 & \selectlanguage{english}%
\selectlanguage{american}%
 & \selectlanguage{english}%
\selectlanguage{american}%
 & \selectlanguage{english}%
$\left|H-1\rightarrow L+1\right\rangle $ $(0.2845)$\selectlanguage{american}%
\tabularnewline
\hline 
\end{tabular}\selectlanguage{english}%
\end{table}